\documentstyle[preprint,aps]{revtex}
\tightenlines
\input{psfig}

\def\ET{$E_T$\ }
\def\MET{$\not\!\!E_T$\ }

\def\METSIG{$\tilde{\not\!\!E_T}$\ }
\def\METS{$\not\!\!E_T/\sqrt{\sum E_T}$\ }

\def\MRSD0{MRSD0$^\prime$ }

\def\RSep{$R_{sep}$\ }
\begin{document}

\draft

\title{Measurement of the Inclusive Jet Cross Section in ${\bar p p}$ 
collisions at
$\sqrt{s}=1.8$ TeV}

\date{\today}

\maketitle

\font\eightit=cmti8
\def\r#1{\ignorespaces $^{#1}$}
\hfilneg
\begin{sloppypar}
\noindent
T.~Affolder,\r {23} H.~Akimoto,\r {45}
A.~Akopian,\r {38} M.~G.~Albrow,\r {11} P.~Amaral,\r 8 S.~R.~Amendolia,\r {34} 
D.~Amidei,\r {26} K.~Anikeev,\r {24} J.~Antos,\r 1 
G.~Apollinari,\r {11}
% ...
\end{sloppypar}
\vskip .026in
\begin{center}
(CDF Collaboration)
\end{center}

\vskip .026in
\begin{center}
\r 1  {\eightit Institute of Physics, Academia Sinica, Taipei, Taiwan 11529, 
Republic of China} \\
\r 2  {\eightit Argonne National Laboratory, Argonne, Illinois 60439} \\
\r 3  {\eightit Istituto Nazionale di Fisica Nucleare, University of Bologna,
I-40127 Bologna, Italy} \\
\r 4  {\eightit Brandeis University, Waltham, Massachusetts 02254} \\
\r 5  {\eightit University of California at Davis, Davis, California  95616} \\
\r 6  {\eightit University of California at Los Angeles, Los 
Angeles, California  90024} \\  
\r 7  {\eightit Instituto de Fisica de Cantabria, CSIC-University of Cantabria, 
39005 Santander, Spain} \\
\r 8  {\eightit Enrico Fermi Institute, University of Chicago, Chicago, 
Illinois 60637} \\
\r 9  {\eightit Joint Institute for Nuclear Research, RU-141980 Dubna, Russia}
\\
\r {10} {\eightit Duke University, Durham, North Carolina  27708} \\
\r {11} {\eightit Fermi National Accelerator Laboratory, Batavia, Illinois 
60510} \\
\r {12} {\eightit University of Florida, Gainesville, Florida  32611} \\
\r {13} {\eightit Laboratori Nazionali di Frascati, Istituto Nazionale di Fisica
               Nucleare, I-00044 Frascati, Italy} \\
\r {14} {\eightit University of Geneva, CH-1211 Geneva 4, Switzerland} \\
\r {15} {\eightit Glasgow University, Glasgow G12 8QQ, United Kingdom}\\
\r {16} {\eightit Harvard University, Cambridge, Massachusetts 02138} \\
\r {17} {\eightit Hiroshima University, Higashi-Hiroshima 724, Japan} \\
\r {18} {\eightit University of Illinois, Urbana, Illinois 61801} \\
\r {19} {\eightit The Johns Hopkins University, Baltimore, Maryland 21218} \\
\r {20} {\eightit Institut f\"{u}r Experimentelle Kernphysik, 
Universit\"{a}t Karlsruhe, 76128 Karlsruhe, Germany} \\
\r {21} {\eightit Center for High Energy Physics: Kyungpook National
University, Taegu 702-701; Seoul National University, Seoul 151-742; and
SungKyunKwan University, Suwon 440-746; Korea} \\
\r {22} {\eightit High Energy Accelerator Research Organization (KEK), Tsukuba, 
Ibaraki 305, Japan} \\
\r {23} {\eightit Ernest Orlando Lawrence Berkeley National Laboratory, 
Berkeley, California 94720} \\
\r {24} {\eightit Massachusetts Institute of Technology, Cambridge,
Massachusetts  02139} \\   
\r {25} {\eightit Institute of Particle Physics: McGill University, Montreal 
H3A 2T8; and University of Toronto, Toronto M5S 1A7; Canada} \\
\r {26} {\eightit University of Michigan, Ann Arbor, Michigan 48109} \\
\r {27} {\eightit Michigan State University, East Lansing, Michigan  48824} \\
\r {28} {\eightit University of New Mexico, Albuquerque, New Mexico 87131} \\
\r {29} {\eightit The Ohio State University, Columbus, Ohio  43210} \\
\r {30} {\eightit Osaka City University, Osaka 588, Japan} \\
\r {31} {\eightit University of Oxford, Oxford OX1 3RH, United Kingdom} \\
\r {32} {\eightit Universita di Padova, Istituto Nazionale di Fisica 
          Nucleare, Sezione di Padova, I-35131 Padova, Italy} \\
\r {33} {\eightit University of Pennsylvania, Philadelphia, 
        Pennsylvania 19104} \\   
\r {34} {\eightit Istituto Nazionale di Fisica Nucleare, University and Scuola
               Normale Superiore of Pisa, I-56100 Pisa, Italy} \\
\r {35} {\eightit University of Pittsburgh, Pittsburgh, Pennsylvania 15260} \\
\r {36} {\eightit Purdue University, West Lafayette, Indiana 47907} \\
\r {37} {\eightit University of Rochester, Rochester, New York 14627} \\
\r {38} {\eightit Rockefeller University, New York, New York 10021} \\
\r {39} {\eightit Rutgers University, Piscataway, New Jersey 08855} \\
\r {40} {\eightit Texas A\&M University, College Station, Texas 77843} \\
\r {41} {\eightit Texas Tech University, Lubbock, Texas 79409} \\
\r {42} {\eightit Istituto Nazionale di Fisica Nucleare, University of Trieste/
Udine, Italy} \\
\r {43} {\eightit University of Tsukuba, Tsukuba, Ibaraki 305, Japan} \\
\r {44} {\eightit Tufts University, Medford, Massachusetts 02155} \\
\r {45} {\eightit Waseda University, Tokyo 169, Japan} \\
\r {46} {\eightit University of Wisconsin, Madison, Wisconsin 53706} \\
\r {47} {\eightit Yale University, New Haven, Connecticut 06520} \\
\r {(\ast)} {\eightit Now at Carnegie Mellon University, Pittsburgh,
Pennsylvania  15213} \\
\r {(\ast\ast)} {\eightit Now at Northwestern University, Evanston, Illinois 
60208}
\end{center}

\newpage

\begin{abstract}
We present results from the measurement of the inclusive jet cross section 
for jet transverse energies from 40 to 465 GeV in the 
pseudo-rapidity range $0.1<|\eta|<0.7$.
The results are based on 87 $pb^{-1}$
of data collected by the CDF collaboration at 
the Fermilab Tevatron Collider.
The data are consistent with previously published results.
The data are also consistent with QCD predictions given the flexibility allowed
from current knowledge of the proton parton distributions.
We develop a new procedure for ranking the agreement of the parton 
distributions with data and find that
the data are best described by QCD predictions using the 
parton distribution functions which have a large gluon contribution
at high $E_T$ (CTEQ4HJ).
\end{abstract}





\section{Introduction}
\label{sec-inclusive jet introduction}

Measurement of the inclusive jet cross section 
is a fundamental test of QCD predictions. 
The Fermilab $p\bar{p}$ collider, with $\sqrt{s}$ = 1.8 TeV,
provides the highest energy collisions of any accelerator and the energies
of the resulting jets cover the widest range of any experiment.
Comparison of the inclusive jet cross section to predictions provides 
information about
parton distribution functions (PDF's) and the strong coupling constant,
$\alpha_s$, for jet energies from 40 - 465 GeV where the jet cross section 
changes by 10 orders of magnitude.
At the highest jet $E_T$, this measurement 
probes a distance scale of the order of $10^{-17}$~cm and has 
traditionally been used to search for new physics. 

In this paper we present a new 
measurement of the inclusive differential
cross-section for jet production at
$\sqrt{s}$ = 1.8 TeV with the CDF detector\cite{CDF-Detector}.  
Our previous measurement of the
inclusive cross section\cite{CDF-Inclusive-Jet-96} using the Run 
1A data sample,
(19.5 pb$^{-1}$ collected during 1992-1993), showed a significant 
excess of the data over the available
theoretical predictions at high $E_T$. 
With substantially smaller data samples,
measurements \cite{{CDF-Inclusive-Jet-92},{CDF-Inclusive-Jet-87}}
of the
inclusive jet cross section prior to the Run 1A result
found good agreement with QCD predictions and provided the best 
limits on quark compositeness~\cite{Eichten}.  
The Run 1A result motivated a reevaluation of the theoretical 
uncertainties from
the PDF's~\cite{cteq-jhuston,walter-98} and 
the derivation of a new PDF which specifically gave higher weight to 
the high $E_T$ CDF data points~\cite{cteq-4}.  
The measurement presented in this report uses the 
87 pb$^{-1}$\cite{CDFLUM} Run 1B data sample (1994-1995)
which is more than 4.5 times larger than for our 
previous result\cite{CDF-Inclusive-Jet-96}.
Comparisons are made to improved theoretical predictions 
and to the results of the D0 Collaboration\cite{D0inc}.

The paper is organized as follows: 
Section II
provides a discussion of the components of the theoretical
predictions and a historical review of
previous jet measurements.
Sections III and IV describe the CDF detector
and the data sample selection respectively. In Section V the 
energy calibration and corrections to the data are presented.
A discussion of the systematic uncertainties follows in 
Section VI.
Section VII describes comparison of this data to previous results.
Section VIII presents quantitative estimates of the theoretical uncertainties
and Section IX shows comparisons of the data to the predictions.
The paper is concluded in Section X. 

\section{Inclusive Jet Cross Sections}

The suggestion that high energy hadron collisions would result in two jets
of particles with the same momentum as the scattered partons~\cite{bj1971}
spawned an industry of comparisons between 
experimental measurements and theoretical predictions.
The initial searches at the ISR ($\sqrt{s}$ = 63 GeV),
provided hints of two-jet structure~\cite{ISR-1}.  Extraction 
of a jet signal was difficult because the sharing of the 
hadron momentum between the constituent partons reduced the effective
available parton scattering energy and the 
remnants of the incident hadrons produced
a background of low transverse energy particles. 
The first clear observation of two jet structure came at
a collision energy of $\sqrt{s}$ = 540 GeV at the
CERN $Sp\bar{p}S$ collider~\cite{UA2-1,UA1-1} along
with the first measurements of the
inclusive jet cross section.  An increased data sample and improved triggering
also led to the measurement of the inclusive jet cross section 
at the ISR~\cite{ISR-2}.

\begin{figure}
\centerline{
{\hbox{
\psfig{figure=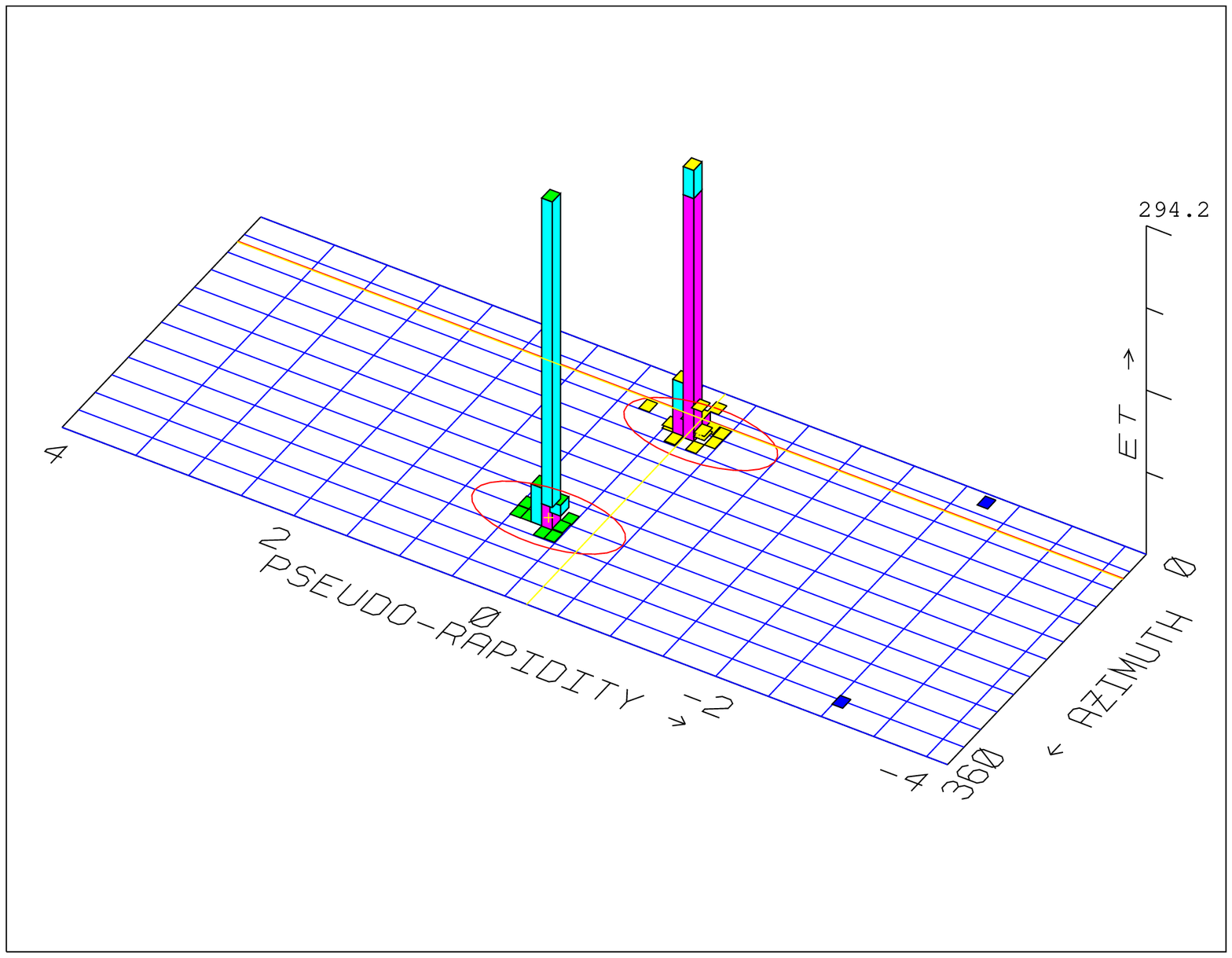,height=7cm,width=7cm}
\psfig{figure=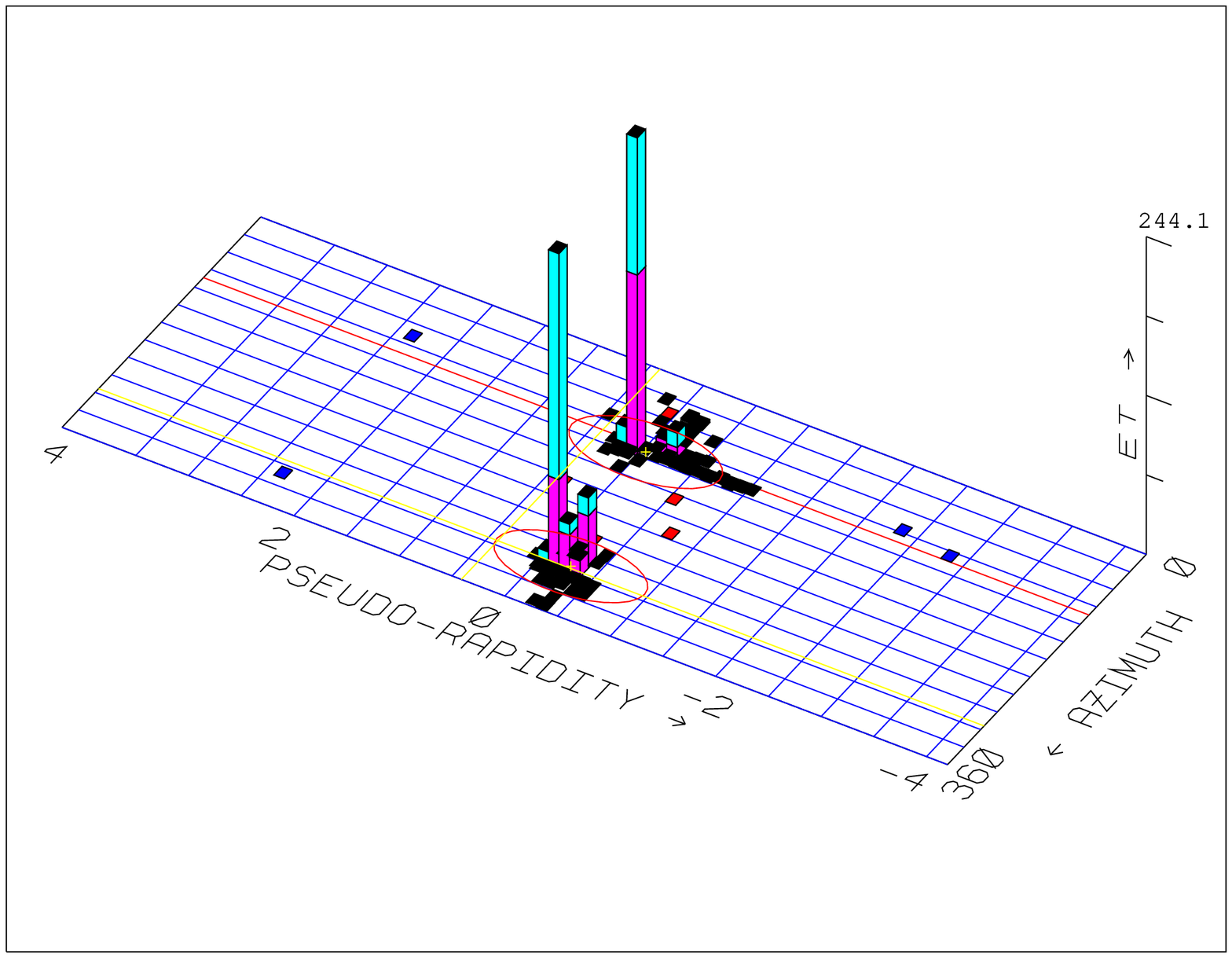,height=7cm,width=7cm}
}}}
\centerline{
{\hbox{
\psfig{figure=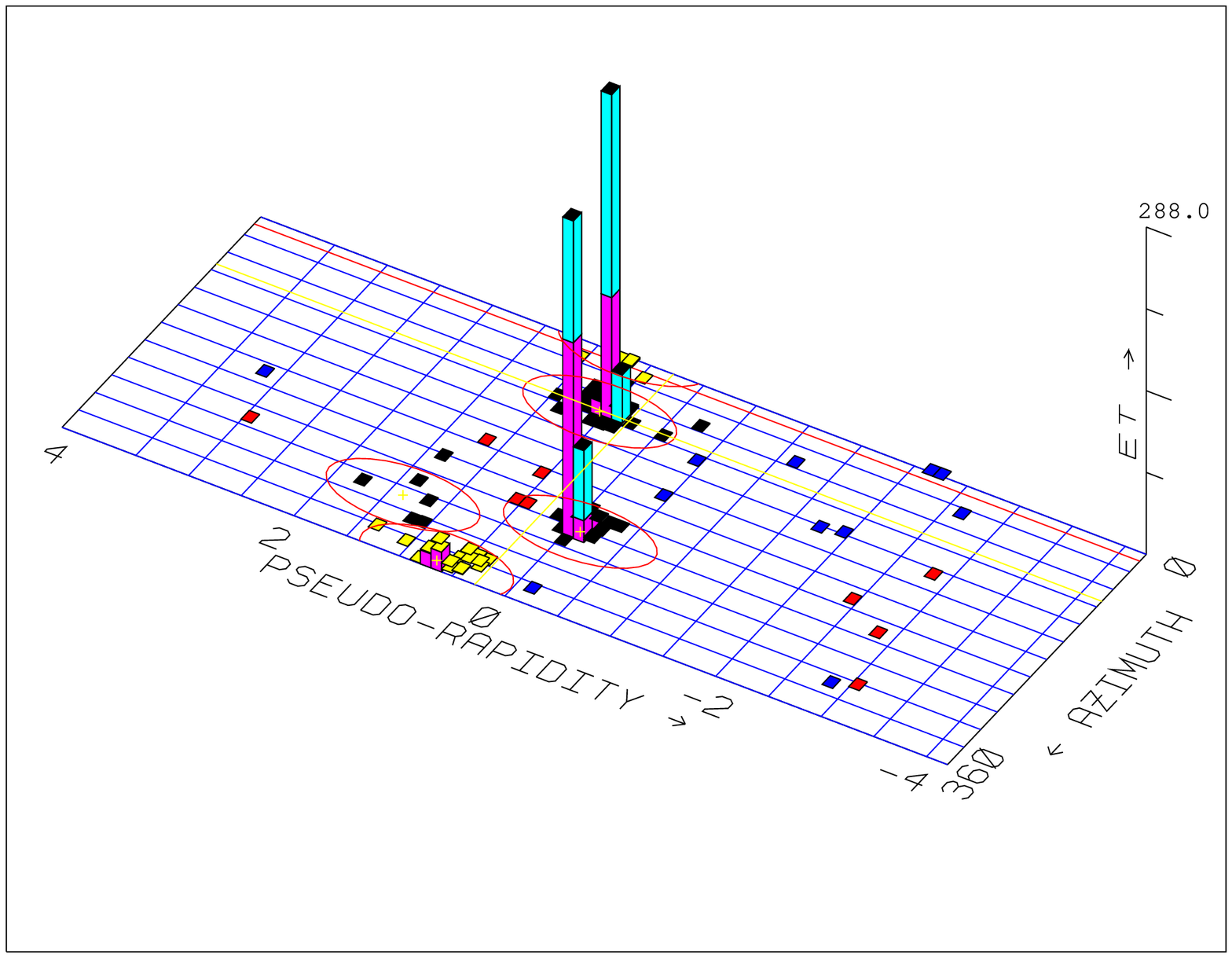,height=7cm,width=7cm}
\psfig{figure=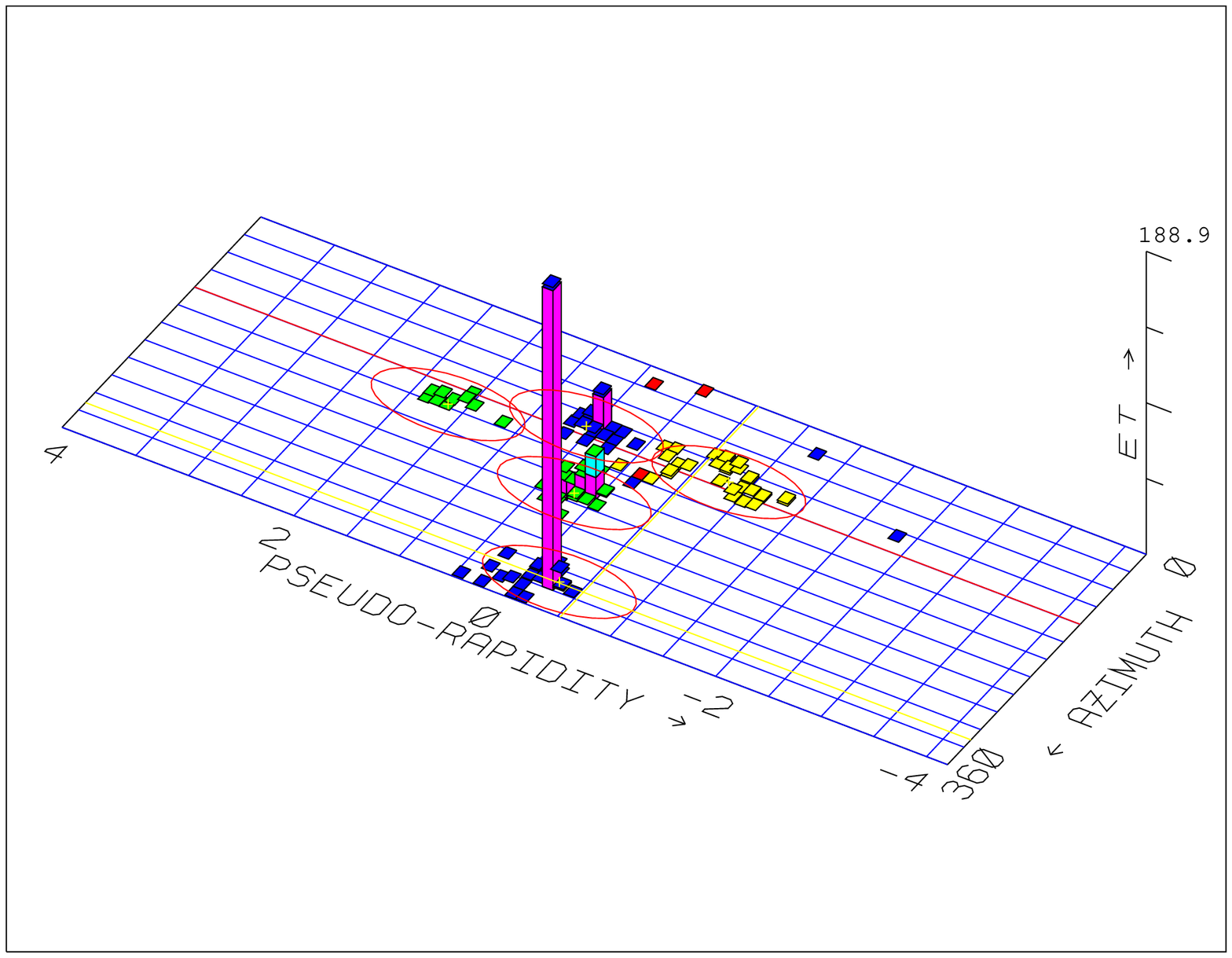,height=7cm,width=7cm}
}}}
\vspace{1cm}
\caption{Jet events in the CDF calorimeter. A jet clustering cone
of radius 0.7 is shown around each jet.  Clockwise from the upper left
they are identified as two-jet, two-jet, five-jet and three-jet.
Tracks for these events
are shown in Figure~\ref{CDFCTC}. }
\label{CDFLEGO}
\end{figure}

\begin{figure}
\centerline{
{\hbox{\hspace{1.0cm}
\psfig{figure=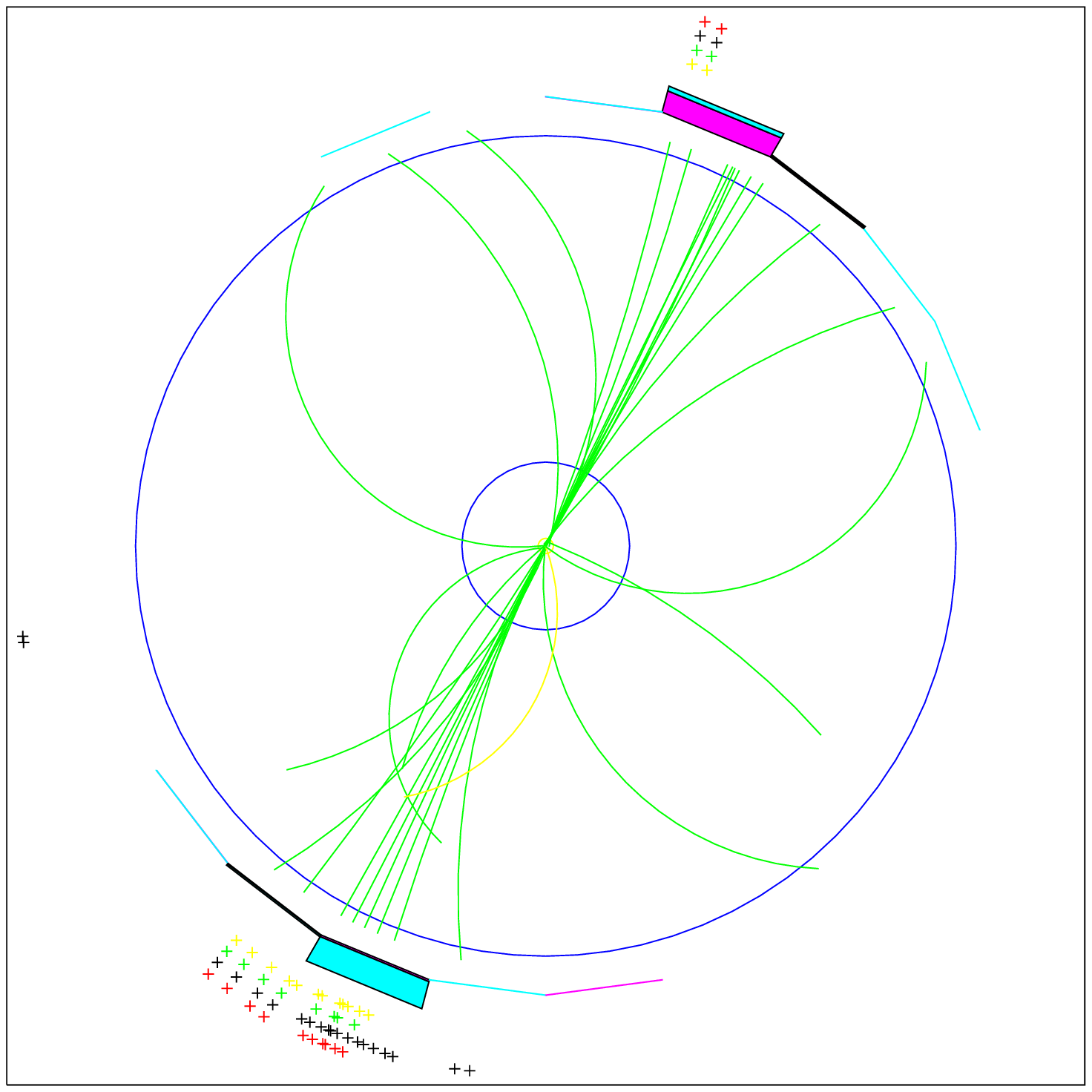,height=7cm,width=8.5cm}
\hspace{-2.0cm}\psfig{figure=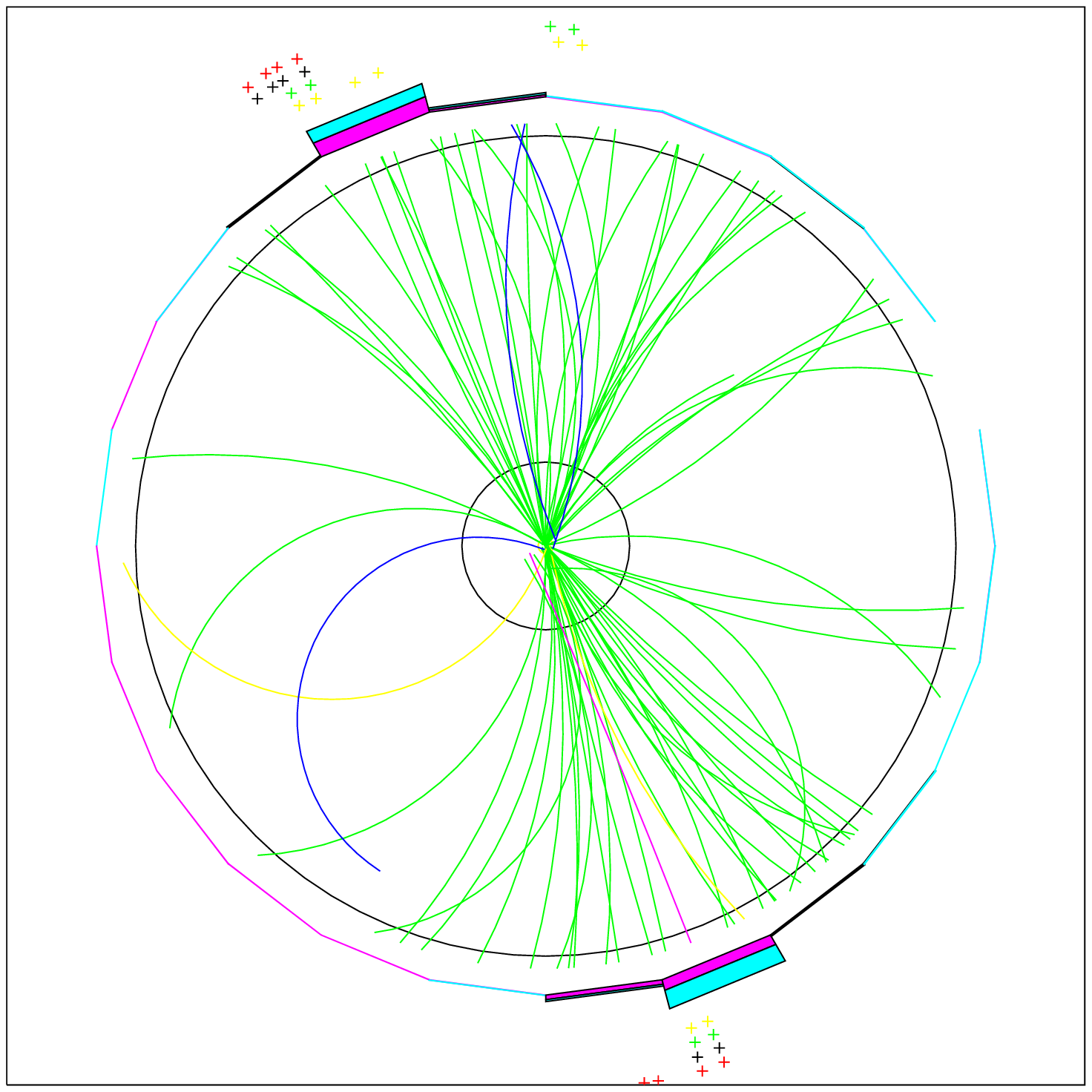,height=7cm,width=8.5cm}
}}}
\centerline{
{\hbox{\hspace{1.0cm}
\psfig{figure=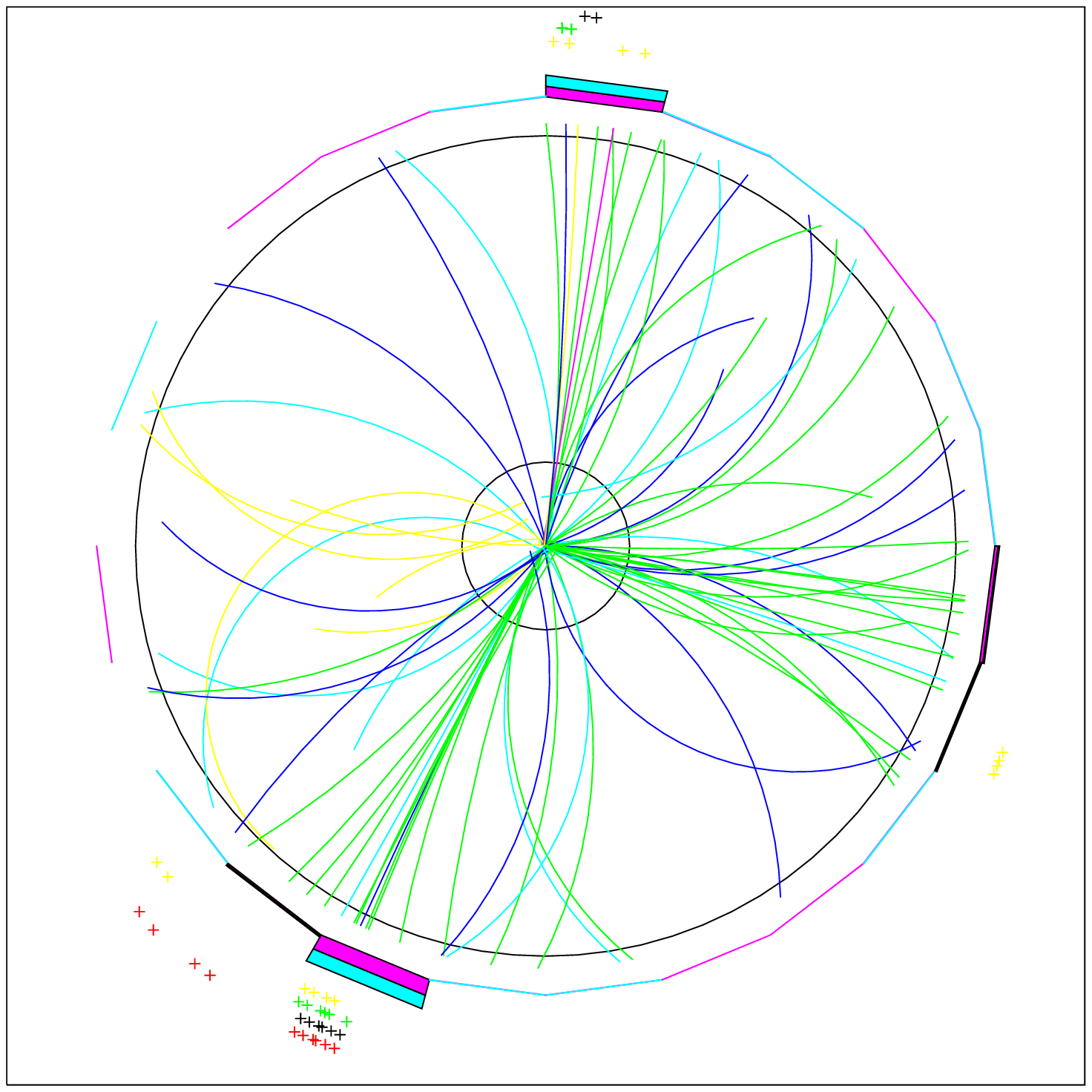,height=7cm,width=8.5cm}
\hspace{-2.0cm}\psfig{figure=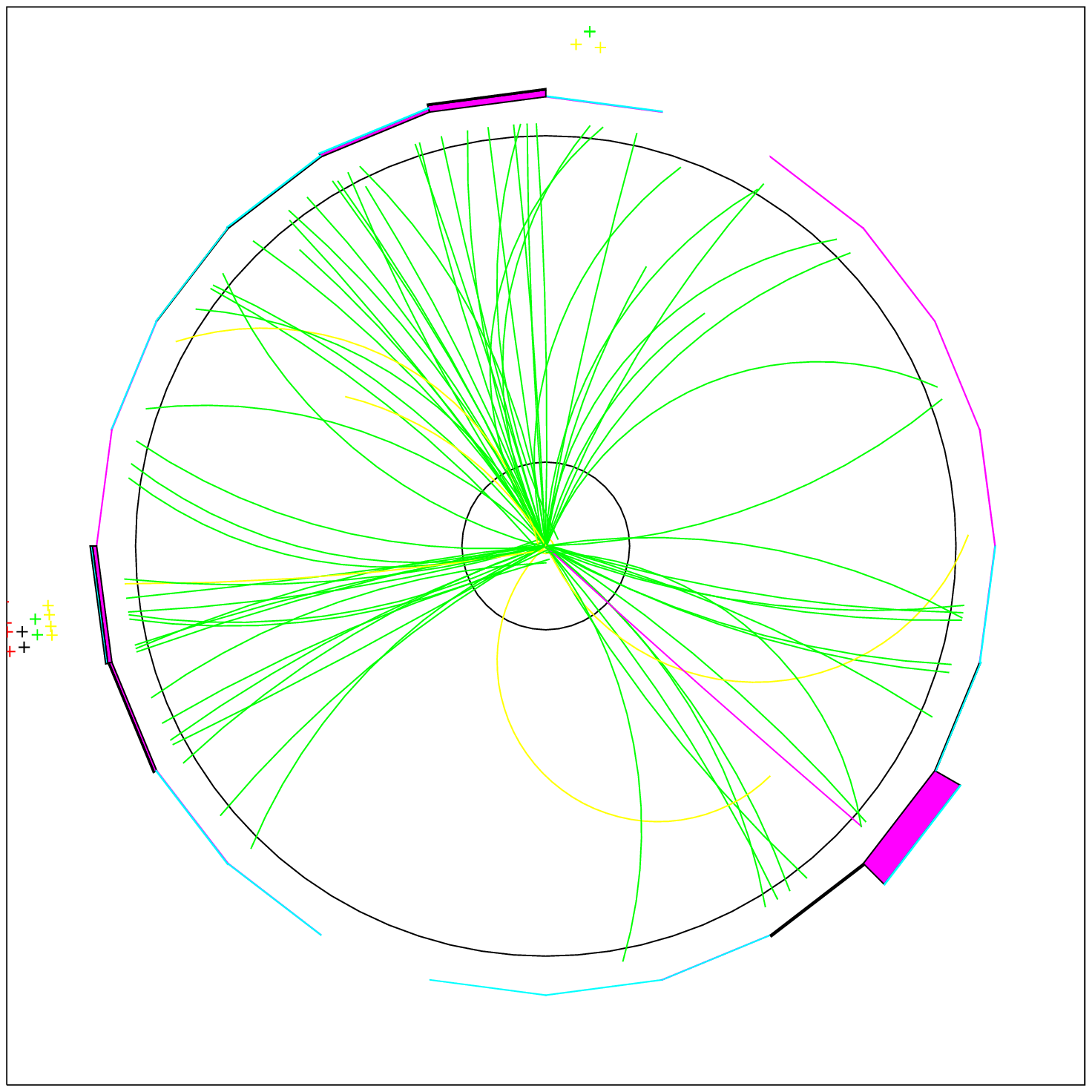,height=7cm,width=8.5cm}
}}}
\vspace{1cm}
\caption{The same jet events in the CDF central tracking chamber.
Clockwise from the upper left
they are identified as two-jet, two-jet, five-jet and three-jet
The calorimeter information for these events is shown in Figure~\ref{CDFLEGO}.
}
\label{CDFCTC}
\end{figure}

Following these early results, improvements in accelerators produced both 
increased sample sizes and increased collision energies.
Higher energy collisions produce jets
of higher energy particles.  This facilitates separation of jet particles from
the remnants of the initial hadrons (called the underlying event)
and reduces the effects of the transverse spreading during 
fragmentation (see for example~\cite{UA2-6,bfthesis}).
Figure~\ref{CDFLEGO} shows some events in the 
CDF calorimeter. In these ``lego'' plots the
calorimeter is ``rolled out'' onto the $\eta$--$\phi$ plane;
$\phi$ is the azimuthal angle around the beam and the pseudo-rapidity
$\eta\equiv-$ln[tan$(\theta/2)$], where $\theta$ is the polar
angle with respect to the incoming proton direction (the $z$-axis).
The tower
height is proportional to the $E_T$ deposited in the tower. The
darker and lighter shading of each tower corresponds to the $E_T$
of the electromagnetic and hadronic cells of the tower
respectively.
The oval around each clump of energy indicates the jet
clustering cone. 
Figure~\ref{CDFCTC} shows the tracks found in the CDF central 
magnetic tracking system for the
same events.
The jet structure in these events is unmistakable.  Note that while the low
and high $E_T$ jets are well contained within the clustering cone, 
the highest $E_T$ jets ($\approx$ 400 GeV) are much narrower than
the 40-60 GeV jets.

As the experimental measurements improved, 
more detailed and precise theoretical predictions
were developed.
When the energy of the collisions increases, the value of the strong
coupling ($\alpha_s$) decreases, improving the validity 
of the perturbative expansion.
At leading order (O($\alpha_s^2$)) one parton from each incoming 
hadron participates in a collision that produces two outgoing partons.
Figures~\ref{CDFLEGO} and~\ref{CDFCTC} clearly show more than 
two jets in some events.  To account for multijet (more than 2) contributions,
leading log Monte Carlo programs were built on the leading 
order tree level predictions by adding parton showers to the
scattered partons. Empirical models for the underlying event were included
along with models for parton fragmentation into hadrons.  
NLO predictions for the inclusive jet cross section emerged in the late 80's
and leading order predictions for multijet events soon followed.
Here we first describe the components of
the theory and then proceed with a discussion of
the development of comparisons between data and theory.
%

\subsection{Theoretical framework}

The cross section for a hard scattering between two incoming hadrons 
(1 + 2 $\rightarrow$ 3 + X) to produce hadronic jets 
can be factorized into components from 
empirically determined PDF's, $f$, 
and perturbatively calculated
two-body scattering cross
sections, $\hat{\sigma}$ . See, for example, reference~\cite{kellisqcd}
for a detailed discussion.  This hadronic cross section is 
written as:
\begin{equation}
\sigma_{1+2 \rightarrow 3 + X} = \sum_{i,j} \int dx_{1}dx_{2}
f_{i}(x_{1},\mu^{2}_{F})f_{j}(x_{2},\mu^{2}_{F})
\hat{\sigma}_{ij} [ x_{1}P, x_{2}P, \alpha_{s}(\mu^{2}_{R})]
\end{equation}

\begin{figure}
\centerline{
\psfig{figure=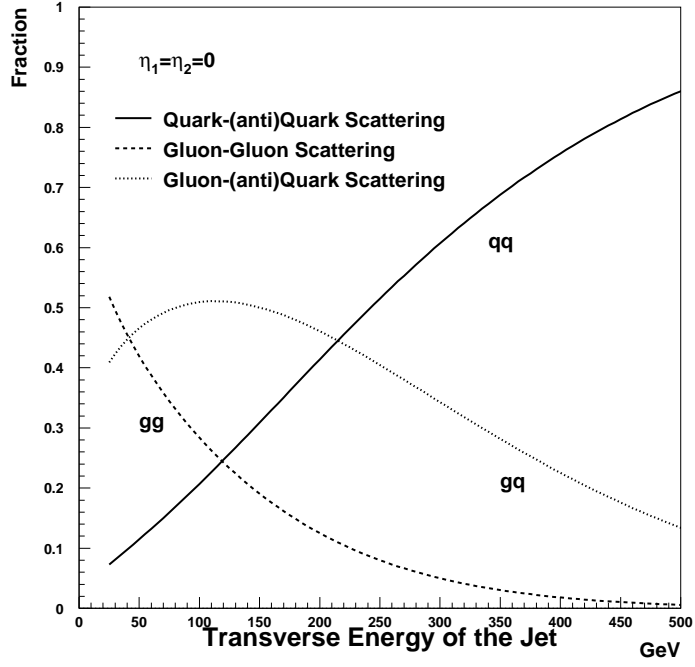,height=10cm,width=10cm}}
\caption{Contributions of the various subprocesses to the inclusive jet 
cross section. This plot was generated with CTEQ4M
and $\mu=E_{T}/2$.
}
\label{fig-subprocesses}
\end{figure}

The PDF's, $f_{i}(x,\mu^{2}_{F})$, describe
the initial parton momentum as a fraction $x$
of the incident hadron momentum $P$ and a function of the 
factorization scale $\mu_{F}$.
The index $i$ refers to the type of parton (gluons or quarks). 
The relative contribution of sub-processes,
based on incoming partons, is shown in Fig.\,\ref{fig-subprocesses} 
for CTEQ4M~\cite{cteq-4} PDF's.
At low $E_{T}$, 
jet production is dominated by $gluon-gluon$ ($GG$) and $gluon-quark$ ($QG$)
scattering. At high $E_T$
it is largely $quark-quark$ ($QQ$) scattering. The $QG$ scattering is about
30\% at $E_T=350$ GeV because of the large color factor
associated with the gluon.
%

One of the essential features of
QCD is that the momentum distributions 
of partons within the proton are universal.
In other words, the PDF's can be
derived from any process 
and applied to other processes.
The PDF's are derived from a global fit to 
scattering experiment data from a variety of scattering processes.
Well defined evolution procedures are used to
extrapolate to different kinematic ranges.  
Uncertainties from the PDF's result from uncertainty in the
input data, the parameterizations of the parton momentum distributions. 
Traditionally, the 
uncertainty in the inclusive jet cross section predictions
from the uncertainty in the PDF's is estimated
by comparing results with different current PDF's.  This is discussed in
detail in Section VIII.

The hard two-body parton level cross section, $\hat{\sigma}$, 
is only a function of the fractional momentum carried by
each of the incident partons $x$, the strong coupling parameter
$\alpha_{s}$, and the 
renormalization scale $\mu_{R}$ 
characterizing the energy of the hard
interaction.
The two body cross sections can be calculated with perturbative QCD
at leading order (LO)~\cite{LO} and more recently at 
next-to-leading order (NLO)~\cite{Greco,EKS}.  
At leading order eight diagrams for the 2$\rightarrow$2 scattering process
contribute. 
The NLO calculation includes the
diagrams which describe the emission of a gluon as an internal loop and 
as a final state parton.  

The scales $\mu_{R}$ and $\mu_{F}$ are intrinsic uncertainties in a fixed
order perturbation theory.  
Typically, as in this paper, they are set 
equal~\cite{kellisqcd} and 
we refer to them collectively as the $\mu$ scale.
Although the choice of $\mu$ scale is arbitrary, a reasonable
choice is related to a physical observable such as the $E_T$ of the jets.
Predictions for the inclusive jet
cross section depend on the choice of scale.
No such dependence would exist if the perturbation
theory were calculated to all orders. 
The addition of higher order terms in the calculation
reduces the $\mu$ dependence.
Typically $\mu$ is taken as a constant
(usually between 0.5 and 2) times the jet $E_T$ resulting in roughly 
a factor of two variation in predicted cross section at LO and 30\% at 
NLO~\cite{RSEP} in the $E_T$ range considered.

Predictions for the jet
cross section as a function of $E_T$ 
are obtained from the generalized cross section
expression above:
\begin{equation}
\frac{ E d^3 \sigma}{d p^3} \equiv \frac{d^3 \sigma}{d P_T^2 d \eta} = 
\frac{1}{2\pi E_T}\frac{d^2 \sigma} {d E_T d\eta}, 
\end{equation}
where the mass of the partons has been assumed to be zero ($P_T = E_T$)
and $\eta$ is the pseudo-rapidity (= rapidity for massless partons).

Experimentally, the inclusive jet cross section is defined as 
the number of jets in a bin of $E_T$ normalized by 
acceptance and integrated luminosity. As an inclusive quantity,
all the jets in each event which fall within the acceptance region
contribute to the cross section measurement.
Typically, measurements are performed in a central ($|\eta|<$1.0) 
rapidity interval.

Although many different 
experiments have measured the inclusive jet cross section,
comparisons between experimental 
measurements and theoretical 
predictions have the same general structure.
A QCD based Monte Carlo program generates partons 
which are then converted into jets of particles via a
process called fragmentation or hadronization.  
The particles resulting from the soft interactions between
the remnants of the collision
(underlying event)
are combined with the particles from the hard scattering.
The fragmentation process and the remnants of the incident protons
are not part of the theoretical cross section calculations.
They are empirically determined from the data. 
The generated particles are
traced through a detector and produce simulated data.
Jet identification algorithms (or clustering algorithms) were developed
to optimize the correspondence between the jets found in the simulated data and
the partons from which they originated. 
Two fundamentally different techniques were developed,
a nearest neighbor algorithm~\cite{UA2-1} and a cone
algorithm~\cite{UA1-1}.  Reference~\cite{SNOWMASS-me} contains a detailed 
comparison.
Corrections to the measured
data are derived based on the correspondence between the
simulated jets and the originating partons.  The
corrected cross section is then compared to a series
of parton level predictions in which parameters of the theory
such as the $\mu$ scale or the PDF's are varied.
Systematic uncertainty in the experimental measurements is dominated 
by the uncertainty associated with producing realistic jets and
underlying events for derivation of these corrections.  
The theoretical uncertainty in parton level predictions is dominated by 
uncertainty in the PDF's.

We present below a brief history of the measurements and predictions of
the inclusive jet cross section.
The experimental and theoretical developments are fundamentally correlated
since the corrections to the raw data depends on accurate modeling of 
the events which in turn depends on data sample size and quality of the data.

\subsection{Measurements and predictions in the 1980's}

The first measurements of the inclusive jet cross section~\cite{UA2-1,UA1-1}
were made by the UA1 and UA2 collaborations. The 
first data sample~\cite{UA2-1} 
included a total of 59 events in the central rapidity region over an $E_T$ 
range of 20 - 70 GeV.
Subsequent measurements by both the UA1 and UA2 
collaborations~\cite{UA1-1,UA2-2,UA1-2,UA2-3} with larger data samples 
found the LO theory
predictions to be compatible with the data.
The uncertainty in the experimental results was dominated by uncertainty 
in the jet energy scale due to the steeply falling shape of the 
cross section.
An estimated 10\% total uncertainty on the jet energy scale resulted
in a factor of two uncertainty on the corrected jet cross section~\cite{UA1-1}.
Both collaborations also performed studies of jet shapes, fragmentation
models, the underlying event 
and different jet identification techniques~\cite{UA2-2,UA1-2}.  
The theoretical predictions for the jet cross section
varied by a factor of two at low $E_T$ (30 GeV)
and about a factor of ten at the highest $E_T$ (100 GeV).
Within these uncertainties, the theoretical predictions were in agreement
with the results of both experiments over the $E_T$ range of
30 to 150 GeV, where the cross section falls by 5 orders of magnitude.

Concurrent with the improved measurements, 
a more complete model of the events was developed.
The Monte Carlo program ISAJET~\cite{ISAJET} included 
a leading log approximation for the effects
of final state gluon radiation and the Feynman-Field independent 
fragmentation scheme.
The leading log approximation generates improved QCD predictions over
tree level calculations by including terms which 
represent the partons 
radiated along, or close to the initial scattered parton direction.
Wide angle, hard emissions are not included.  
The independent Feynman-Field fragmentation model was used to convert 
the parton shower into a jet of hadrons.  Note that the 
fragmentation and parton shower schemes are closely coupled in 
the transformation of partons into hadrons.
If the parameters of the parton shower scheme 
are changed then the parameters in the
fragmentation functions must also change to maintain overall consistency 
and agreement with data.
Detailed studies of jet shapes, fragmentation and particle multiplicities 
found that the ISAJET program provided an improved description of the data 
over simple fragmentation functions (e.g. cylindrical phase space),
but did not produce the correct amount of underlying event energy or 
energy at the jet edges~\cite{UA1-2}.

Significant deviations from the predictions at high $E_T$ might 
indicate the presence of quark substructure~\cite{comp}. 
A new contact interaction was
characterized in terms of the energy
scale $\Lambda_c$ which represented the strength of this new interaction.
Most of the theoretical and experimental uncertainties were in
the normalization while the presence of quark compositeness would
produce a change in the shape of the spectrum at high $E_T$.
To avoid the largest theoretical
uncertainties, the QCD predictions were normalized to the data
in the low $E_T$ region,
where the effects of the contact interaction
were expected to be small.  
A model dependent limit of $\Lambda_c>$275 GeV was
obtained~\cite{UA2-2}.

Studies of two-jet production properties such as the dijet mass and
angular distributions were also 
performed~\cite{UA2-2,UA1-2,UA2-3,UA1-3,UA1-6,UA2-5,UA2-4,UA2-7} 
along with measurements of the structure and
number of multijet (3 or 4 jets) 
events~\cite{UA1-4,UA1-7,UA2-8}.  

With the increase in the collision energy of the CERN $Sp\bar{p}S$ to 
$\sqrt{s}$ = 630 GeV and the collection of additional data,
new measurements of the inclusive jet cross section
~\cite{UA1-5,UA2-5} pushed the limits on quark 
compositeness to $\Lambda_c>$415 GeV~\cite{UA1-5}.  
Uncertainties on the measurements and predictions were still
large.  Typically the predictions varied by a factor of two 
due to the dependence on the $\mu$ scale, PDF's, and higher 
order corrections~\cite{altarelli}. 
The experimental uncertainty was estimated at 70\% with the largest 
component (50\%) coming from the uncertainty in modeling
the events (e.g. fragmentation, underlying events)~\cite{UA1-5}.
The ratio 
of the cross sections at $\sqrt{s}$=540 and 630 GeV provide a test of 
scaling~\cite{UA2-5,UA1-5}.  
Although many of the uncertainties canceled in the ratio,
the remaining uncertainties were large enough that the data was consistent
with both perfect scaling and with the non-scaling QCD effects~\cite{UA1-5}.

In the late eighties significant improvements in the comparisons between
data and theory came from a variety of sources.  From the theoretical
front, NLO QCD predictions for the inclusive jet cross section
became available~\cite{Greco,EKS} and
the LO shower Monte Carlo programs were more sophisticated.
The ISAJET program was upgraded to include the effects of initial state
radiation. Two new leading log Monte Carlo 
programs (PYTHIA~\cite{PYTHIA} 
and HERWIG~\cite{HERWIG}) were also developed with improved
fragmentation schemes and both included initial and final state radiation. 
PYTHIA was based on a string fragmentation
model, while HERWIG used cluster fragmentation to generate
the parton and hadron showers associated with the jets.
On the experimental 
front the CDF collaboration 
began collecting data at a higher center of mass energy, $\sqrt{s}$ = 1.8 TeV,
and the CERN $Sp\bar{p}S$ delivered larger data samples.

The final measurement of the inclusive jet cross section from the CERN
$Sp\bar{p}S$ used data collected by the UA2 Collaboration~\cite{UA2-10}.
Statistical uncertainties
were of order 10\%, while the overall normalization uncertainty
was 32\%.  Comparisons to QCD predictions with a plethora of PDF's
showed shape variations of order 30\%. The corrections to the
cross section used the
PYTHIA Monte Carlo~\cite{PYTHIA} to generate the partons (with initial 
and final state radiation) and the JETSET\cite{JETSET} program for 
fragmentation .
The largest component of the systematic uncertainty came from
the model dependence of the acceptance and fragmentation corrections (25\%).
The underlying event was adjusted to agree with the data and
contributed roughly 10\% to the uncertainty at 60 GeV and 5\% at 130 GeV.
A pseudo-cone algorithm was used to identify jets.  The standard nearest
neighbor algorithm was used to form preclusters. Then nearby preclusters
within a large cone 
$\Delta R$ = $\sqrt{\Delta\eta^2 + \Delta\phi^2}$ and $\Delta\eta$ = 1.3
of each other were merged.
Only at the highest $E_T$ ($>$100 GeV) were the statistical uncertainties 
dominant.  The cross sections were also measured in forward rapidity regions.
The ability of the theory to describe the data in these regions was
marginal.
A limit on the compositeness scale of $\Lambda_c>$ 825 GeV was derived from the 
central region data using the most pessimistic PDF and systematic uncertainties.

The first measurement of the inclusive jet
cross section at $\sqrt{s}$= 1800 GeV was performed by the CDF collaboration and
consisted of 16,300 clusters~\cite{CDF-Inclusive-Jet-87}.
It spanned the $E_T$ range from 30 to 250 GeV for the central rapidity region.
The systematic uncertainties were largest at low $E_T$, 70\% at 30 GeV 
compared to 34\% at 250 GeV. Comparisons were made to 
LO predictions.  The range of theoretical predictions using different PDF's,
and $\mu$ scales was roughly a factor of three.  
The data was also compared to the results from other 
experiments~\cite{ISR-2,UA2-5,UA1-5}.  
Uncertainties in the comparisons
arose due to different clustering algorithms, different
corrections for underlying events, showering outside the jet as well as
overall normalization uncertainties.  The non-scaling effects of QCD 
could not be confirmed with the comparison to the $\sqrt{s}$= 630 data.
However, the effects of QCD scale breaking could be observed by 
comparison to the $\sqrt{s}=$ 63 GeV data~\cite{ISR-2}.

\subsection{Jet measurements and predictions in the 1990's}

The NLO parton level predictions ushered in a new era of comparisons 
between data and theory.  The inclusion of the O($\alpha_s^3$) contributions
to the scattering cross section reduced the
uncertainty due to the choice of $\mu$ scale~\cite{RSEP}
from roughly a factor of two to approximately 30\% for $\mu$=2-0.5 times jet
$E_T$~\cite{RSEP}.  More significantly however, the NLO calculations
produce events with 2 or 3 partons in the final state.  These partons
could be grouped together (clustered) to produce a parton level approximation
to a jet of hadrons.  Details of both these issues are discussed below.

\subsubsection{Parton clustering}

Jet identification is a fundamental step in measurement of the inclusive
jet cross section.  
With LO predictions there are two partons in the final state and 
each one is equated to a jet. 
These predictions have no dependence on jet finding algorithms or 
on jet shapes or size. 
However, the NLO
predictions can have three partons in the final state and thus 
dependences on clustering can be investigated.
To minimize the difference between NLO parton level predictions
and measured jet properties, a clustering algorithm was defined
which could be implemented for both situations~\cite{SNOWMASS}.   
In this algorithm (called the Snowmass algorithm), two partons which fall
within a cone of radius R in $\eta$--$\phi$ space (R =
$\sqrt{\Delta\eta^2 + \Delta\phi^2}$ and $\Delta\eta$ and
$\Delta\phi$ are the separation of the partons in pseudo-rapidity
and azimuthal angle) are combined into a ``jet".
With this algorithm, two partons must
be at least a distance of 2R apart to be considered as separate
jets.  If two partons are contained in a cone, then the $E_T$ of
the resulting jet is the scalar sum of the $E_T$ of the individual partons.
A similar algorithm (described later) with  $R=$ 0.7 is implemented in
the experimental data analysis by using
calorimeter towers (shown in Figure~\ref{CDFLEGO}) in place of the partons.  

Comparison of data to NLO predictions for jet shapes 
and the dependence of the cross section on cone size
found that a consistent description of the cross section
could only be obtained through the introduction of an additional 
parameter, $R_{sep}$ into the theoretical calculations ~\cite{RSEP}. 
The $R_{sep}$ parameter was intended to mimic the effects of 
cluster merging and separation employed for analysis of experimental
data.  This will be discussed in more detail in the description of the
experimental algorithm and in the treatment of theoretical
uncertainty. It is remarkable, however, that the 
NLO predictions, with only 2 or 3 partons in the final state, and
the simple introduction of
the $R_{sep}$ parameter can give a reasonable description of 
the hadronic energy distribution within jets ~\cite{RSEP}, although
each jet consists of 10's of hadrons.

The NLO predictions also changed the way the jet energy is corrected.
In contrast to the LO predictions,
the effect of parton energy lost outside the jet cone 
is modeled at the parton level.
The corrections for this out-of-cone (OOC) energy 
which were used for comparison to LO predictions
were highly dependent on the non-perturbative fragmentation models and
were a large contributor to uncertainty in the corrected cross sections.
When data are compared to NLO predictions, no correction for OOC
energy is necessary. 

\subsubsection{Choice of the $\mu$ scale}

The NLO predictions for the inclusive jet cross section
significantly reduced the dependence of the cross section on
the choice of scale.
For the usual range of $\mu=$ $2E_T$ to $E_T/2$ the variation in 
the prediction 
was reduced from a factor of two to about 20\%~\cite{RSEP,EKS}.
However, a subtlety in the choice of scale also arose. 
At LO there are only two partons of equal $E_T$.  At NLO 
the partons may or may not be grouped together to form parton level jets,
and $E_{T1}$ and $E_{T2}$ are not necessarily equal.  Thus, if
the scale is to be the $E_T$ of each jet, there may be  more than one scale
for each event in the NLO calculations.

In previous 
publications~\cite{CDF-Inclusive-Jet-96,CDF-Inclusive-Jet-92,CDF-Inclusive-Jet-87}, 
and in the following
chapters, the CDF data is compared to the NLO predictions 
of Reference~\cite{EKS}.  This program analytically calculates the inclusive 
jet cross section at a specific $E_T$.
In the evaluation of the cross section, the
PDF's and subprocess cross sections and $\alpha_s$ are all calculated
at that $E_T$.   As a result, the cross section as a function of $E_T$ can
be directly related to $\alpha_s$ and even used as a 
measurement of the running of $\alpha_s$~\cite{walter-alphas}.

More recently a NLO event generator, JETRAD, was developed~\cite{JETRAD}.  
This program produces the energy-momentum four vectors 
for the two or three final state partons.
These partons can be clustered together and treated as jets in 
a manner similar to the analytic predictions.  For this
program, it is necessary to have one weight per event, or in other
words, one scale per event, rather than one scale per jet.
The $E_T$ of the leading parton ($E_T^{max}$) was chosen to set the scale since
it is never the one to be clustered with the emitted gluon.

In contrast to the normalization shifts associated with 
changing the $\mu$ scale from 0.5$E_T$ to 2$E_T$, 
the effect of the using $E_T^{max}$ instead of $E_T$ jet introduces 
a small change in shape.
The size of the effect ranges from about 4\% (smaller for $E_T^{max}$)
at 100 GeV to $<$ 1\% at 465 GeV.  Below 100 GeV the cross section with
$E_T^{max}$ decreases more quickly; at 50 GeV the difference is about 6\%.
All of the predictions presented here use
$E_T$. Comparisons of the theoretical 
predictions will be discussed in Section VIII.

\subsubsection{Experimental measurements}

CDF measured the inclusive jet cross section with 30 $nb^{-1}$ of data 
collected in 1987~\cite{CDF-Inclusive-Jet-87}, 4 $pb^{-1}$  
from 1989~\cite{CDF-Inclusive-Jet-92} and 19 $pb^{-1}$  from 
1992-1992 (Run 1A)~\cite{CDF-Inclusive-Jet-96}.
With each measurement the statistical and systematic uncertainties
were reduced.  The dijet angular distribution and the dijet mass
spectrum were also compared to LO and NLO predictions~\cite{cdf-dijet1,%
cdf-dijet2,cdf-dijet3,cdf-dijet4,cdf-dijet5,cdf-dijet6,%
cdf-dijet7,cdf-dijet8,cdf-dijet9}.  These data were analyzed using 
clustering algorithms and corrections which were influenced by 
the intention to compare to NLO rather than LO predictions (e.g. no
correction of energy out side the jet cones).
Comparisons to data from UA1 and UA2 were complicated
by the different clustering algorithms and corrections schemes;
CDF used a cone of R= 0.7 and did not correct for OOC while UA1 and UA2 used
jet sizes of order R= 1 - 1.3 and made OOC corrections.  Measurement
of the QCD scale breaking effects was possible with CDF data
at 546 and 1800 GeV~\cite{CDF-XT-analysis}.  Measurements 
of multijet events showed that the
newest shower Monte Carlo, HERWIG, could predict multijet rates and
event properties up to 6 jets, but still lacked some contributions
from wide angle scattering~\cite{CDF-Clustering,SUMET}

\subsection{Summary}

The NLO predictions significantly improved the agreement 
between data and theory for the inclusive cross section.  
Two of the largest uncertainties were
substantially reduced.  One remaining issue is the modeling
of the underlying event.  Typically the amount of background energy is
estimated from minimum bias data (data collected using only minimal
requirements).  However, no QCD based prediction, or even prescription
is available.  

\section{The CDF Detector}

\begin{figure}
\centerline{
{\hbox{
\psfig{figure=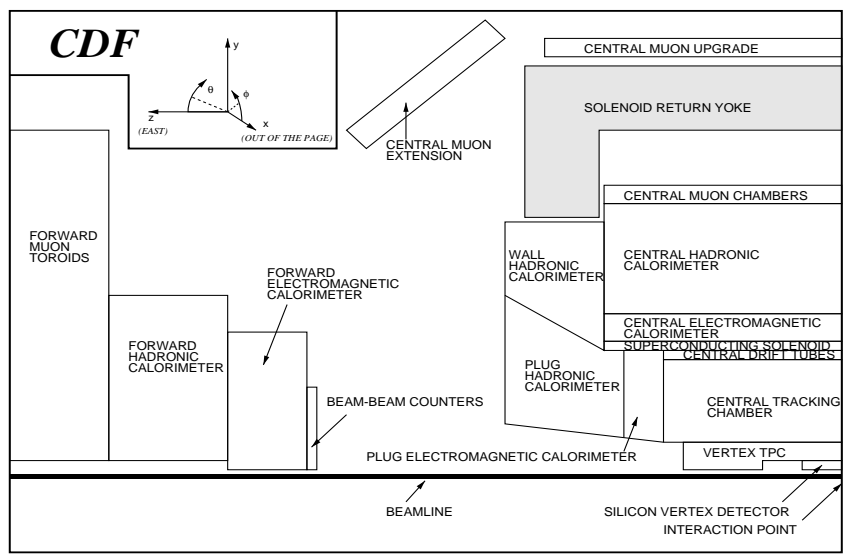,height=20cm,width=15cm}
}}}
\vspace{-2in}
\caption{One quarter section of the CDF  detector.}
\label{cdfpic}
\end{figure}

The Collider Detector at Fermilab (CDF) \cite{CDF-Detector} 
is a combination of tracking systems inside a
1.4 T solenoidal magnetic field and surrounded by
electromagnetic and hadronic calorimeters and muon detection
systems. Figure~\ref{cdfpic}
shows a schematic view of one quarter of the CDF detector.
The measurement of the inclusive jet
cross section uses the calorimeters for measurement of the jet 
energies.
The tracking systems provide the location
of the $p\bar{p}$ collision vertex 
and in-situ calibration of the calorimeters.


Closest to the beampipe is the silicon vertex detector (SVX)~\cite{SVX-det}.
It is roughly 60 cm long and covers the radial region from 3.0 to 7.9 cm.
The $r$--$\phi$ tracking information provided by the SVX
allows precise determination of the transverse position of the
event vertex and contributes to the track momentum resolution.
Surrounding the SVX is the vertex drift chamber (VTX).  This device
provides $r$--$z$ tracking information and is used to determine the position of
the $p\bar{p}$ interaction (event vertex) in $z$.
Both the SVX and the VTX are mounted inside a 3.2 m long
drift chamber called the central tracking chamber (CTC). 
The CTC extends from a radius of 31 to 132 cm.
The momentum resolution~\cite{topmass} of 
the SVX--CTC system is 
$\delta P_{T}/P_{T}^2 = [(0.0009P_{T})^{2} + (0.0066)^{2}]^{1/2}$ 
where $P_{T}$ has units of GeV/c. 
Measurement of the response of the calorimeter to isolated tracks 
provides an {\it in--situ} measurement of
the calibration of the calorimeter.  This is particularly important
for low energy
particles (where test beam information is not available).
The CTC is also used to study 
jet fragmentation properties~\cite{frag} and to tune the fragmentation
parameters of the Monte Carlo simulations.
Figure~\ref{CDFCTC} shows four events in the CTC.

Outside the solenoid a combination 
of three electromagnetic and hadronic calorimeter systems provide 
$2\pi$ coverage in azimuth and extends to $|\eta| = 4.2$.
The rapidity coverage of 
each calorimeter is given
in Table~\ref{table-calo}.
The calorimeters are segmented into projective towers. Each tower
points back to the center of the nominal interaction region and is
identified by its pseudo-rapidity and azimuth.

The central electromagnetic (CEM) calorimeter 
is followed at larger radius by the central hadronic
calorimeters (CHA and WHA). 
The CEM
absorber is lead and the CHA/WHA absorber is 4.5 interaction lengths of iron;
scintillator is the active medium in both.
These calorimeters
are segmented into units of 15 degrees in azimuth and $\approx$ 0.1
pseudo-rapidity. Two phototubes bracket each tower in $\phi$ and the
geometric mean of the energy in the two tubes is used to determine the
$\phi$ position of energy deposited in a tower. 
Electron energy
resolution in the CEM is $13.7\% / \sqrt{E}$ plus 2\% added in
quadrature. For hadrons the single particle resolution depends on
angle and varies from roughly $50\% / \sqrt{E}$ plus 3\% added in
quadrature in the CHA to $75\% / \sqrt{E}$ plus 4\% added in
quadrature in the WHA. In the forward regions 
calorimetric coverage is provided by gas proportional chambers: the
plug electromagnetic (PEM) and hadronic calorimeters (PHA) 
and the forward electromagnetic (FEM)
and hadronic calorimeters (FHA).
Figure~\ref{CDFLEGO} shows jet events in CDF calorimeter.
%

The luminosity, or beam exposure, is measured with 
scintillation hodoscopes located near the beam pipe on both sides
of the  interaction point.  A coincidence of hits
in both the up and down stream sides indicates the presence of 
a $p\bar{p}$ collision.  The integrated luminosity of a given time period
is calculated from the number of collisions observed, normalized by acceptance
and efficiency of the counters and by the total $p\bar{p}$ cross 
section~\cite{CDFLUM,WZXSEC,zpt}.  

\section{Data Set}

\subsection{Trigger}

The data were collected using a multilevel trigger system.
The lowest level trigger, Level 1, required a single 
trigger tower (roughly 0.2 x 0.3 in
$\eta$-$\phi$ space) to be above an $E_T$ threshold.
These thresholds were typically $\leq$ 20\% of the Level 2 (L2) cluster $E_T$ 
requirement
and thus had negligible effect on the combined trigger efficiency.
The most significant trigger requirement for the jet sample
was for a L2 trigger cluster.
This trigger used a nearest 
neighbor cluster algorithm with 
a seed tower threshold of 3 GeV $E_T$ and a single tower threshold of 1 GeV.
The $E_T$ of the calorimeter towers
were calculated assuming the interaction occurred at the center of
the CDF detector ($z$= 0).
To avoid saturating the L2 trigger bandwidth while spanning
a wide range of $E_T$, three low $E_T$ trigger
samples were collected using $E_T$ thresholds of 70, 50, and 20 GeV
and nominal prescale factors of 8, 40, and 1000 respectively.  
These samples are referred to as jet-70, jet-50, and jet-20, respectively.
In Run 1A the $E_T$ thresholds were the same and
the prescale factors were 6, 20, and 500.
The highest $E_T$ clusters
came from either of two unprescaled paths at L2: 
a single cluster of $>$ 100 GeV $E_T$ or a sum over all clusters 
$>$ 175 GeV $E_T$.  We will refer to the high $E_T$ sample as jet-100.

For these samples, the third level trigger was used primarily to
remove backgrounds such as phototube breakdowns or coherent detector noise
which produced clusters for the L2 trigger.
Level 3 (L3) reconstructed jets using the standard offline
algorithm~\cite{CDF-Clustering} and made lower requirements 
on the jet $E_T$ than
were used in L2.  For the L2 triggers of 70, 50, and 20 GeV
the L3 requirements were 55, 35, and 10 GeV respectively.  The highest
$E_T$ jet sample was collected with a cut at L3 of 80 GeV.  
In the Run 1A analysis the events passing the L3 cut of 80 GeV 
were required to have passed a L2 cut at 100 GeV.  In Run 1B 
this requirement was removed. The efficiency of the jet triggers 
will be discussed in section IV.D.

In addition to the jet data described, a sample of
minimum bias data was collected.  The trigger for this sample 
was a coincidence of hits in scintillation hodoscopes surrounding the 
beampipe.  This sample is used to measure the luminosity~\cite{CDFLUM} 
and to study backgrounds which contribute to the jet energies.

\subsection{Z vertex and multiple interactions}

The protons and antiprotons are distributed in 
bunches which extend of order 50 cm along the beamline.
As a result, 
$p\bar{p}$ interactions occur over a wide range in $z$.
For each event, vertex reconstruction is performed using primarily
the information provided by a set of time projection chambers (VTX).
The vertex 
distribution is roughly a Gaussian with width 30 cm and a mean within
a few centimeters of the center of the detector ($z$=0).
To ensure good coverage 
each event was required to have
a vertex within $|z|<$ 60 cm.
The efficiency of this cut, 93.7$\pm$1.1\%, was determined 
from fits of the z vertex distribution in 
minimum bias data to the beam shape parameters
and averaged over the Run 1B sample~\cite{zpt}. 

In Run 1A, the number of events with more than one $p\bar{p}$
interaction was small ($<$10\%).  An algorithm which ranked the
found vertices on the basis of the number of tracks
associated with each vertex picked the correct 
vertex for the jet event 98\% of the
time.
In Run 1B, the instantaneous luminosity was
higher and thus the number of events with multiple interactions increased.
Studies which associated tracks with individual jets found that
the standard vertex selection algorithm picked the
correct vertex 88\% of the time. For the remaining  12\%
of events, the correct vertex was identified using the
tracks pointing to the individual jets. 
The mis-assignment of the z vertex smears the measured
$E_T$ of the jets with an rms which depends on the jet $E_T$; for
the jet-20 sample the rms is 9\% while for the high $E_T$ jet sample it
is 14\%. When the correct vertex is used for all the events, instead of the
standard vertex selection algorithm, the
measured jet cross section is $\approx$ 1\% lower, except for the 
highest $E_T$ bin where 2 out of 33 events move out of the bin, giving a 6\% 
decrease.

\subsection{Jet clustering}

The CDF clustering algorithm\cite{CDF-Clustering}
uses a cone similar to the Snowmass parton clustering algorithm~\cite{SNOWMASS}.
The CDF algorithm groups together calorimeter towers within
a cone of radius $R=(\Delta\eta^2+\Delta\phi^2)^{1/2}=0.7$
and identifies them as jets.  Enhancements of the 
Snowmass algorithm were necessary for identification,
separation and merging of 
nearby clusters of energy in the calorimeter.
The final definition of the $E_T$ of the jet also differs from
the Snowmass definition and is detailed below.

In the central region, the calorimeter segmentation (towers) 
is roughly 0.1 x 0.26 in $\eta-\phi$ space.  
The $E_T$ of a
tower is the sum of the $E_T$'s measured in the
electromagnetic and hadronic compartments of that tower.
These are calculated by assigning a
massless four--vector with magnitude equal to the energy deposited
in the compartment
and with direction defined by the unit vector pointing from the
event origin to the center of the compartment.
To be included in a cluster, towers were required 
to contain
at least 100 MeV $E_T$.  To start a new cluster, 
a seed tower with $E_T>$ 1 GeV was required.

The clustering has four stages.  The first is a rough clumping together
of neighboring towers.  The second involves iterating until the
list of towers assigned to a cluster does not change.
Next merging/separation criteria
are imposed on overlapping jets and finally the jet four-vector is 
determined from the towers assigned to the cluster.  The detailed steps 
are:
1) an $E_T$ ordered list of towers with $E_T>$1.0
GeV is created; 2) beginning with the highest $E_T$ tower,
preclusters are formed from an unbroken chain of
contiguous seed towers 
provided the towers are within a 0.7x0.7 window 
centered at the seed tower;
if a tower is outside this window 
it is used to form a new precluster; 3) the preclusters are
ordered in decreasing $E_T$ and 
grown into clusters by finding the \ET weighted centroid and
collecting the energy from all towers with more than 100 MeV within
R=0.7 of the centroid; 4) a new centroid is calculated from the set
of towers within the cone and a new cone drawn about this position;
steps 3 and 4 are repeated until the set of towers contributing to
the jet remains unchanged; 5) clusters are reordered in decreasing 
$E_T$ and overlapping jets are merged if they
share $\geq$75\% of the smaller jet's energy;  if they share less
the towers in the overlap region are assigned to the nearest jet.

The final jet energy and momentum is computed from the final list of towers:

\begin{eqnarray}
E^{jet} &=& \sum_{i} E^{i}  \\
P_{x} &=&\sum_{i} E_{i}\sin(\theta_{i})\cos(\phi_{i})\\
P_{y} &=&\sum_{i} E_{i}\sin(\theta_{i})\sin(\phi_{i})\\
P_{z} &=&\sum_{i} E_{i}\cos(\theta_{i})\\
\phi_{jet} &=& \tan^{-1} [P_{y}/ P_{x}] \\
\sin\theta_{jet} &=& \frac{\sqrt{P^{2}_{x} + P^{2}_{y}}}
{\sqrt{P^{2}_{x} + P^{2}_{y} + P^{2}_{z}}} \\
E_T^{jet} &=& E^{jet}\sin\theta_{jet}.
\end{eqnarray}

\begin{figure}
\centerline{\psfig{figure=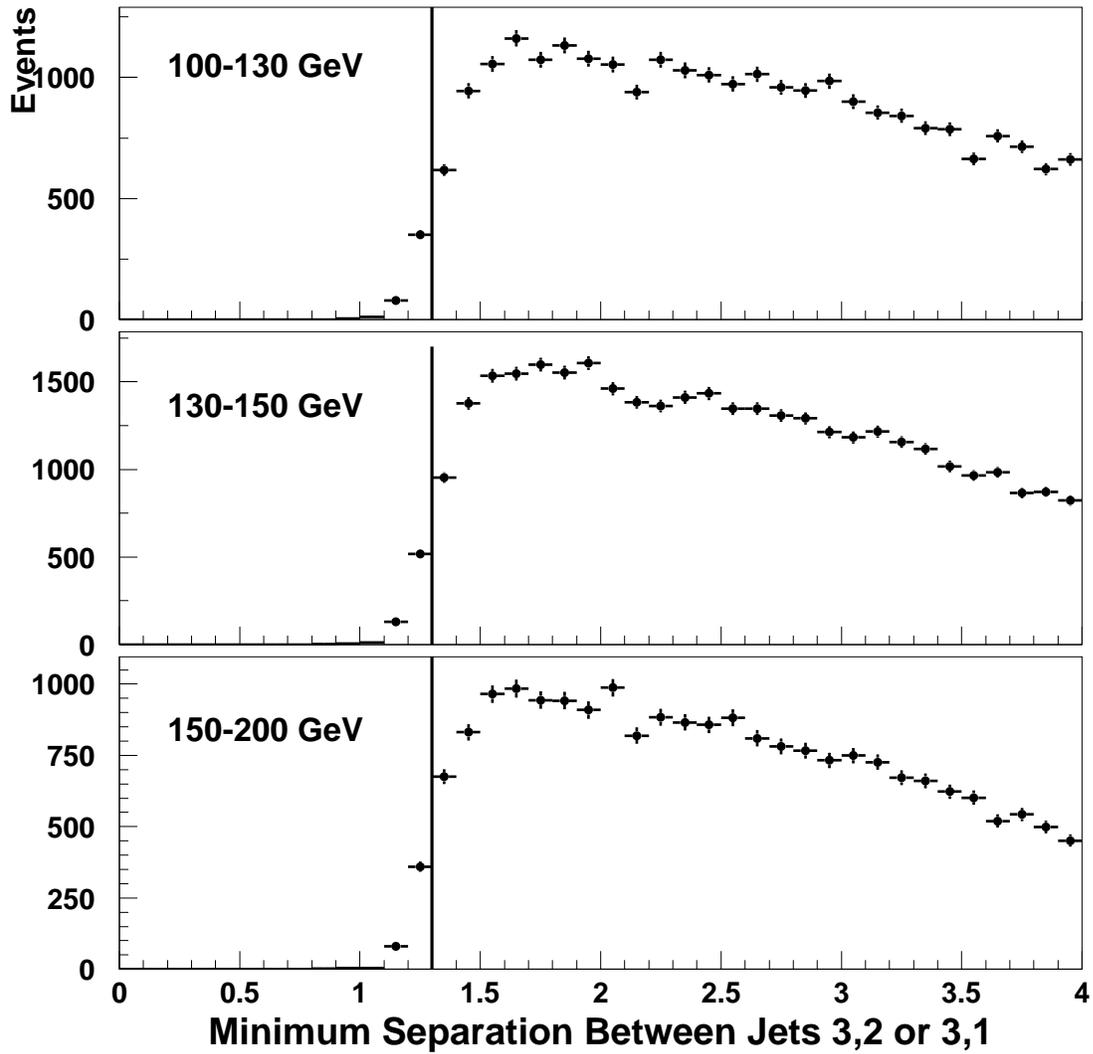,height=16cm,width=16cm}}
\caption{ Minimum separation (in units of cluster radius) 
between the 3rd jet and the 1st or 2nd jet 
in different bins of jet $E_T$. At a separation of 1.3R
at least 50\% of the clusters are separated.}
\label{fig-rsep-data}
\end{figure}

Studies of this algorithm with different cone sizes found
that it will separate two clusters whose
centroids are 1.3R apart in $\eta-\phi$ space roughly 50\% of
the time. Figure~\ref{fig-rsep-data}
shows distribution of $R_{sep}$, the separation between 
the 3rd jet and the 1st or 2nd jet (which ever is smaller) divided by
the clustering cone radius of 0.7, for three 
bins of $E_T$: 100-130 GeV, 130-150 GeV, and 150-200 GeV.


The algorithm used in the
NLO predictions (Snowmass) defines the $E_T$ of a jet as the scalar sum of the
$E_T$'s of the individual towers (or partons).  With this algorithm the
jets are massless ($E_T = P_T$).  In the data however, we observe that
the jets do have a width and thus a mass~\cite{SNOWMASS}.  
Rather than ignore this
information we adopted the four-vector definition of the jet $E_T$
as described above.  With the CDF definition, the jet mass is defined
as $E^2-{\bf \vec P^2}$.
Studies~\cite{SNOWMASS} found that
the CDF clustering algorithm and the Snowmass algorithm
were numerically very similar.

\subsection{Trigger efficiency}

\begin{figure}
\centerline{
{\hbox{
\psfig{figure=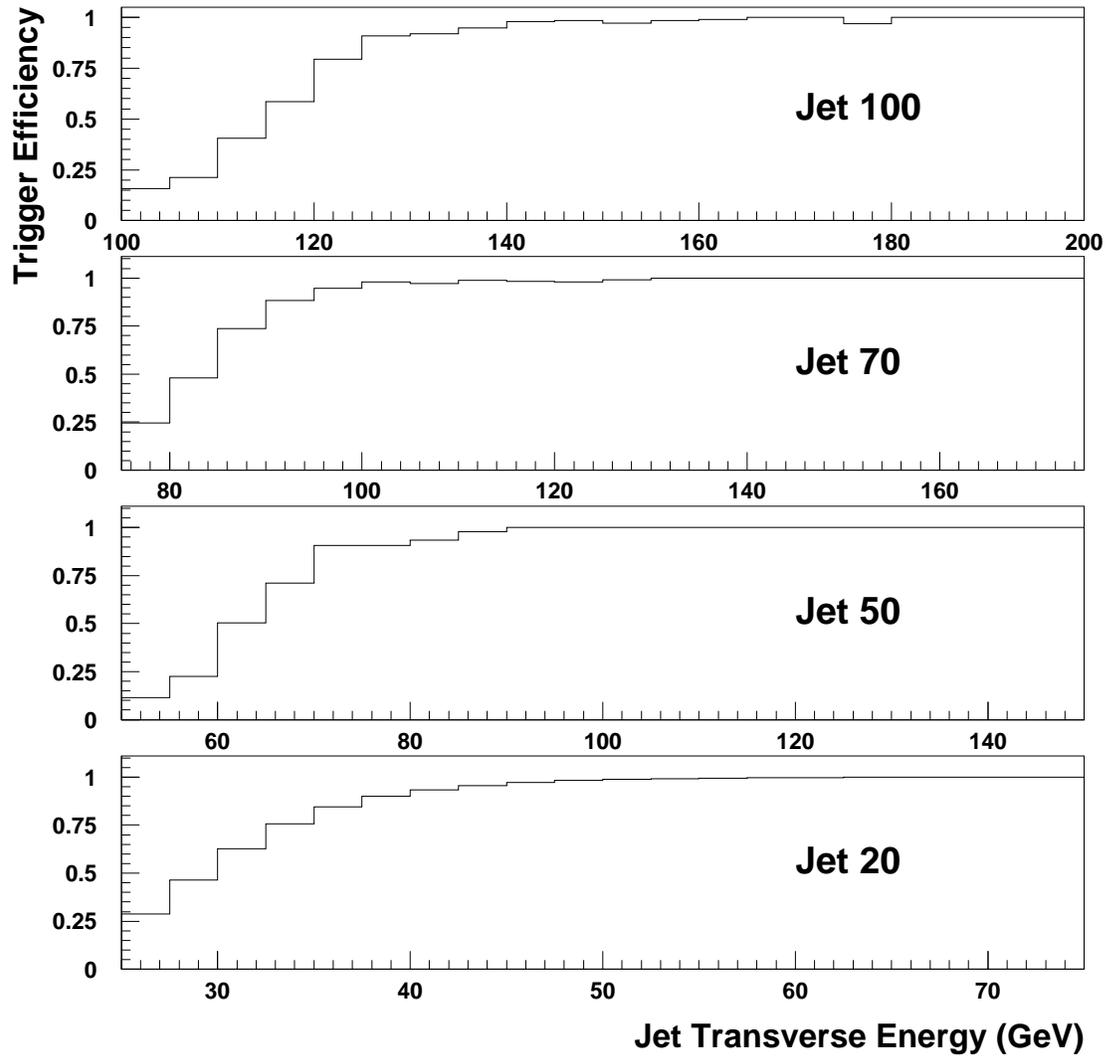,height=16cm,width=16cm}
}}}
\caption{Trigger efficiency for the 100, 70, 50, and 20 GeV L2 triggers.
The 100, 70, and 50 GeV triggers use overlap with the next lower trigger 
to determine the efficiency.
The jet-20 plot uses the $2^{nd}$ jet in the event.}
\label{teff}
\end{figure}

As mentioned earlier (section IV.A) the efficiency for jet triggering
was dominated by the L2 trigger.
The L2 clustering and the standard CDF algorithm are quite different.
For each trigger sample the efficiency of the L2 cluster $E_T$ cut 
is measured as a function
of the jet $E_T$ derived using the standard algorithm.
The overlap of the separate trigger samples allows derivation of
trigger efficiency curves.
For example, for the jet-50 efficiency curve 
the jet $E_T$ spectrum of events from the jet-20 sample
which contain a L2 cluster with 
$E_T>$50 GeV is divided by the $E_T$ spectrum of all the jet-20 events.
This technique was used for the jet-50, jet-70, and jet-100 samples
and the results are shown in Figure~\ref{teff}.
The uncertainty on the trigger efficiency is determined
using binomial statistics.
The slow turn on in efficiency, shown in Figure~\ref{teff}, in 
all samples is primarily due to the
difference in single tower threshold between the L2 trigger clustering and the
standard CDF jet algorithm combined with the use of the reconstructed
interaction vertex instead of $z$=0.
To ensure trigger efficiency $>$ 95\%, jet $E_T$ thresholds of 
130, 100, and 75 were applied to the 100, 70, and 50 GeV trigger
samples respectively.

The efficiency for the 20 GeV threshold
was determined from the $2^{nd}$ highest $E_T$ jet in the event
because no lower threshold sample was available.
Two different methods of selecting events for this study were
tried. Method (a)
required that the highest $E_T$ jet offline match the
highest $E_T$ L2 jet in $\eta-\phi$ space to $\Delta R<$0.5. 
Method (b)
required that both the 1st and 2nd jets in the event match the 
1st and 2nd L2 clusters to $\Delta R<$0.5. 
To simulate the effect of the trigger, these events were
required to have a $2^{nd}$ L2 cluster
with $E_T >$ 20 GeV.  The ratio of $E_T$ spectra for events which
passed the cut to the full samples (defined by a or b) shows the
efficiency.
Both methods were tested on the 50 GeV trigger. 
Compared to the trigger overlap method, method (b)
gave systematically larger efficiency estimates 
while method (a) found good agreement with the trigger
overlap method.  For the jet-20 trigger efficiency, method (a) 
was used and the uncertainty was taken as half the difference between
the two methods.

Studies of the events which passed the jet-100 GeV and the
$\sum E_T$-175 GeV trigger found that the 175 GeV trigger
was more efficient than the jet-100 GeV trigger. In addition, the
efficiency determined from the overlap from the 100 and 175 samples
agreed with the efficiency of the overlap with the 70-GeV sample to
within 1\%. Based on these results we conclude that the combination
of 175 and 100 triggers is 100\% efficient 
for jet $E_T>$ 130 GeV. We assign a trigger efficiency 
uncertainty of 0.5\% to the first point (130-140 GeV), 
to cover the differences between
the two methods.  Above 140 GeV the trigger efficiency uncertainty is 
negligible.

Finally, an effective prescale factor was
determined for each of the low $E_T$ samples by normalization to the 
next highest 
$E_T$ sample in the bins which overlapped.  The uncertainty in these
effective prescale factors was taken as half the difference between the 
measured factor and the nominal value.
Table~\ref{table-treq} summarizes, for all bins below 140 GeV, the 
low edge of jet $E_T$ bin with the standard CDF clustering algorithm, 
the requirements of the L2 trigger,
the trigger efficiency, and the uncertainty in the trigger efficiency. 

\begin{figure}
\centerline{
{\hbox{
\psfig{figure=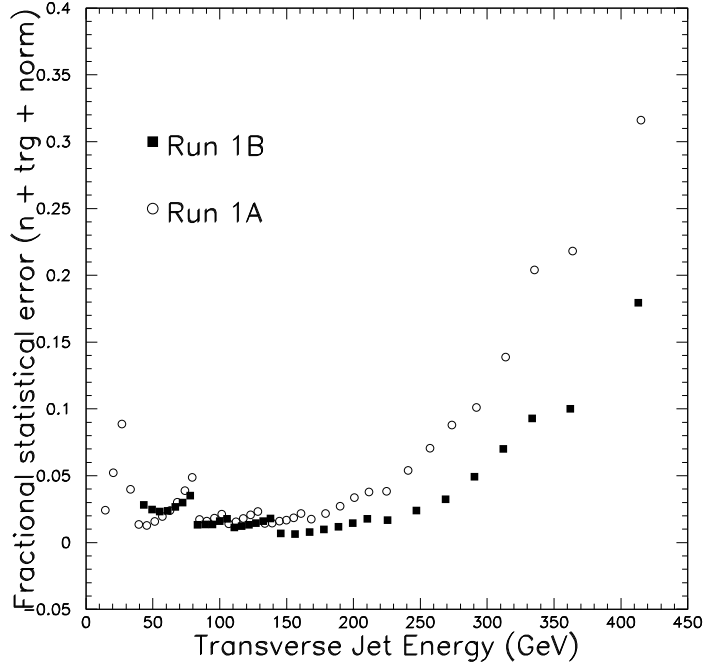,height=10cm,width=10cm}
}}}
\caption{Percentage uncorrelated uncertainty on the Run 1A and 1B data
sets.}
\label{fig-stat-1a1b}
\end{figure}

In section V.C the corrected cross section will be presented.  The
uncertainty on each point will be the 
quadrature sum of the trigger efficiency, 
the uncertainty in the prescale factor and 
the statistical error from the number of events in the bin.  
These uncertainties 
are treated as uncorrelated from point to point and this combination is 
treated as statistical error for the remainder of the analysis.
Figure~\ref{fig-stat-1a1b} shows the percentage uncorrelated uncertainty 
on each data point for the Run 1A and 1B data sets. Note that below 150 GeV,
the precision of the data is roughly the same due to the 
factor of two increase in the prescale factors.


\subsection{Backgrounds}

\begin{figure}
\centerline{
{\hbox{
\psfig{figure=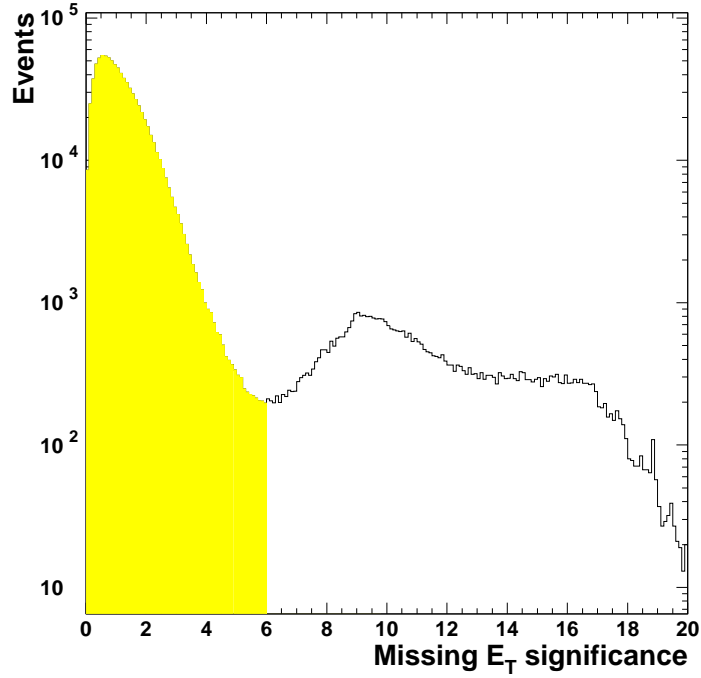,height=10cm,width=10cm}
}}}
\caption{Distribution in \METSIG  after timing cut.  The shaded region shows
the events kept by the \METSIG cut.}
\label{metsig-ncut}
\end{figure}

\begin{figure}
\centerline{
{\hbox{
\psfig{figure=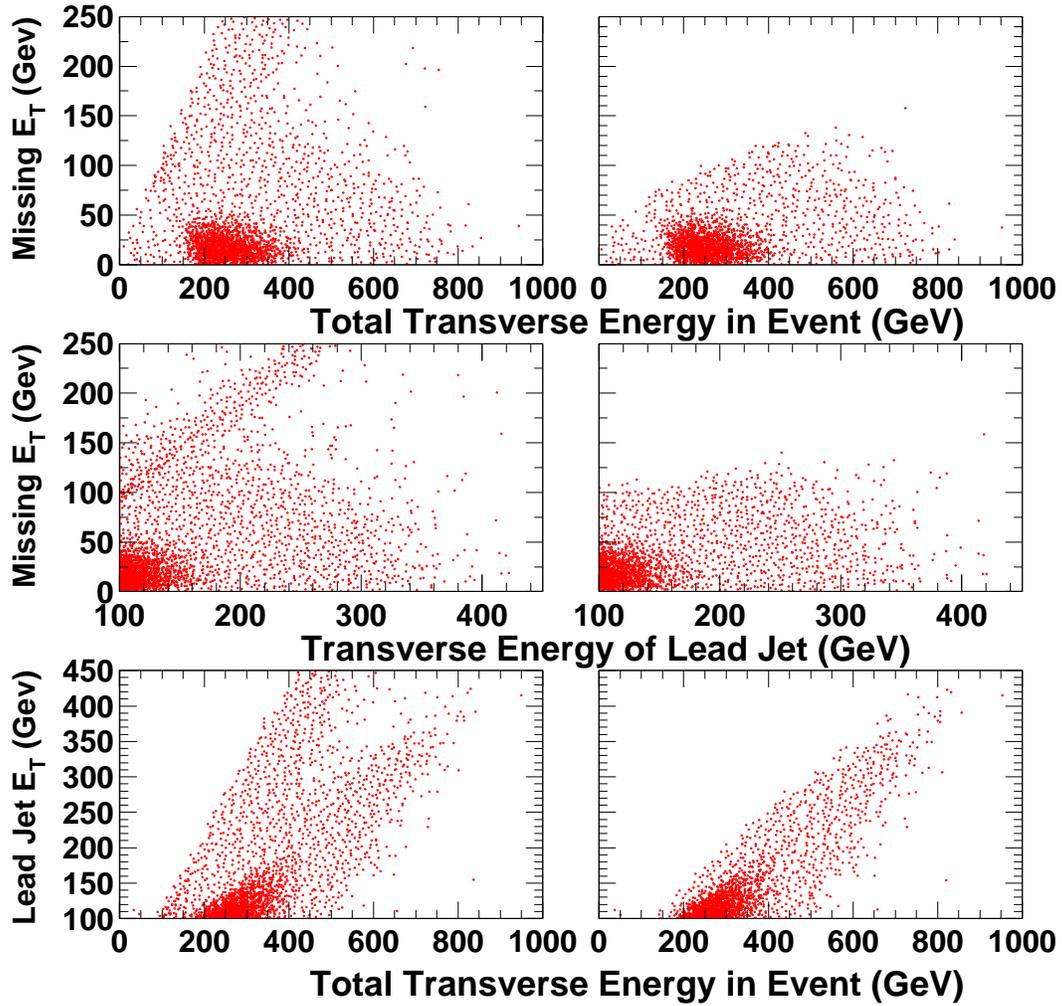,height=15cm,width=15cm}
}}}
\caption{ Raw data distributions before(left) and after(right) \METSIG cut.}
\label{metvset}
\end{figure}

\begin{figure}
\centerline{\psfig{figure=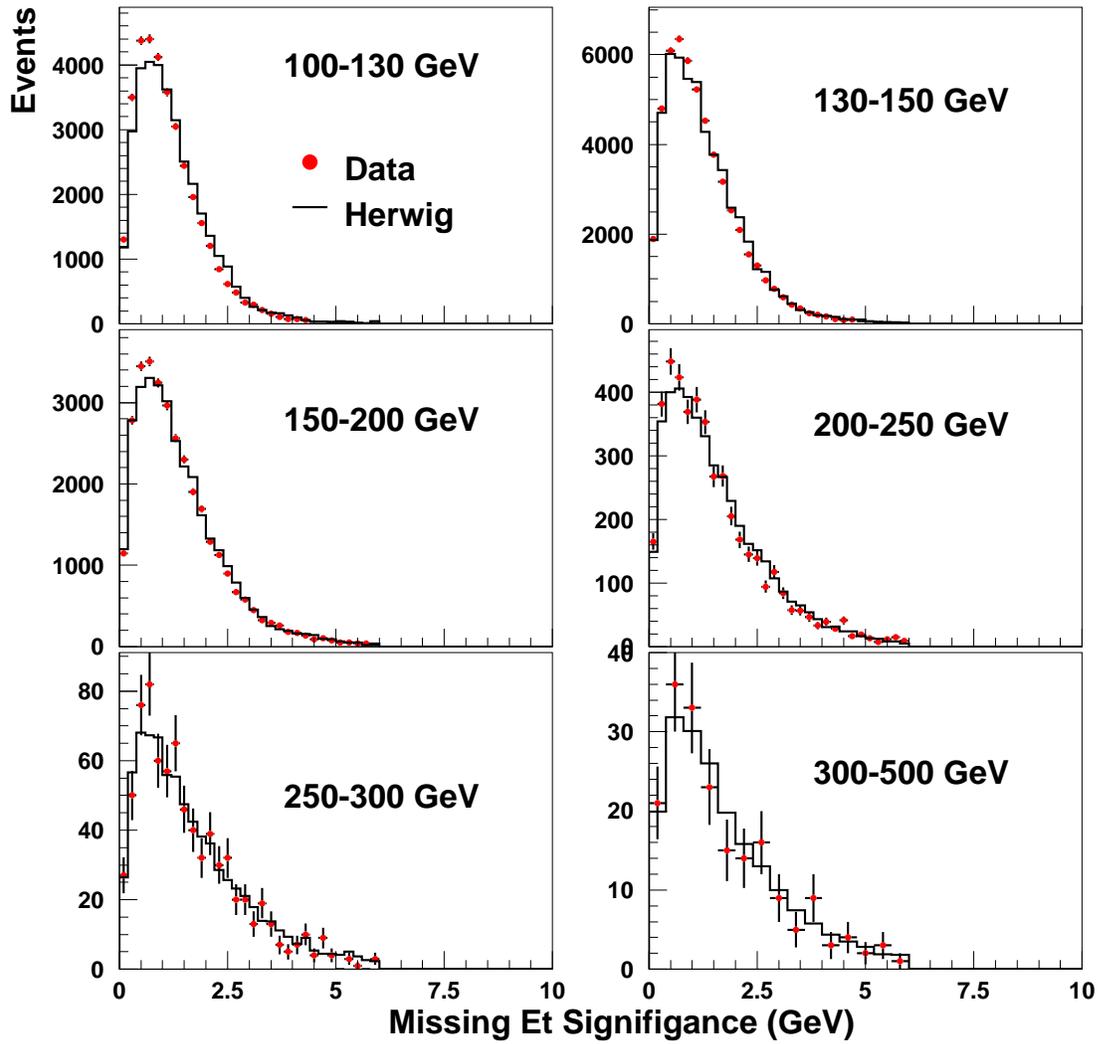,height=16cm,width=16cm}}
\caption{Distributions of missing $E_T$ significance from data (points) and
HERWIG (histogram). The labels on the individual plots (e.g. 100-130 GeV)
indicate the $E_T$ range of the leading jet.}
\label{fig-metsig}
\end{figure}

As discussed in previous 
papers~\cite{CDF-Inclusive-Jet-96,CDF-Inclusive-Jet-92,CDF-Inclusive-Jet-87}, 
cosmic rays, accelerator loss backgrounds and detector noise
were removed with cuts on
timing and on missing $E_T$ significance, \METSIG =
\METS where the sum is 
over all towers in the calorimeter.  
Events with more than 8 GeV of energy in the hadron calorimeter 
out of time with respect to the 
$p\bar{p}$ interaction were rejected. Scans of events failing 
this cut indicate that $<$0.1\% per jet $E_T$ bin are real jet events.
Figure~\ref{metsig-ncut} shows the \METSIG distribution after the timing cut.
As in previous analyses, the \METSIG
was required to be less than 6~GeV$^{1/2}$.
Figure~\ref{metvset} shows scatter plots of \MET versus $\sum{E_T}$,
\MET versus lead jet $E_T$ (highest $E_T$ jet) and 
lead jet $E_T$ versus $\sum{E_T}$
before (left side) and after(right side) the \METSIG cut.
The efficiency of the \METSIG cut, 100 $^{+0}_{-1}$\%, was determined from
event scanning and the study of the properties of the events which
fail the cuts.
All these cuts are identical to those used in the previous 
analysis~\cite{CDF-Inclusive-Jet-96}.
In addition, events resulting from errant beam particles 
were more numerous in Run 1B 
than in previous measurements.  These were rejected by requiring the total
energy seen in the calorimeter to be $<$1800 GeV.  No jet events were 
rejected by this cut.
Remaining backgrounds are conservatively estimated to be
$<$0.5\% per bin with $E_T<260$ GeV. All the events containing a
cluster with $E_T>260$ GeV
were scanned and were found to be typical jet events. 
Figure~\ref{fig-metsig} shows the \METS after all the cuts
compared to the expected distributions from the HERWIG~\cite{HERWIG} Monte 
Carlo + CDF detector simulation.  The distributions are in good agreement.

%

\subsection{Additional checks}

The raw data are corrected for 
calibration, acceptance, and efficiency.  For these corrections we rely on
a detector simulation which has been tuned to the data as described 
in later sections.  The ultimate comparisons are to NLO parton level QCD
predictions.  These contain at most 3 partons which are identified as jets.
The fragmentation/hadronization of partons is well modeled for LO QCD
predictions, but complications and double counting would occur if these
models were used for the NLO predictions.  Thus for a study of
general event properties 
we use the HERWIG shower Monte Carlo to generate jets.  HERWIG uses
LO matrix elements, plus a leading log approximation for the parton shower
and then applies a cluster hadronization to convert the partons
to particles.
The resulting particles 
are passed through the detector simulation.
In the comparisons that follow, HERWIG 5.6 was used with CTEQ3M PDF's.
The data is divided into 6 $E_T$ bins shown in Table~\ref{tab-raw-plots},
based on the leading jet $E_T$. In the following series of Figures,
the lowest $E_T$ bin is plotted
is in the upper left corner, the next highest $E_T$ bin is to its right, etc.
The highest $E_T$ bin is the lower right corner.
The Monte Carlo output (histogram) is normalized to 
the CDF data in each bin.  There are at least 2500 MC events in each bin.

Figure~\ref{fig-metsig} shows the MC \METSIG distributions
in the six bins compared to the data. This quantity is sensitive 
to the simulation
of both the hard and the spectator interactions.  The agreement between 
the data and the MC improves with increasing jet $E_T$.  The cut on this 
quantity is used only to reject
background.  The MC distributions imply that this cut may have rejected 
1-2\% of the events above 300 GeV, although visual scans
of events with 6$<$ \METSIG $<$8 indicated that none were lost.

\begin{figure}
\centerline{\psfig{figure=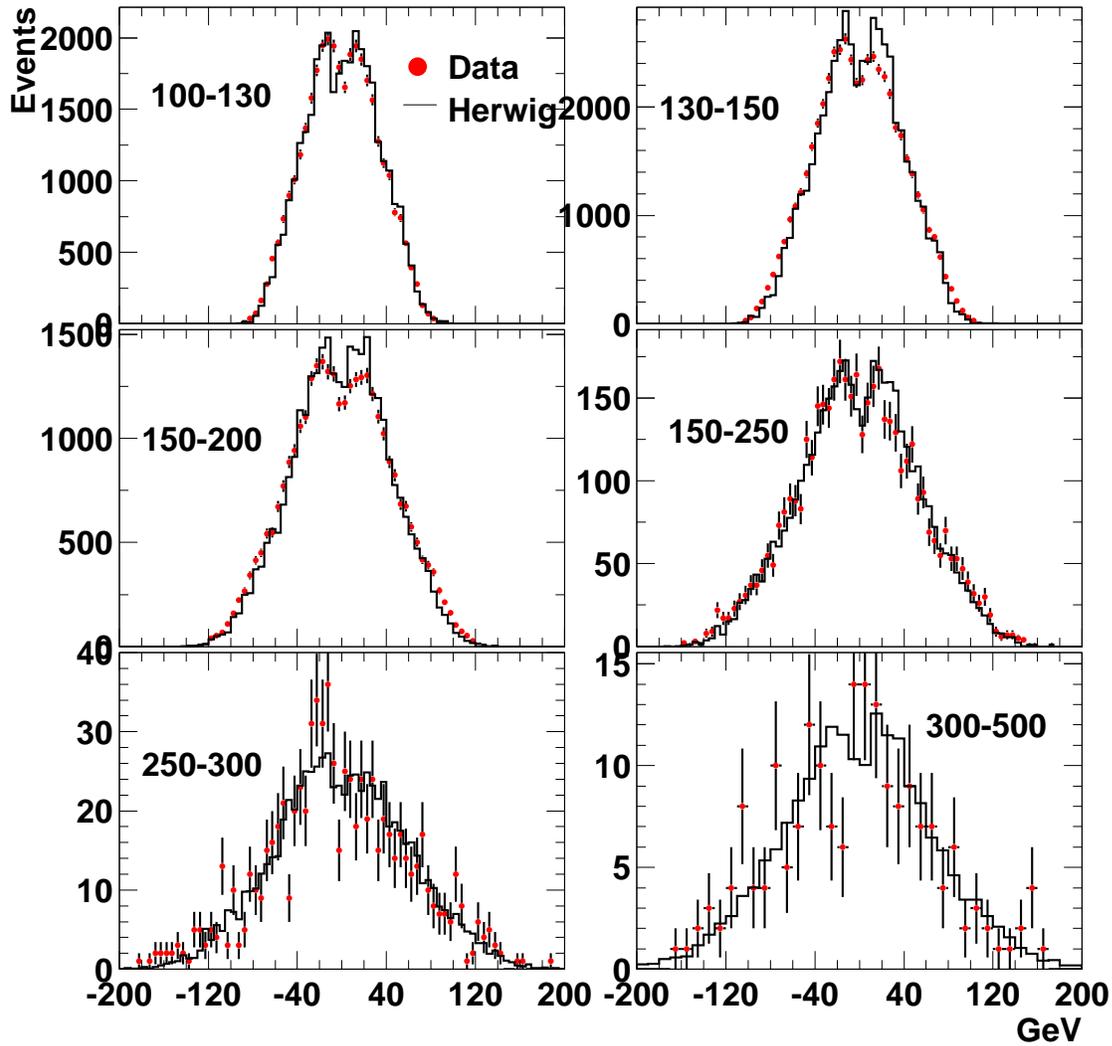,height=16cm,width=16cm}}
\caption{$E_T$ difference between leading two jets for data (points)
and HERWIG (histogram).
The sign of the difference
is chosen based on sign($\phi_1 - \phi_2$).
The labels on the individual plots (e.g. 100-130 GeV)
indicate the $E_T$ range of the leading jet.}
\label{fig-et12}
\end{figure}

Figure~\ref{fig-et12} shows the difference in 
the transverse energies of the two leading jets.  The sign of the difference
is chosen based on sign($\phi_1 - \phi_2$).
The $E_T$ difference is from a) energy resolution of the detector
and b) additional jets produced from the hard scattering.
As a shower MC, HERWIG has been found to model this additional jet
activity quite well up to jet multiplicities of six~\cite{SUMET}.
The agreement between data and HERWIG shown is this plot indicates that both
the energy resolution and the production of additional jets is well
modeled.

\begin{figure}
\centerline{\psfig{figure=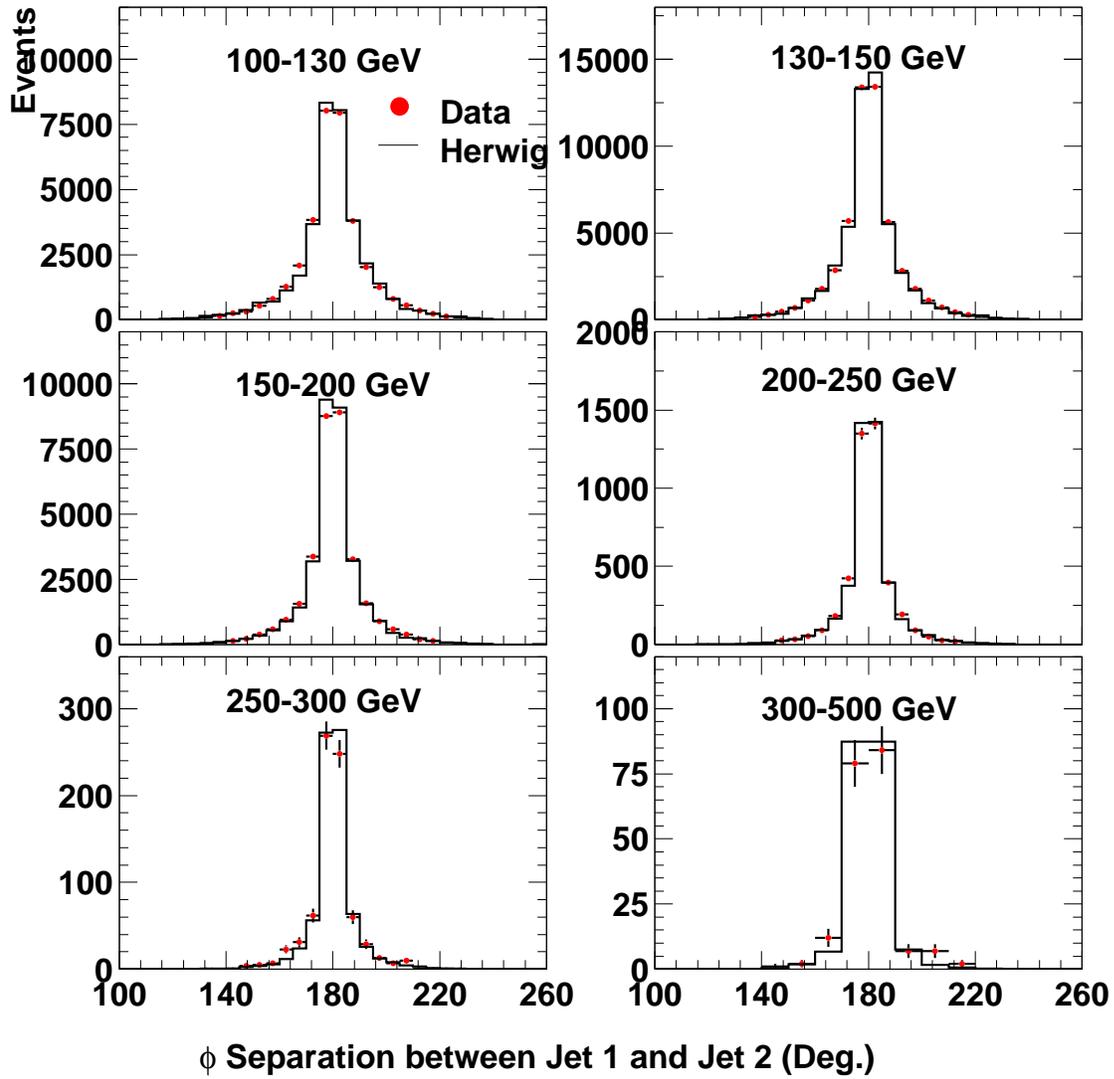,height=16cm,width=16cm}}
\caption{Difference in $\phi$ between leading two jets 
for data (points) and simulation (histogram).
The labels on the individual plots (e.g. 100-130 GeV)
indicate the $E_T$ range of the leading jet.}
\label{fig-phi12}
\end{figure}

Figure~\ref{fig-phi12} shows the difference in azimuthal angle of the 
two leading jets in the event.  As with the $E_T$ imbalance of the
2 leading jets, this quantity depends on the number of jets produced in
the hard collisions and on the non-uniformities and
resolution (this time in $\phi$ not $E_T$) of the detector.
Good agreement is observed.

\begin{figure}
\centerline{\psfig{figure=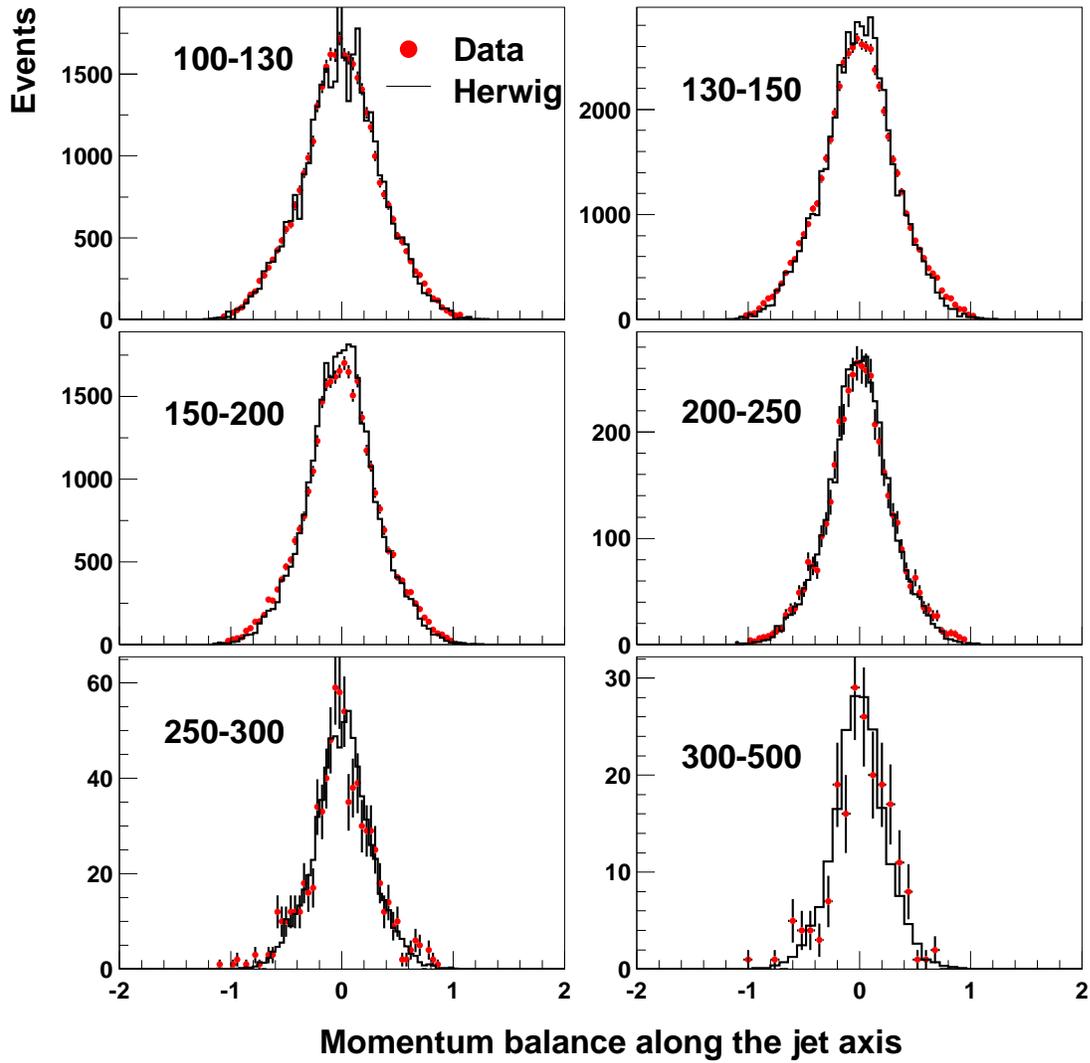,height=16cm,width=16cm}}
\caption{Fractional $E_T$ imbalance along dijet axis ($k_{||}$) 
for data (points) and simulation (histogram).
The labels on the individual plots (e.g. 100-130 GeV)
indicate the $E_T$ range of the leading jet.}
\label{fig-kpara}
\end{figure}

The effect of additional jets can be minimized by measuring
the energy mismatch parallel to the axis defined by the leading two jets.
We call this quantity $k_{||}$.
The direction of the projection axis
${\bf \hat{n}}$ is 
defined as perpendicular to the
bisector, ${\bf \hat{t}}$, of the two jets:
\begin{equation}
{\bf \hat{t}}  =  \frac{ {\bf \hat{n_1}}+{\bf \hat{n_2}}}
{|{\bf \hat{n_1}}+{\bf \hat{n_2}}|} 
\end{equation}
where ${\bf \hat{n}_{1,2}}$ are unit vectors along two leading jets in
the x-y plane.
Then  $k_{||}$ is given by
\begin{equation}
{{\vec E_t^1}\cdot{\bf \hat{n}}+ {\vec E_t^2}\cdot{\bf \hat{n}}}.
\end{equation}
Figure~\ref{fig-kpara} shows the normalized $k_{||}$ distributions 
($\frac{2k_{||}}{E_T^1+E_T^2}$)
for the
data and the MC simulation.  The good agreement indicates that
the jet energy resolution is well modeled by the detector simulation.

\begin{figure}
\centerline{\psfig{figure=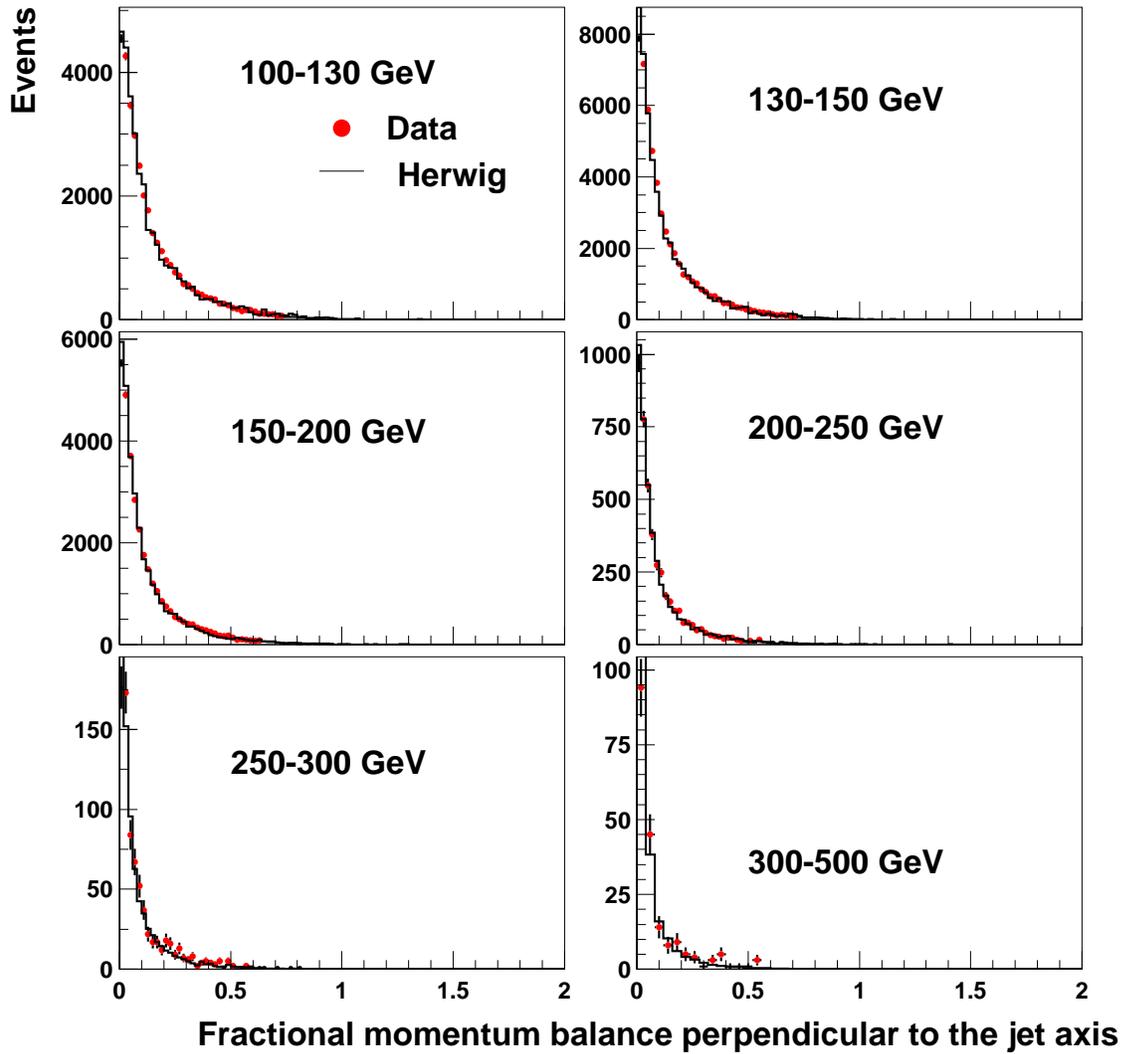,height=16cm,width=16cm}}
\caption{Fraction $E_T$ imbalance perpendicular to dijet axis ($k_{\perp}$)
 for data (points) and simulation (histogram).
The labels on the individual plots (e.g. 100-130 GeV)
indicate the $E_T$ range of the leading jet.}
\label{fig-kperp}
\end{figure}

The energy imbalance along the  ${\bf \hat{t}}$ direction, $k_{\perp}$,
is sensitive to both the energy resolution and to additional jet production.
Figure~\ref{fig-kperp} shows the normalized $k_{\perp}$ distributions. 
There is good agreement 
between the data and the Monte Carlo predictions.

\begin{figure}
\centerline{\psfig{figure=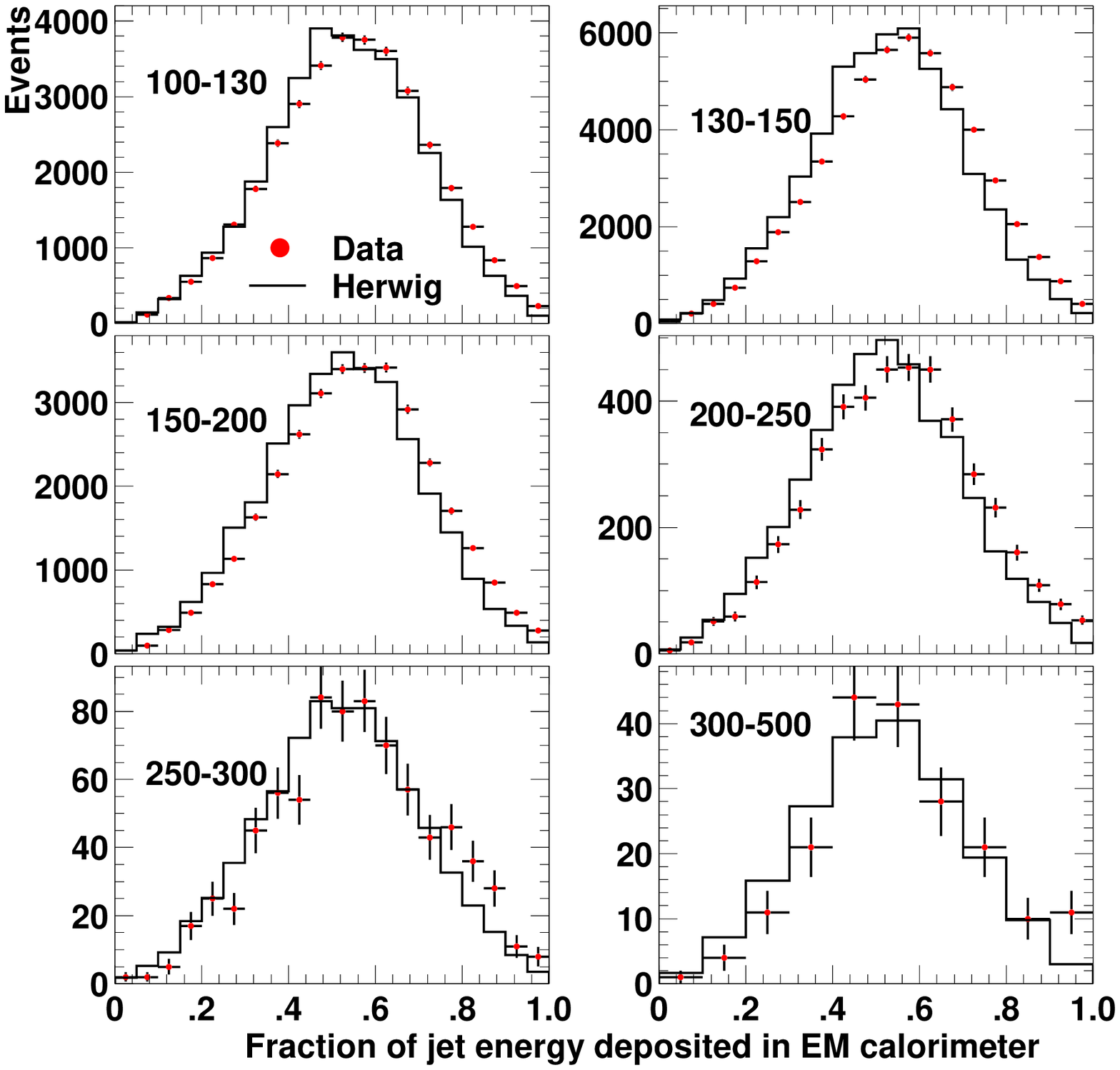,height=16cm,width=16cm}}
\caption{Fraction of electromagnetic energy in jets
 for data (points) and simulation (histogram).
The labels on the individual plots (e.g. 100-130 GeV)
indicate the $E_T$ range of the leading jet.} 
\label{fig-emf}
\end{figure}

The CDF calorimeter measures the energy in two depth segments.
The EM calorimeter is located in front of the hadronic calorimeter and
measures the energy of the electromagnetic particles (primarily 
$\pi^{0}$'s) in the jets, along with some energy from the
hadronic particles. Figure~\ref{fig-emf} shows 
the fraction of jet energy deposited in the EM calorimeter for events in
the six $E_T$ bins.
There is good agreement
between data and MC.
The discrepancies 
have a very small effect on jet energy calibration.

\begin{figure}
\centerline{\psfig{figure=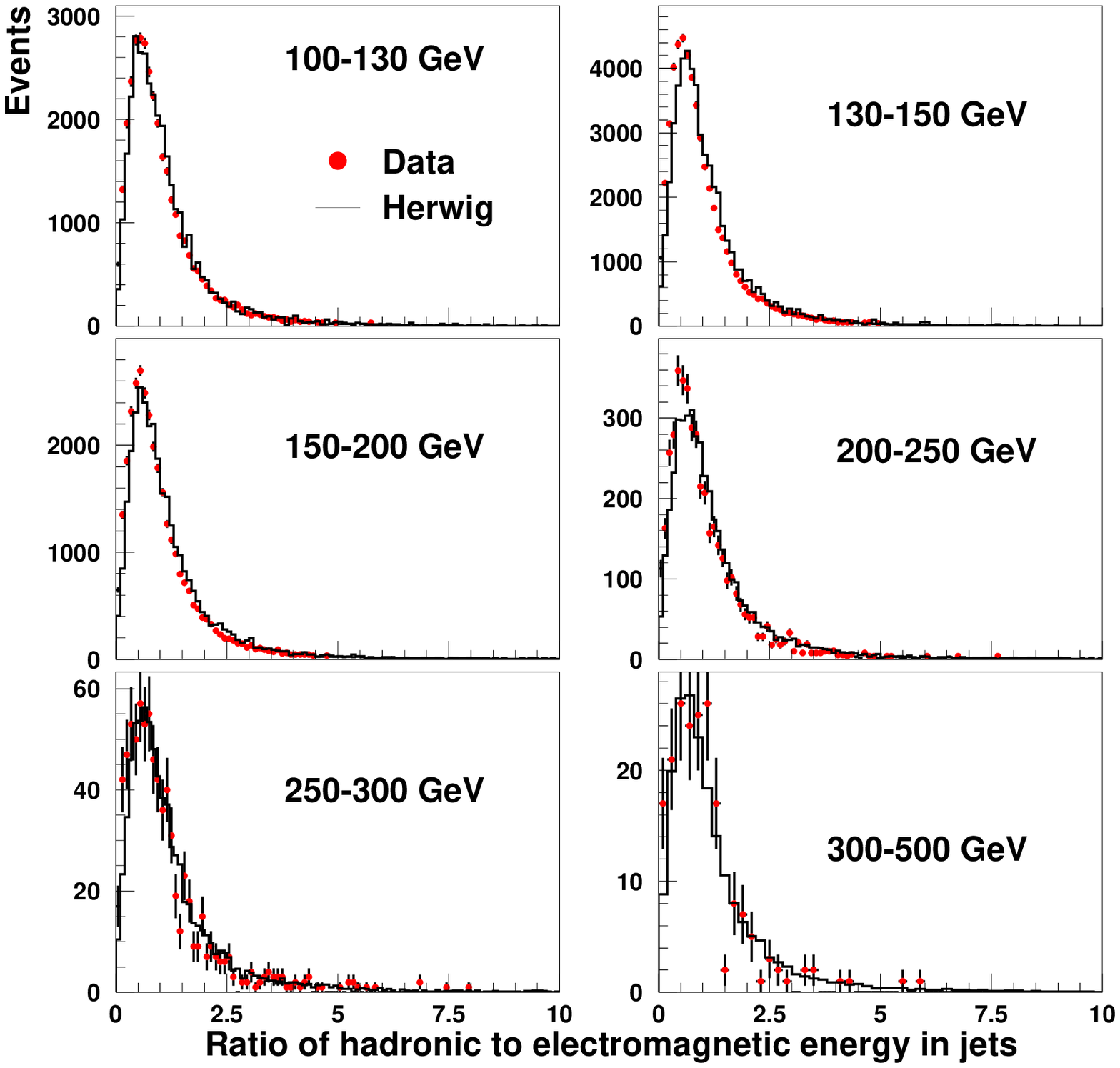,height=16cm,width=16cm}}
\caption{Ratio of hadronic to electromagnetic energy in
jets for data (points) and simulation (histogram).
The labels on the individual plots (e.g. 100-130 GeV)
indicate the $E_T$ range of the leading jet.}
\label{fig-hadem}
\end{figure}

Higher $E_T$ jets fragment into higher $P_T$
particles which 
sample the calorimeter at greater depths. 
The scintillator response 
might not be constant as a function
of depth due to radiation damage from the beam exposure.
This effect is not included in the detector simulation.
The electromagnetic section is calibrated using electrons from 
collider data and thus reduced response due to aging is already accounted for.
The ratio of the jet energy measured in the 
hadronic and  electromagnetic calorimeters,
(1-emf)/emf, would be sensitive to this effect.
Figure~\ref{fig-hadem} shows that the agreement between
data and MC predictions is good.
We conclude that 1) there is no detectable
depth-dependent effect and 2) there is no detectable extra leakage for
high $E_T$ jets. 

These checks reveal no systematic
problems with the high $E_T$ data which are not modeled by 
the detector simulation or included in our systematic uncertainties.


\section{Corrections to the Raw Cross Section}

The raw cross section must be corrected for energy mismeasurement and for the
smearing caused by finite $E_T$ resolution.
An ``unsmearing procedure"~\cite{CDF-XT-analysis} is used to 
simultaneously correct for both effects. A consequence of this
technique is that the corrections to the jet cross section are directly coupled 
to the corrections to the jet energy.  
The unsmearing procedure involves
three steps.
First, the response of the calorimeter to jets is measured and parameterized 
using a jet production model plus a detector simulation 
which has been tuned to the CDF data.  Specifically, particles 
produced by a leading order dijet MC plus fragmentation are clustered
into cones in ($\eta - \phi$) of radius 0.7. This defines the
corrected (or true) jet energy.  To estimate the response of the detector to
jet events, particles from an underlying event are added to the jet 
fragmentation particles and all the particles are traced through 
the detector and then
clustered with the standard CDF algorithm.  Fluctuations in the
underlying event and in the detector response are included in this process.
The distribution of measured jet $E_T$ for a given true jet $E_T$ is
called the response function.

Second, a trial spectrum is convoluted (smeared) with 
the response functions and
fit to the measured data.  The parameters of the trial spectrum
are adjusted to find the minimum $\chi^2$.
Finally the correspondence between 
the trial spectrum, and the smeared spectrum is used to derive bin-by-bin
corrections to the measured spectrum.  The statistical fluctuations
present in the raw data are preserved in the corrected spectrum.
The details of these three steps are discussed below.

\subsection{Response Functions}

The response functions give the relationship between the energy 
measured in a jet cone in the calorimeter and the true $E_T$ of the 
originating parton (e.g. the sum of the particles in a cone of 0.7 around the
original parton direction).
If the calorimeter were perfectly linear the
response functions would be derived simply from sum of the 
energy of the jet particles within a cone of R=0.7.
However, since our calorimeter is non-linear below 10 GeV,
the response to a jet depends on the $P_T$ spectrum of the particles 
in the jet.  As a simple example, the response to a 30 GeV
jet is different if it is made of two 15 GeV particles 
compared to six 5 GeV particles. Thus, to understand the
calorimeter response to jets, we measure both the response to 
single particles (calibration) and the number and 
$P_T$ spectrum of the particles within a jet.

Corrections for the effect of 
the underlying event energy are included in the response functions:
the true $E_T$ is defined before the underlying event is added
while the measured $E_T$ contains the underlying event contribution.
The amount of underlying
event energy is measured in the data and is described later.
As in previous analyses, no correction is applied for the 
energy from the partons or fragmentation which falls outside the jet
cone. Estimates of this 
energy are fundamentally dependent
on assumptions in theoretical models and are partially included 
in the NLO predictions.
In the next two sections we describe how the detector calibration and the
jet fragmentation are measured
in the data and used to tune the Monte Carlo simulations.

\subsubsection{Calibration}
 
\begin{figure}
\centerline{
{\hbox{
\psfig{figure=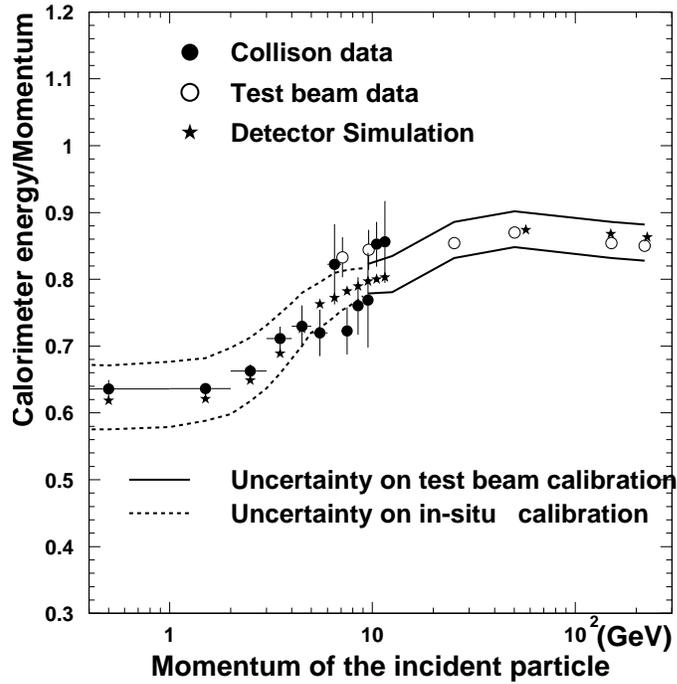,height=10cm,width=10cm}
}}}
\caption{
In situ and test beam single particle 
response as a function of particle momentum.
The stars indicate the response in the detector simulation.}
\label{fig-tb-e-over-p}
\end{figure}

\begin{figure}
\centerline{
{\hbox{
\psfig{figure=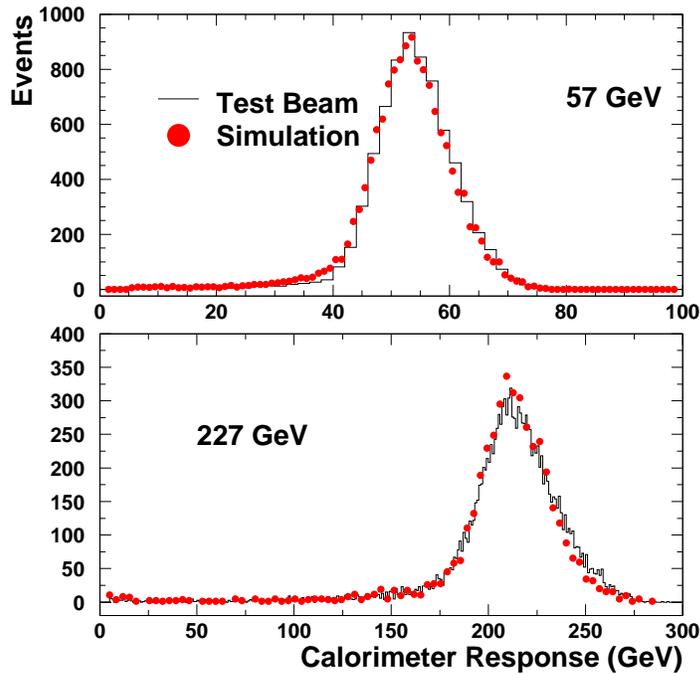,height=10cm,width=10cm}
}}}
\caption{
$E_{cal}/P_{\pi}$ for test beam pions
and detector simulation.}
\label{fig-tb-e-over-pb}
\end{figure}

The calorimeter response was measured using 10, 25, 57, 100 and
227 GeV electrons and pions from a test beam. 
Figure~\ref{fig-tb-e-over-p} shows the calorimeter response compared to the 
simulation for various pion energies. The band around the 
mean values shows the systematic uncertainty which includes
the  uncertainties in the testbeam momenta, the variation
of the calorimeter response over the face of tower and the tower-to-tower
variations.
At high $P_T$ the calorimeter is found to be linear up to the last measured
point (227 GeV). No evidence of photo-tube saturation or additional 
leakage of showers for high $P_T$ pions is observed. 
The shape of the calorimeter response to 57 and 227 GeV pions compared with the
simulation is shown in Figure~\ref{fig-tb-e-over-pb}. 

At low $E_T$ the response of the calorimeter was measured by selecting
isolated tracks in the tracking chamber.  The tracks were extrapolated to
the calorimeter and the corresponding energy deposition was compared to the
track $P_T$.  This technique allowed the response of
the calorimeter from 0.5 to 10 GeV to be measured in situ during the
data collection periods.
Figure~\ref{fig-tb-e-over-p} shows the measured E/P distribution.  The 
band around the points represents the systematic uncertainty which 
is primarily due to neutral pion background subtraction. 
The CDF hadronic response is non-linear at low 
$P_T$, decreasing from 0.85 at $P_T=10$ GeV to 0.65 at $P_T=1$ GeV.


The central electromagnetic calorimeter was calibrated  
using electrons from the collider data and with periodic radioactive
source runs.
This calorimeter is linear over the 
full $P_T$ range.  The response of the calorimeter was found
to decrease slowly with time (roughly 1\% per year).  
This reduction is monitored with the electron
data and an average response for the data sample is derived from the
$Z$ mass.  Each jet is corrected for this scale change according to 
the electromagnetic energy (neutral pions) of the jet.

\subsubsection{Jet Fragmentation}

The $P_T$ spectrum of the charged particles
in a jet (fragmentation functions) was measured from CDF data using 
tracking information. 
The shower MC program ISAJET + a detector simulation were used to 
study the jet response.  ISAJET has 
a Feynman-Field fragmentation model which allows easy tracing of particles to
their parent partons.  The fragmentation functions can also be tuned to 
give excellent agreement with the data.  The agreement is limited only
by the statistical precision of the data~\cite{CDF-XT-analysis}.
Our tuned version of this fragmentation function is called CDF-FF.
The uncertainty on the fragmentation functions
was derived from the uncertainty in the track reconstruction.

\begin{figure}
\centerline{
{\hbox{
\psfig{figure=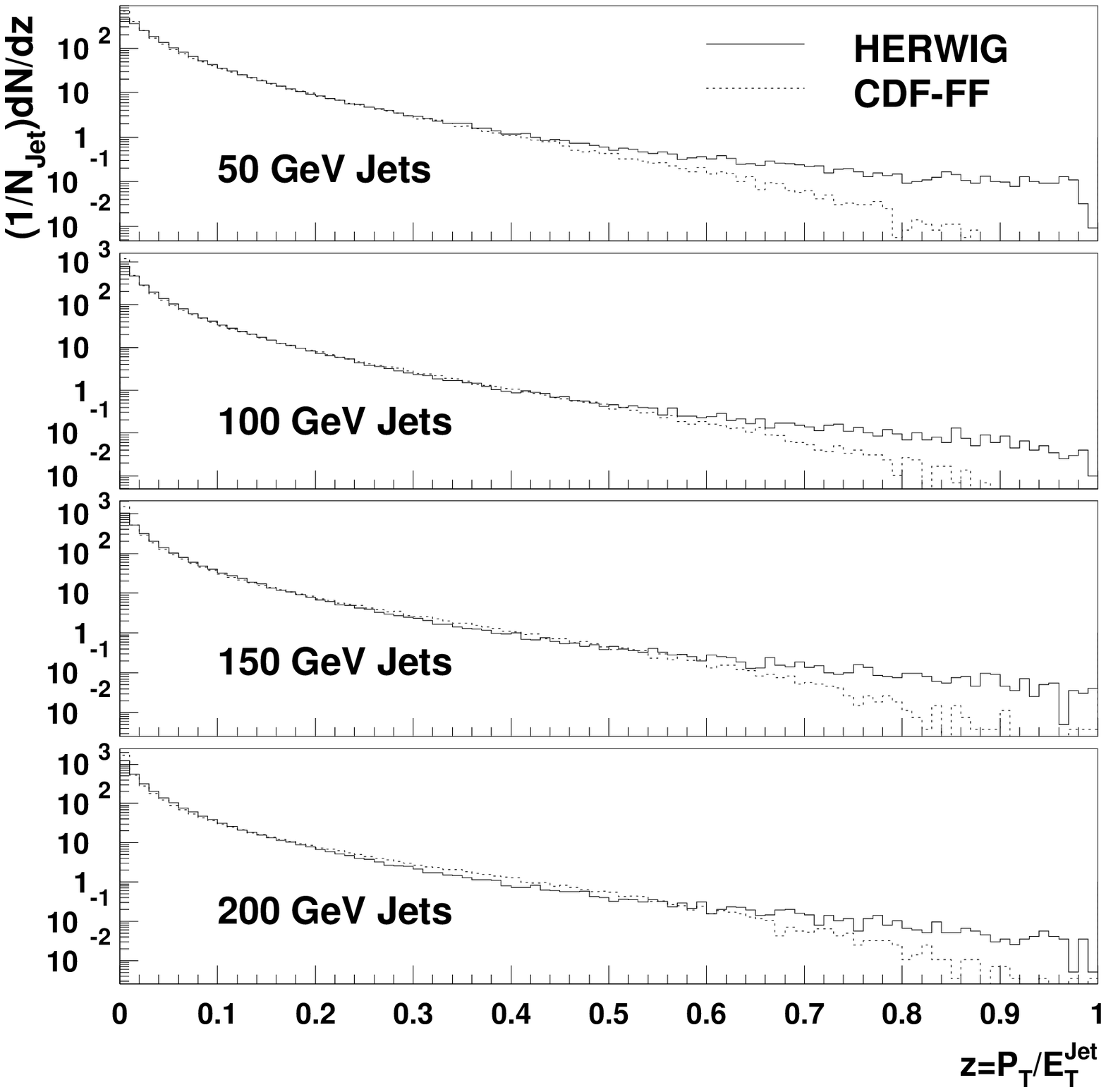,height=8cm,width=8cm}
\psfig{figure=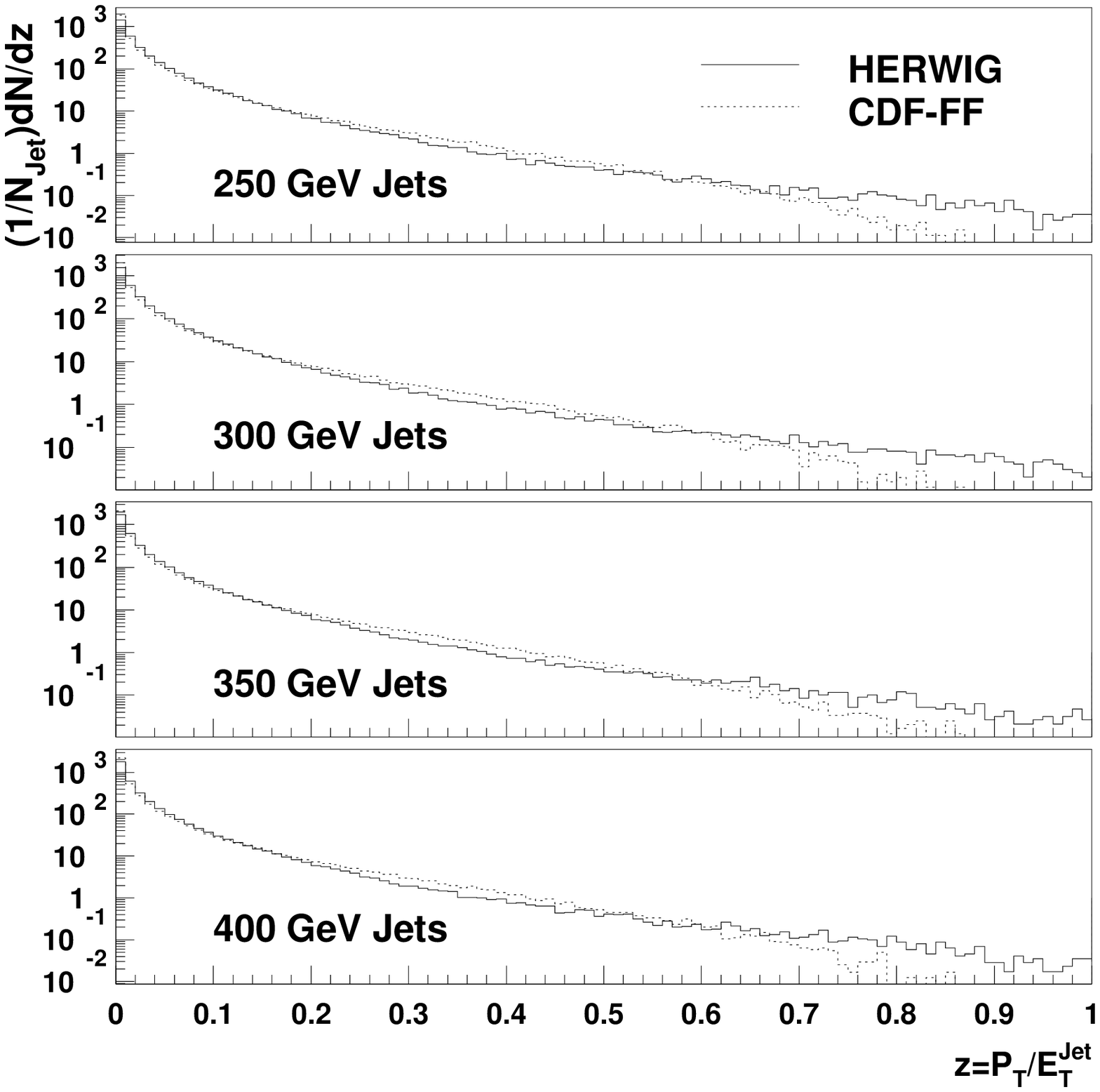,height=8cm,width=8cm}
}}}
\caption{The jet fragmentation properties for different $E_T$ jets
using CDF-FF and HERWIG fragmentation functions}
\label{fig-fragmentationab}
\end{figure}

As a cross check, jet response functions were also derived using the
fragmentation in HERWIG Monte Carlo.  This fragmentation is similar to
a string fragmentation and was tuned to the LEP data, but not to the CDF data.
The HERWIG
fragmentation is compared with the CDF fragmentation
(without any detector simulation) in 
Fig.\ref{fig-fragmentationab}.  The agreement between the two sets is very good.
The change in the cross section when the HERWIG fragmentation functions
were used instead of the CDF-FF functions
is smaller than the uncertainty attributed to
fragmentation functions (see below).

\begin{figure}
\centerline{
{\hbox{
\psfig{figure=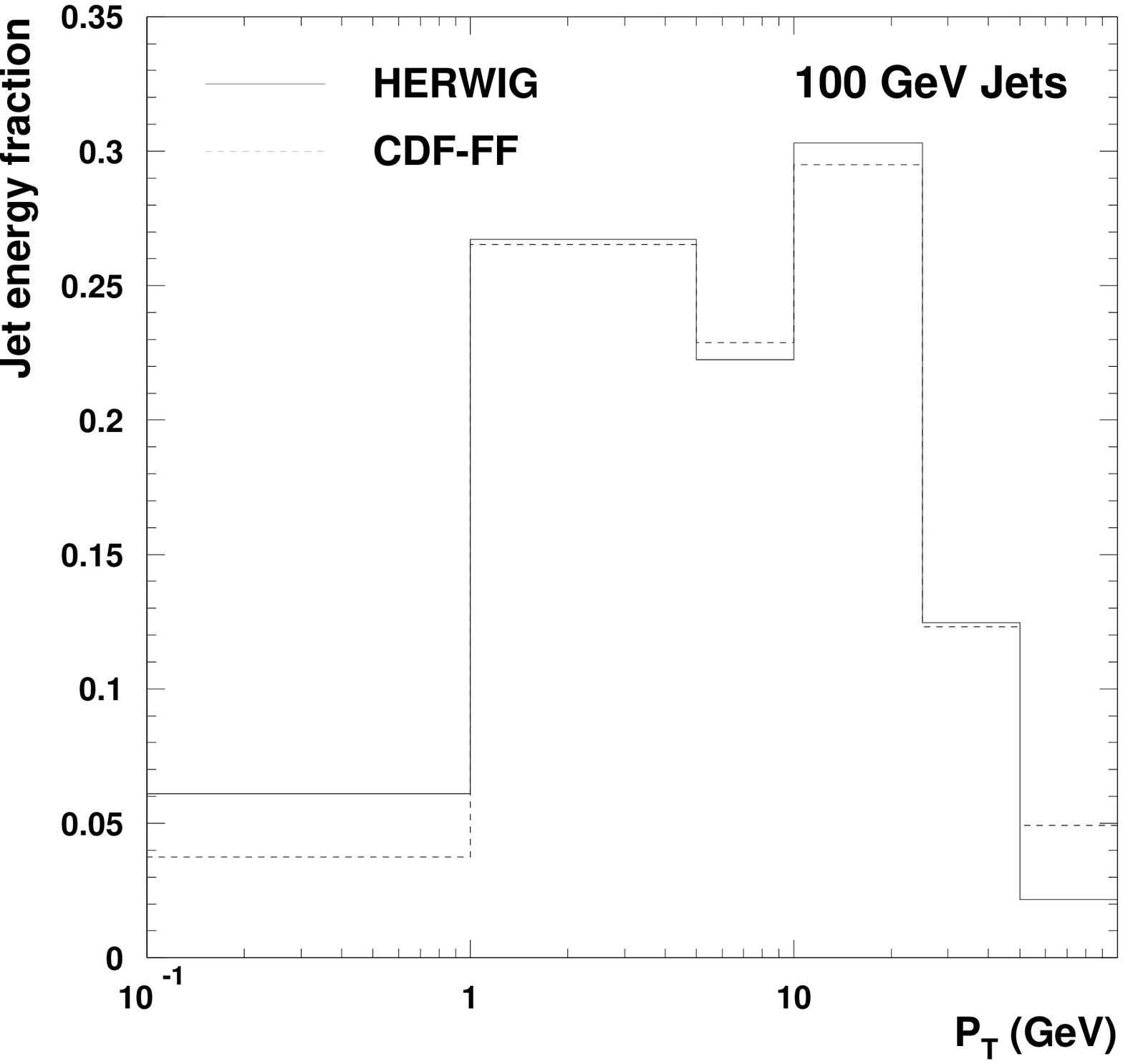,height=8cm,width=8cm}
\psfig{figure=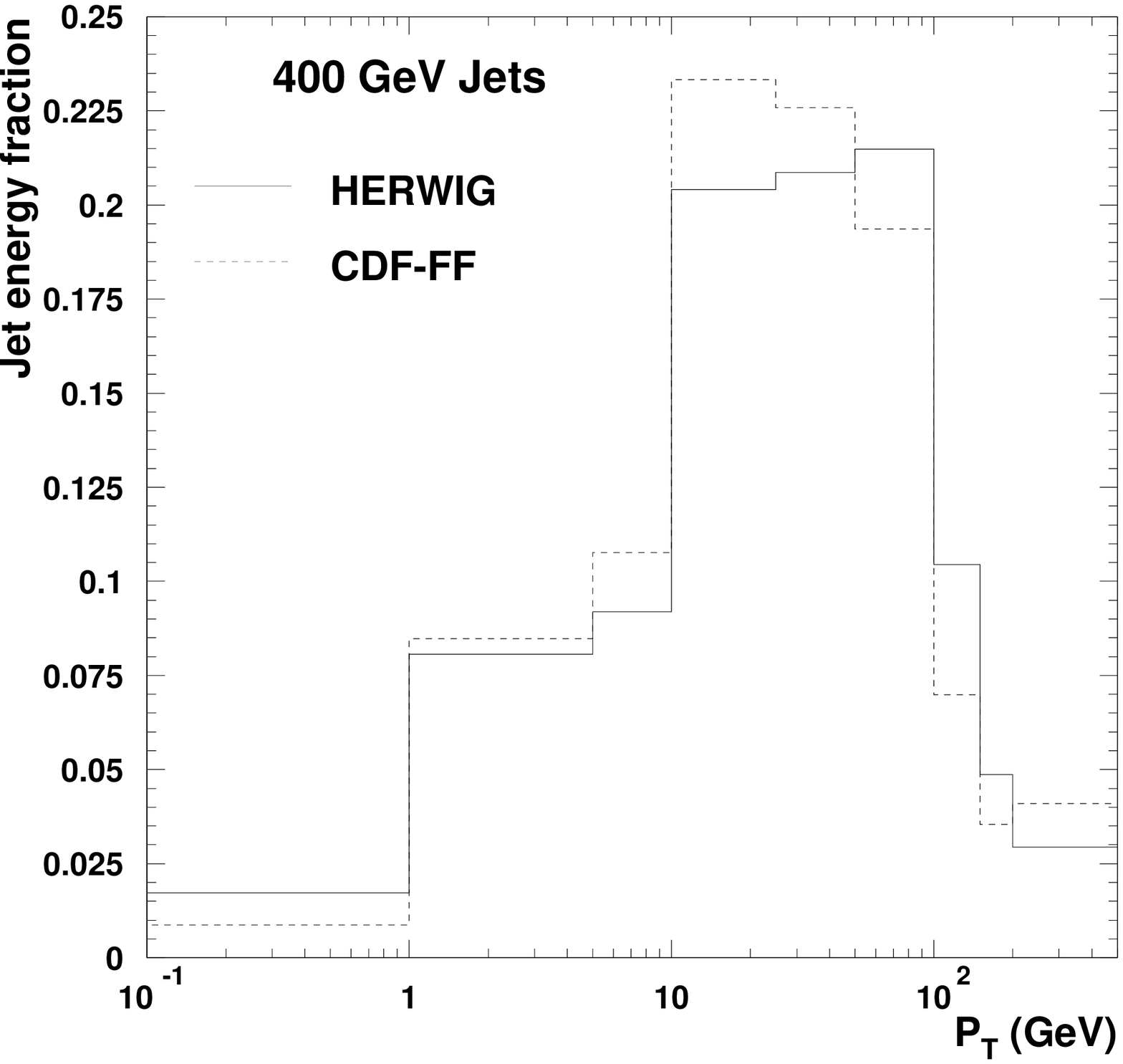,height=8cm,width=8cm}
}}}
\caption{Fraction of jet energy in particles of different $P_T$ }
\label{ptfrac}
\end{figure}
\begin{figure}
\centerline{
{\hbox{
\psfig{figure=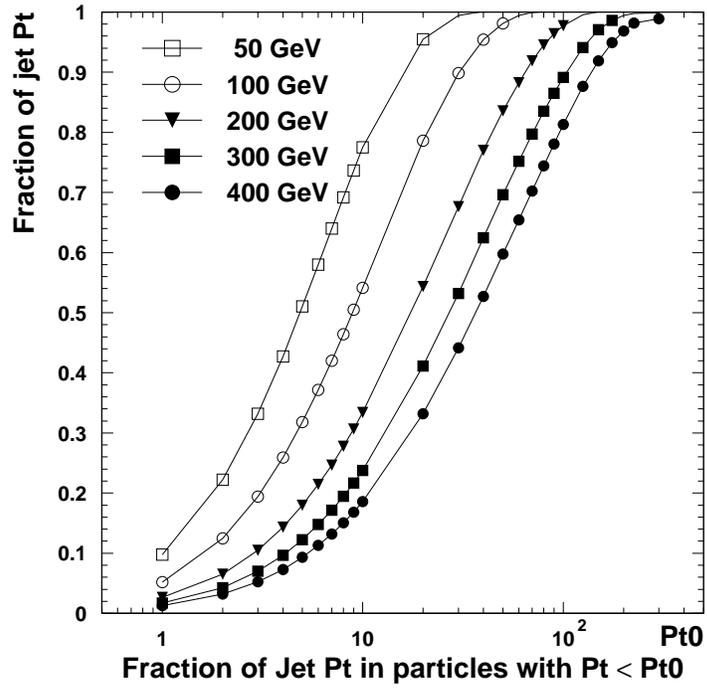,height=10cm,width=10cm}
}}}
\caption{The 
fraction of jet $E_T$ carried by the (true) particles with $P_T < P_{T_0}$
using HERWIG}
\label{fig-fragmentationc}
\end{figure}

In addition to the low energy non-linearity mentioned above, one might
be concerned about potential non-linearity at very high $E_T$, beyond 
the reach of the testbeam calibration (227 GeV).
Figure~\ref{ptfrac} shows the percent of jet energy carried by
different $P_T$ particles for 100 GeV jets and 400 GeV jets.
Both the CDF-FF model and HERWIG are shown and
are in good agreement.  Note that even in 400 GeV jets,
less than 4\% of the jet energy 
is carried by particles with $P_T>$ 200 GeV.
Fig.\,\ref{fig-fragmentationc} shows the HERWIG prediction for the 
fraction of jet energy 
carried by particles of different $P_T$. 
For jets with $E_T>200$ GeV, 
only a few percent of energy goes in the non-linear low $E_T$ region 
and in the region above the last test beam point.


\subsubsection{Underlying event and multiple interactions}

The underlying energy in the jet cone (i.e. the ambient energy from
fragmentation of partons not associated with the hard scattering)
is not well defined theoretically. 
We thus develop our own estimates of 
the amount and effects of this energy.  
Two techniques have been used in the past. 
In the first, energy was measured in cones perpendicular in $\phi$ to the 
dijet axis. In the second, ambient energy was measured in
soft collisions (e.g. the minimum bias sample discussed in section IV.A).
Comparison of these energy levels found that the jet events
were significantly more active than the minimum bias events.
Studies with jets in different regions of the detector and
with the HERWIG Monte Carlo indicated that about half the increased energy 
in the jet events was due to radiation from the jets
and that there was roughly a 30\% variation in the energy perpendicular
to the jet axis depending on event selection criteria~\cite{bfthesis}.
For comparison to NLO predictions (where the effects of gluon radiation
are included at some level) it is appropriate to subtract only 
the energy from the soft collision.  One subtly is that since jets 
arise from collisions with small impact parameters, the interaction
of the hadron remnants might be more energetic than in the average 
minimum bias event.  For these reasons, all jet analyses at CDF assume
an uncertainty of 30\% on the underlying event energy which contributes to
a jet cone.  This should be kept in mind when comparing to 
measurements from other experiments~\cite{d0ue}.

For the analysis in this paper, the primary method we use to estimate the
underlying event energy is
based on the minimum bias data sample.
An alternative method, which
uses the energy in a cone perpendicular to the leading jet direction
gives similar 
results and is described at the end of this section.
Both the minimum bias data sample and the jet data include 
events which have multiple soft $p\bar{p}$ collisions.  Corrections
for this effect are also derived.

To estimate an average underlying event contribution to the jet energy 
from the minimum bias data,
a cone of radius 0.7 was placed at random locations in the region 
of our measurement.
The energy in the cone is 
measured as a function of the number of 
vertices.   For the minimum bias data the average number of 
vertices is 1.05.  The energy as a function of the number of
found vertices is shown in Table~\ref{table-ue}.  
In the jet samples the average number of found vertices was 2.1.
An average correction for the jet data is found by combining the energy
measured in the cone in the minimum bias data and
the number of interactions in the jet data.  For a cone of 0.7 the
correction to the raw jet $E_T$ is 2.2 GeV.
This correction is applied as a shift in the mean of the jet response 
functions and the tails of the response function are scaled appropriately.

An alternative method for estimating the
underlying event energy was also investigated.  The energy deposited at
$\pm$ 90$^0$ in $\phi$ from the jet lead axis in a cone of 0.7 was measured.  
The cones 
at 90$^0$ will contain energy from jet activity, energy from the proton 
remnants and energy from any additional $p\bar{p}$ collisions in the same event.
To estimate the contribution of the ``jet activity", 
we compared the energy in the cones at +90 and -90$^0$.  Jet activity 
can contribute to both cones, however, one cone is usually closer to
a jet since the jets are not exactly 180$^0$  apart.
Separate averages of minimum and maximum 90$^0$ cone energies in each event 
were formed. The mean $E_T^{max-cone}$ was found to depend on the average
$E_T$ of the jets in the events while the mean $E_T^{min-cone}$ was
independent of the jet $E_T$.  The mean $E_T^{min-cone}$ for each of 
the jet trigger samples was 2.2 $\pm$ 0.1 GeV.   This is in good agreement
with the estimate based on the number of vertices in the jet data and the
minimum bias data result.
Additional studies were performed varying the tower threshold 
for inclusion in the clusters. The single tower threshold used for
jet clustering is 100 MeV.
Lowering the tower threshold from 100 to 50 MeV increased the
measured energy in a cone by 140 MeV.  

While a measurement of the energy in a cone either in minimum bias data, or
the jet data can be made precisely (few percent), there is a large 
uncertainty in
the definition of the underlying event.  To cover 
definitional differences and threshold effects
we assign an uncertainty of 30\% (0.66 GeV) to the underlying event energy.
This is the dominant uncertainty for the low $E_T$ 
inclusive jet spectrum.

\subsubsection{Cross checks of the jet energy scale}

As discussed earlier, the
jet energy scale is set by the in-situ calibration 
with single particles at low $E_T$ and by the test beam data at high $E_T$.
The validity of the resulting corrections 
can be cross-checked using 
events with a leptonically decaying Z boson and one jet.
The transverse momentum balancing of the jet and the $Z$ was 
measured and compared to the Monte Carlo simulations used in
this analysis~\cite{topmass}.
The ratio of $[P_T(Z) - P_T(jet)]/P_T(Z)$ observed in the data
was 5.8\%  $\pm$ 1.3(stat.)\%, compared to the 4.0\%  $\pm$ 0.3(stat.) \% 
in the Monte Carlo simulation for jets with a cone size of 0.7.
The actual value of the imbalance is influenced by the presence
of additional jets in the events, and the 
transverse boost of the $Z$-jet system.  This measurement 
required that any jets other than the leading jet have less than 6 GeV $E_T$
and that the $P_T$ of the reconstructed $Z$ boson be greater than 30 GeV.
Without any cut on the second jet, the $P_T$ imbalance between
the $Z$ and the leading jet rises to roughly 11-12\% 
in both the data and the Monte Carlo simulation.  
This imbalance was also separated 
into components parallel and perpendicular to the $Z$-jet axis and 
both were found to be in reasonable agreement with the data.
The imbalance was also studied for different jet cone 
sizes (R=0.4, 0.7 and 1.0).  In general,
the magnitude increased with larger cone sizes and the agreement
between data and Monte Carlo improved.
The uncertainty on the imbalance due to 
the uncertainty in the jet energy scale corrections is 3-4\% and covers and
difference between the data and MC simulation.
Thus, we do not attempt to correct the jet energy scale or 
tune the Monte Carlo based on these results.  Rather, we take the agreement
between the data and the detector simulation as an indication that the
simulation does a good job reproducing the response of the detector
to jets.

The jet energy scale can also be verified by reconstructing the $W$ mass 
from the two non-b jets in top events~\cite{topmass2}.  
The measured $W$ mass is 
consistent with the world average $W$ mass.
From these checks we conclude that the jet energy scale and corrections
are well understood and that the Monte Carlo
simulations are in good agreement with the data.

\subsubsection{Parameterization of the Response functions}

\begin{figure}
\centerline{
\psfig{figure=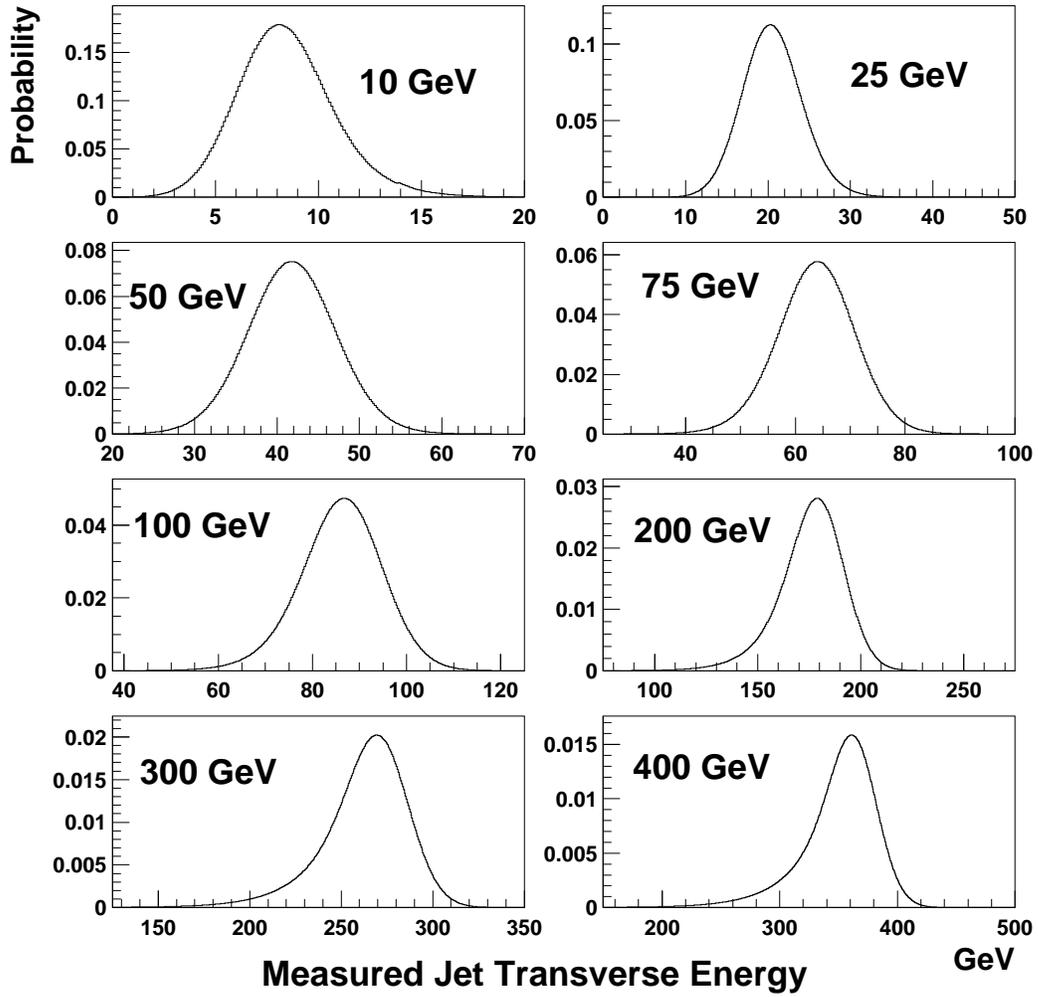,height=15cm,width=15cm}}
\caption{CDF calorimeter response for different $E_T^{True}$ jets}
\label{fig-resp}
\end{figure}

Using the Monte Carlo + detector simulation described above,
the response of the calorimeter to jets of various
true $E_T$ is simulated.  
We call $E_T^{True}$ the 
sum of the $E_T$ of all particles in a cone of R=0.7 around the
jet axis which originated from the scattered parton.
We denote $E_T^{smeared}$ to be the $E_T$ of the jet after the
detector simulation. The $E_T^{smeared}$ distribution for a given
$E_T^{True}$ is fit using four parameters (mean, sigma and
the upward and downward going tails). This function is called 
the ``response function".  
The shape of the response functions for different $E_T^{True}$ are shown in 
Fig.\,\ref{fig-resp}. The low-$E_T$-tails increase with
increasing $E_T^{True}$ because the jets become narrower
and hence the effects of the detector cracks become
more prominent.

\subsection{Unsmearing the measured spectrum}

Armed with the response functions, we can now determine the true
spectrum from the measured distribution through the following steps.

We parameterize the true (corrected) inclusive jet spectrum with
functional form
\begin{eqnarray}
\frac{d\sigma(E_T^{True})}{dE_T^{True}} =
P_0\times (1-x_T)^{P_6} \times 10^{F(E_T^{True})}
\end{eqnarray}
where 
$F(x)= \sum_{i=1}^{5} P_{i}\times \left[\log(x)\right]^{i}$,
$P_0 ... P_6$ are fitted parameters
and $x_T$ is defined as $2E_T/\sqrt{s}$.

\begin{figure}
\centerline{
\psfig{figure=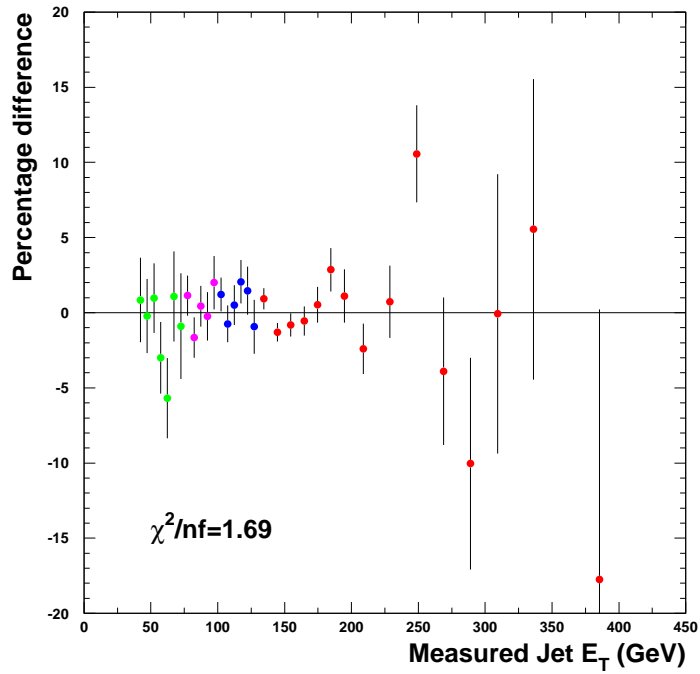,height=10cm,width=10cm}}
\caption{Residuals of the best fit curve (standard curve) and 
the measured cross section}
\label{fig-resid}
\end{figure}

The smeared (i.e corresponding to the measured cross section) 
cross section in a bin is
then given by
\begin{eqnarray}
\sigma^{smeared}(bin) =
\int_{L}^{H}d E_T\int_{5}^{600} d E_T^{True}
\{\frac{d\sigma(E_T^{True})}{dE_T^{True}}\}
Response(E_T^{True},E_T)
\end{eqnarray}
where H,L are the upper and lower edges of the measured $E_T$ bins.
To obtain the parameters of the true spectrum, we fit the smeared spectrum,
$\sigma^{smeared}(bin)$, to the measured cross section. 
The parameters of the input true spectrum $P_{1...6}$ are adjusted until
a good fit is obtained.  The $P_0$ parameter is determined by 
requiring the total smeared
cross section to equal the total measured cross section.
For the Run 1B data sample, the best fit parameters 
of the true cross section are given in Table~\ref{Table-syserr}.
We refer to this as the ``standard curve".
The residuals ($\sigma^{measured}(bin)$
-$\sigma^{smeared}(bin)$)/(data stat. unc.) as a function 
of $E_T$ for the standard curve are shown in 
Fig.\,\ref{fig-resid}. The $\chi^2$/DOF for the fit is 43.88/(33-7)
corresponding to a confidence level (CL) of 4\%.
No systematic biases in the fit are observed.  The errors on the points are
the sum in quadrature of the statistical uncertainty in the measured 
cross section and the uncertainty in the trigger efficiency and 
normalization factors.
Note that the integration is over the full spectrum and thus the best-fit true 
spectrum does not depend on the binning of the data.  Finer and coarser binning
were tried and did not affect the results or conclusions.

\begin{figure}
\centerline{
\psfig{figure=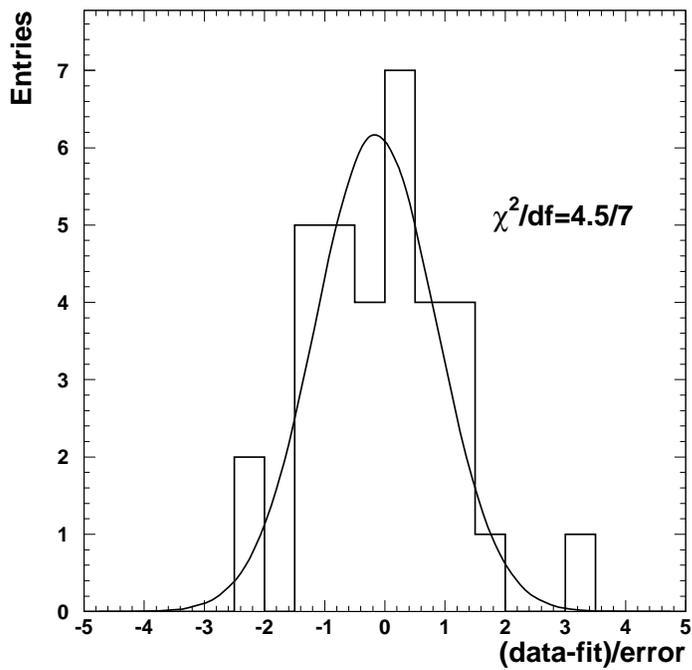,height=10cm,width=10cm}}
\caption{Residuals of the best fit curve (standard curve) and 
the measured cross section.  Distribution is fit to a Gaussian of 
width 1.0}
\label{fig-resid1-raw}
\end{figure}

To further investigate the significance of the large total $\chi^2$,
we histogram the residuals of the fit as shown in 
Figure~\ref{fig-resid1-raw}. The RMS width of the distribution
is 1.16 instead of the expected value of 1.0, a reflection
of the large total $\chi^2$, but the distribution is fairly Gaussian.
Figure~\ref{fig-resid1-raw} also shows a fit to a Gaussian
of width 1 gives a $\chi^2/DOF$ of 5.9/10.  More explicitly, 20
out of 33 points
(60\%) are within $\pm 1 \sigma$. 
We have carried out numerous checks that our errors were not underestimated
and could find no indication of such.
We conclude that the large $\chi^2$ and low probability
for the fit to the standard curve is due to a statistical fluctuation.

\subsubsection{$E_T$ and Cross Section Corrections}

Given the true spectrum, we can correct the measured data.
The $<E_T^{corrected}>$ for a bin is defined as
\begin{eqnarray}
<E_T^{true}>\times \frac{<E_T^{measured}>}{<E_T^{smeared}>}
\end{eqnarray}
where averaging is done on the raw bins.
The corrected cross section for the bin at the
$<E_T^{corrected}>$  then given by
\begin{eqnarray}
\sigma^{true}(E_T^{corrected}) \times 
\frac{\sigma^{measured}(bin)}{\sigma^{smeared}(bin)}
\end{eqnarray}

\begin{figure}
\centerline{
\psfig{figure=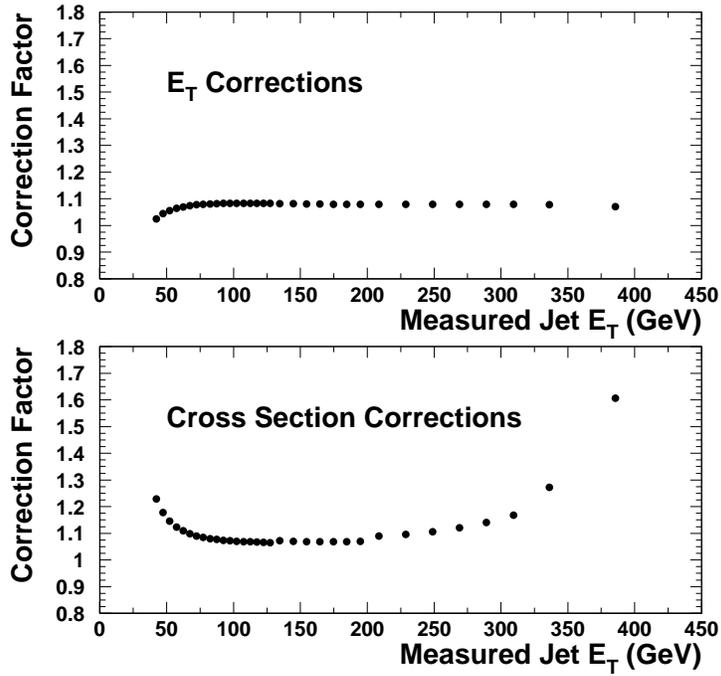,height=10cm,width=10cm}}
\caption{
The ratio of corrected $E_t$ and corrected cross section to the measured
$E_T$ and measured cross section}
\label{fig-cor}
\end{figure}

Thus, the corrected cross section values are the true
spectrum evaluated at a particular $E_T$ value (i.e. $<E_T^{corrected}>$),
and the $E_T$ and cross section correction factors 
are correlated.
The $E_T$ and cross section correction factors are given in
Fig.\,\ref{fig-cor}.  The correction factors are almost constant
except at extremely low $E_T$ and high $E_T$ where the spectrum is
very steep. 

The unsmearing procedure was extensively
tested with simulated event samples based on $E_T$ spectra
from the current data and the NLO QCD theory predictions.
The corrected cross section is
stable at better than a $5\%$ level to different choices of the
functional forms of true spectrum even for the highest $E_T$ points.
However, it should be noted that the uncertainty increases substantially
if the curve is extrapolated beyond the
last data point.

\subsection{Corrected inclusive jet cross section}

\begin{figure}
\centerline{\psfig{figure=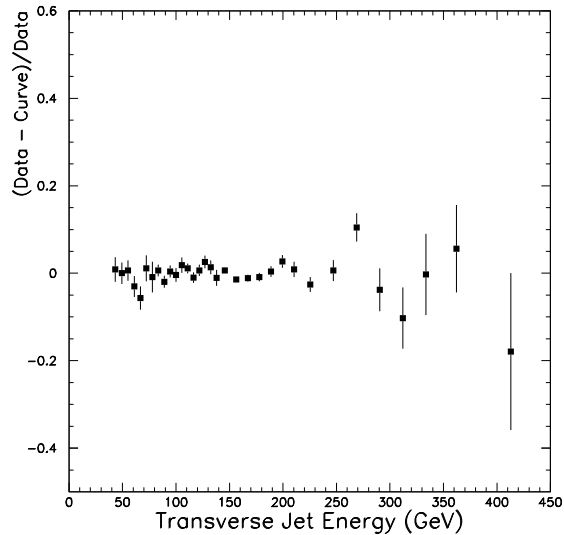,height=8cm,width=8cm}}
\caption{
Percentage difference between the corrected inclusive cross section data 
and the standard curve which was determined in the unsmearing process 
(see text) and represents the best smooth fit to the data.}
\label{fig-nom}
\end{figure}

\begin{figure}
\centerline{
\psfig{figure=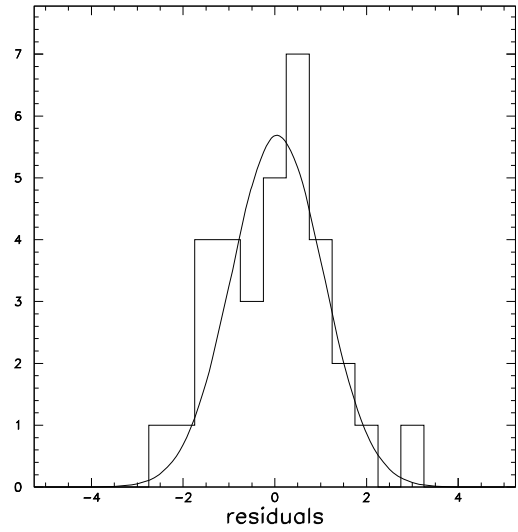,height=8cm,width=8cm}}
\caption{Histogram of the residuals, (Data-curve)/error, of the 
corrected data compared
to the standard curve.  The curve is the result of a fit to a Gaussian 
of width 1.}
\label{fig-resid1}
\end{figure}

\begin{figure}
\centerline{
\psfig{figure=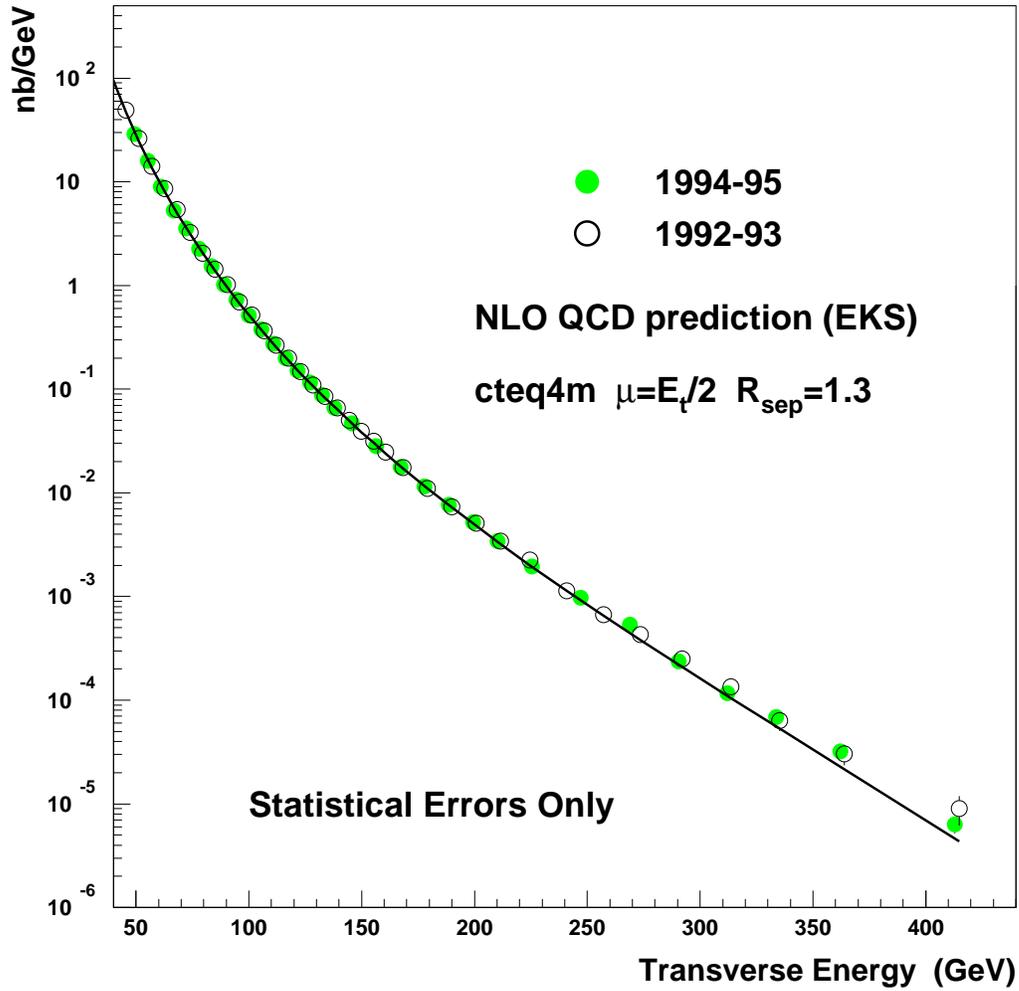,height=15cm,width=15cm}}
\caption{Inclusive jet cross section from the Run 1B data (94-95)  compared to
a QCD prediction and to the published Run 1A data (92-93).}
\label{run1b-log}
\end{figure}

The Run 1B corrected cross section is given in Table~\ref{tab-cor-xsec}
and is shown in 
Fig.\,\ref{fig-nom} (top) compared
to the standard curve determined from the unsmearing.
The uncertainties on the data points, uncorrelated bin-to-bin,
are from counting statistics, trigger efficiency and prescale corrections
and are collectively referred to as the uncorrelated uncertainty.
The correction procedure preserves the 
percentage uncorrelated uncertainty 
on the measured cross section for the corrected cross section.
The total $\chi^2$ between the corrected data and the
standard curve is 44.1 for 33 points. 
The lower panel shows the contribution of each bin to the
total $\chi^2$. 
Large contributions to the $\chi^2$
are observed for a few points which have small uncorrelated uncertainty.
For example, bin 20 (150 GeV) has 0.6\% stat. unc. 
and is -1.4\% from the smooth curve and
bin 28 (270 GeV) has 3.2\% stat. unc. and is 10.5\% from the smooth curve.
Neither of these points is on a trigger boundary; we have investigated the data
in these bins and find no anomalies.
In Figure~\ref{fig-resid1}
we plot the residuals of the corrected data to the standard curve.
The residual is defined as 
(corrected data - standard curve)/(uncorrelated error on the data).
As with previous comparisons between the raw data and 
the smeared standard curve we observe that
although the width of the residual distribution is somewhat 
larger than 1, it is still a reasonable fit to a Gaussian of width 1.
Figure~\ref{run1b-log} shows the corrected Run 1B cross section compared
to a QCD prediction and to the published Run 1A cross section.


\section{Systematic Uncertainties}

The majority of the uncertainty associated with the inclusive jet
cross section arises from the uncertainty in the simulation of the
response of the detector to jets.  As discussed above, the simulation is
tuned to the data for charged 
hadron response, jet fragmentation, 
and $\pi^0$ response. 
Additional uncertainty
is associated with the jet energy resolution,
the definition of the underlying event, the stability
of the detector calibration over the long running periods and 
an overall normalization uncertainty from the luminosity determination.

\subsection{Components of systematic uncertainty}

\begin{figure}
\centerline{\psfig{figure=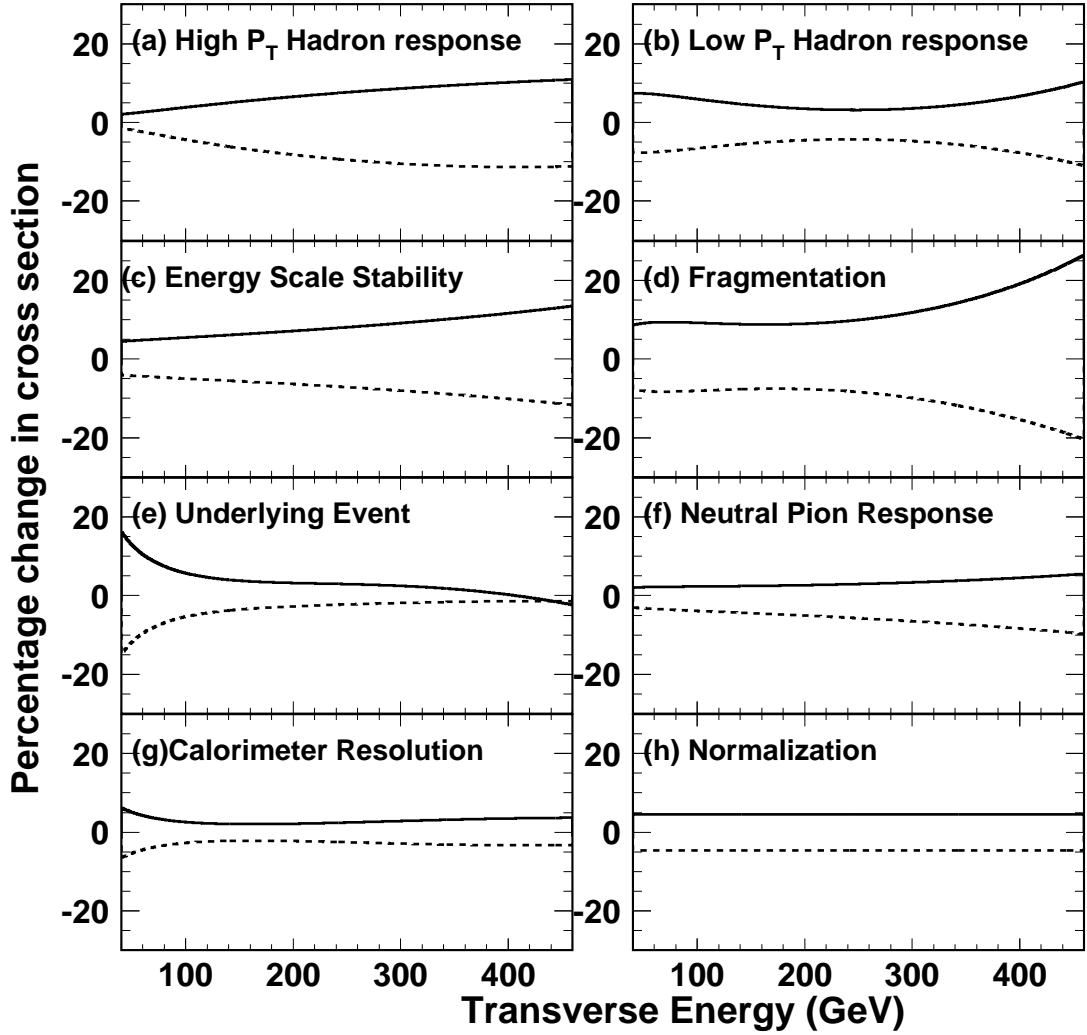,height=16cm,width=16cm}}
\caption{The $\pm1\sigma$ fractional change in cross section due to 
the dominate sources of systematic uncertainty.}
\label{fig-sys-uncertainties}
\end{figure}

The uncertainty on the jet cross section associated with each source
is evaluated through shifts to the response functions.
For example, to evaluate the effect of a ``$1\sigma$" shift
in the high $P_T$ hadron response, the energy scale in the 
detector simulation was changed by 3.2\% and new response 
functions were derived.
These modified response functions were then used to 
repeat the unsmearing procedure and find the modified corrected cross
section curve. 
The difference in the modified cross section curve and the standard curve
(nominal corrections)
is the ``$1\sigma$" uncertainty.
This uncertainty is 100\% correlated from bin to bin.
The parameters of the curves for the
``$1\sigma$" changes in cross section for the eight independent 
sources of systematic uncertainty are given in
Table~\ref{Table-syserr}. 
For each of the uncertainties the 
percentage change from the standard curve
is shown in Fig.~\ref{fig-sys-uncertainties}.

Fig.~\ref{fig-sys-uncertainties}(a) shows the uncertainty from the 
charged hadron response
at high $P_T$.  The $+$3.2\%, $-$2.2\% uncertainty on the hadron 
response includes 
the measurement of pion momenta in the test beam calibration and
variation of calorimeter response near the tower boundaries.
Fig.~\ref{fig-sys-uncertainties}(b) 
shows the uncertainty from the 5\% uncertainty in 
calorimeter response to low-$P_T$\
hadrons. The simulation was tuned to isolated single track
data. The largest contribution to the uncertainty came from
the subtraction for energy deposited by neutral pions
which may accompany a charged track.
Studies of calorimeter response to muons and to
low energy isolated charged hadrons indicate
that absolute calibration was maintained
with an estimated uncertainty of $\pm 1\%$\ (upper limit
$\pm 2.5\%$) from the 1989 run to this run (1994-95).
Fig.~\ref{fig-sys-uncertainties}(c) shows the uncertainty on the cross section
due to this estimate of the energy scale stability.
Jet fragmentation functions used in the simulation were determined from
CDF data with uncertainties derived from tracking efficiency.
Fig.~\ref{fig-sys-uncertainties}(d) shows the uncertainty in the cross section
from the fragmentation function, including our ability to extrapolate the
form of the fragmentation function into the high $E_T$ region where it is
not directly measured from our data.
The determination of the underlying energy from data is sensitive to 
thresholds and event selection. 
We assign a 30\% uncertainty to cover a range of reasonable variations.
Fig.~\ref{fig-sys-uncertainties}(e) shows the uncertainty in the cross section
from this assumption.
Fig.~\ref{fig-sys-uncertainties}(f) shows the uncertainty from 
the electromagnetic calorimeter response to neutral pions and
Fig.~\ref{fig-sys-uncertainties}(g) shows the 
uncertainty associated with the modeling of the jet energy resolution.
Fig.~\ref{fig-sys-uncertainties}(h) represents the 4.6\% normalization 
uncertainty 
from the luminosity measurement (4.1\%) and the efficiency of 
the $z_{vertex}$ cut (2.0\%).
The uncertainties shown in Fig.~\ref{fig-sys-uncertainties} and parameterized in
Table~\ref{Table-syserr} are very similar in size and shape to the 
uncertainties quoted
on the Run 1A result~\cite{CDF-Inclusive-Jet-96}.  The primary difference 
comes from the increased precision of the data at high $E_T$ providing
tighter constraints on the curves.

\section{Comparison to other data}

\subsection{Comparison to Run 1A}

To compare the Run 1B data to the Run 1A result we use the 
smooth curve from Run 1A to calculate the Run 1A cross section
at the Run 1B $E_T$ points (the Run 1A and Run 1B 
results used different binning).
Note that the statistical 
uncertainty on the Run 1A measurement is roughly equivalent to
the Run 1B data
below 150 GeV due to the increased prescale factors in Run 1B.
Above 150 GeV,
where no prescale factors were used,
the uncertainty in the Run 1B data is a factor of 
two smaller.

For a comparison between the corrected cross sections for 
Run 1A and Run 1B results we 
introduce a procedure that will later be used to compare our data 
with theoretical predictions.  Here we use the MINUIT~\cite{MINUIT}. 
program to 
minimize the $\chi^2$ between the Run 1B data and the Run 1A 
standard curve (treated as "theory").
We allow each systematic uncertainty to shift the data 
independently to improve the agreement between the data and the theory. 
The resulting systematic shifts are added to the $\chi^2$. 
In contrast to a more traditional covariance matrix approach,
this technique reveals which systematic uncertainties are producing the
most significant effects on the total $\chi^2$.
For completeness, the covariance matrix 
technique and results are discussed in Appendix A.  

The $\chi^2$ between data and theory is defined as

\begin{equation}
 \chi^2 = \sum_i^{nbin}  \frac { (T_i/F_i - Y_i)^2}{(\Delta Y_i)^2} +
\sum_k S_k^2, 
\end{equation}
where
\begin{equation}
F_i = 1 + \sum f_i^k S_k,
\end{equation}
and
\begin{equation}
f_i^k= |C_i^k-C_i^{STD}|/C_i^{STD} .
\end{equation}

The $Y_i$ are the corrected cross section, $\Delta Y_i$ are the statistical
uncertainty in the cross section, $T_i$ are the theory predictions,
$C_i^{STD}$ is the standard curve and 
$C_i^k$ are the curves for each of the $k$ systematic uncertainties
(in cross section), evaluated
for the $ith$ bin.  The $S_k$ are up to eight parameters (one for 
each systematic uncertainty)
that are adjusted in the fit to give good agreement between the data $Y_i$ and
the theory curve, $T_i$.
Figure~\ref{fig-sys-uncertainties}
shows the systematic uncertainty curves, e.g. the $f^k$.
In the fitting process, the systematic uncertainties 
can be chosen individually or combined.

A number of choices have led to this definition.
1) The error curves represent the fractional change in cross section
which results from 1$\sigma$ shift in one of the inputs,
e.g low $P_t$ hadron response,
to the detector simulation, as discussed in section 6.
Each of the uncertainty curves comes from an independent source. 
Thus, the $\chi^2$ is increased by the quadrature sum of the shifts.
2) The denominator is taken as the uncorrelated uncertainty in the data.
This avoids complications in translating from the theoretical 
prediction (which is produced as a cross section) to the 
theoretical number of events. 
3) The shifts to the theory from the systematic uncertainties are
computed as factors which multiply the theory predictions,
as are the corrections from the raw cross section to the 
corrected cross section.   When multiple
systematic effects are considered, the net systematic shift is the sum 
of the individual shifts.

\begin{figure}
\centerline{
\psfig{figure=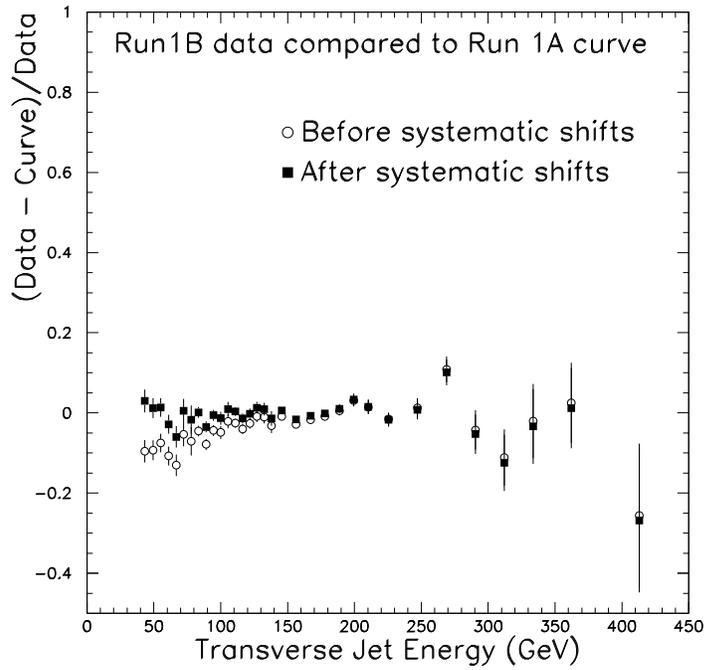,height=10cm,width=10cm}}
\caption{Run 1B data compared to Run 1A smooth curve 
before (open) and after (solid) fitted shifts due to 
underlying event, energy scale 
stability and relative normalization have been included.
Only the statistical uncertainty on the 1B data is shown.}
\label{fig-1a-a}
\end{figure}

The open circles in Figure~\ref{fig-1a-a} show the fractional difference
between the 1B data 
points and the 1A curve ((1B cross section - 1A curve)/1B cross section).
The difference at low $E_T$ comes mainly from
the different definition of the underlying event energy. 

For the $\chi^2$ comparison between the Run 1A and Run 1B results, 
the uncorrelated uncertainty in both the Run 1A and 1B measurements
must be included.  
To estimate the uncertainty in the 1A measurement
at the Run 1B $E_T$ points we scale the corresponding 1B uncertainty.
Below 150 GeV, since the uncorrelated uncertainties are similar,
we simply use the 1B uncertainty for the 1A cross section.  
Above 150 GeV, the ratio of the
luminosities for the data samples (87/19.5) indicates that the
1A uncertainty is a factor of 2.12 larger than the 1B uncertainty at the
same $E_T$ point. 
Using the quadrature sum of the Run 1A and Run 1B uncertainties
has the effect of
increasing the local (uncorrelated) uncertainty  
and produces lower a $\chi^2$ to a smooth curve.
With only the uncorrelated uncertainties the 
$\chi^2$ between the 1B data and the 1A curve is 96.1.
If the  relative normalization uncertainty between 1A and 1B is
included (1.5\% for 1A in quadrature with 2\% for  1B) the
total  $\chi^2$ is 42.9 for the 33 Run 1B data points.

The procedure presented above allows us to study the effects of the 
individual contributions to the comparison between data and theory.
For example, 
the Run 1A definition of the underlying event resulted in
a smaller subtraction than was used for the Run 1B data.
If the underlying event uncertainty is included on the Run 1B data,
but no relative normalization uncertainty,
the fit finds a total $\chi^2$ of 18.5 which includes
a 0.7$\sigma$ shift in the jet transverse 
energy from the underlying event. In other words, a change in the
underlying event correction of 0.7$\sigma$ (= 0.46 GeV) results
in a $\chi^2$ of 18.5.
Between Run 1A and 1B the relevant uncertainties are 
the underlying event, the long term energy 
scale stability and the relative normalization.
If these three are used
then the total  $\chi^2$ is 15.0.
The other uncertainties are
derived from tuning of the detector simulation and are common between
the two measurements.
The solid points in
Figure~\ref{fig-1a-a} show the fractional difference between the 
1B data and the 1A curve after the shifts resulting from a fit which included
the underlying event, the long term energy 
scale stability and the relative normalization uncertainties.
We conclude that the Run 1A and 1B measurements are in good agreement.
\subsection{Comparison to the D0 measurement}

\begin{figure}
\centerline{\psfig{figure=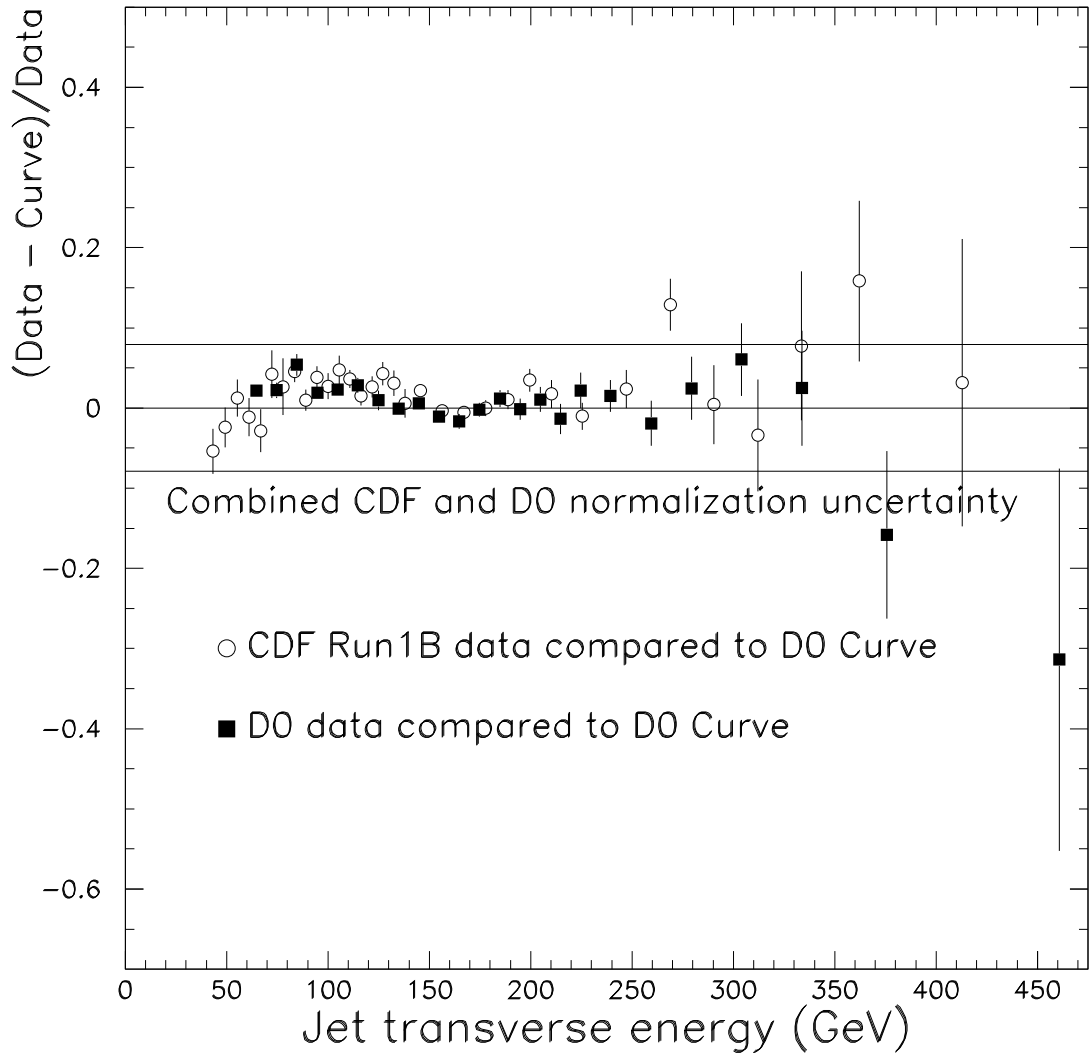,height=16cm,width=16cm}}
\caption{Comparisons of D0 and CDF data to D0 smooth curve
in the region 0.1$<|\eta|<$0.7.} 
\label{cdf-d0-d0curve}
\end{figure}

We now compare the CDF data with the cross section reported by the 
D0 collaboration.  As in the comparison to the Run 1A CDF measurement it
is necessary to use a parameterized curve for this comparison
since the cross section is measured at different points in $E_T$.
Since the lowest $E_T$ point measured by D0 is at $E_T$= 64.6 GeV,
the lowest 4 CDF points 
will not be included in the fits.
We estimate the D0 uncorrelated uncertainty at the CDF $E_T$ points
with a linear interpolation between the uncertainty on two D0 points
which bracket the CDF $E_T$ point.
Before the data sets can be directly compared it is also necessary to
take into account the different assumptions in the determination of the
total luminosity of each sample.  D0 uses a world average 
total $p\bar{p}$ cross section while CDF uses its 
own measurement~\cite{CDFLUM}.
As a result, the D0 inclusive jet
cross section is 2.7\% systematically lower than CDF.
Figure~\ref{cdf-d0-d0curve} shows the CDF and D0 data compared to the fit to
the D0 data~\cite{D0inc1}, after the relative normalization has
been taken into account.  Note that the low $E_T$ CDF points
are plotted but not included in the following fit results.
The $\chi^2$ between the CDF 1B data and D0 curve using 
only the statistical uncertainty from both experiments 
and the 2.7\% normalization shift is 64.7 for the 29 CDF points.  
This drops to 35.6 when the 
combined normalization uncertainty on CDF (4.6\%) and D0 (6.1\%)
is included in the fit.  If all the systematic uncertainties
on the CDF data are also included the total $\chi^2$ is 28.7.
We conclude that the CDF and D0 data are in good agreement.

The D0 collaboration has published a comparison between the
D0 data and the CDF curve from Run 1A using
a covariance matrix
technique to include the CDF and D0 systematic uncertainties.  
The rather large $\chi^{2}$ (63.3 for 24 degrees of freedom, a
confidence level (CL) of 0.002\%)
obtained when the CDF curve was "treated as theory" is not
surprising when one considers that no statistical uncertainties
are included with the CDF curve
and for the comparison to the highest $E_T$ point, the CDF
curve is extrapolated 50 GeV above the last CDF data point.
In addition, the relative normalization difference between the
two data sets is not included.

More recently the covariance matrix method was used to
compare the D0 data and CDF 1B curve~\cite{annrev}.
The $\chi^2$ was 41.5 for 24 degrees of freedom
including both statistical and systematic uncertainties on 
the D0 data and no uncertainty on the CDF curve.
When only the uncorrelated uncertainty on both CDF and D0 are included
(no systematic uncertainty for either data set),
and the 2.7\% relative normalization
difference~\cite{CDFLUM} is removed, the $\chi^2$ is 35.1 for 24 
degrees of freedom, with a CL of 5.4\%.
When the systematic uncertainties in the covariance matrix are
expanded to include both the D0 and CDF systematic uncertainties 
the $\chi^{2}$ equals 13.1 corresponding to a CL 
of $96$\%.

\section{Theoretical Uncertainty}

The predictions for the inclusive jet cross section depend on 
input parameters such as the parton distribution functions,
the choice for the value of $\alpha_s(M_Z)$, the
choice of renormalization and factorization scales and the
method of grouping partons into jets.  Of these, the uncertainty
from the parton distribution functions is the largest.

As in previous publications, 
the primary program used by CDF for comparison with the data is due to
Ellis, Kuntz and Soper\cite{EKS,DSoper-mu-scale}.  We refer to this 
program as EKS and use it to determine the uncertainty in the predictions.

\subsection{Uncertainty from Parton Clustering}

As discussed earlier, clustering at the parton level and
clustering in the experimental data should be the same.
In contrast to the parton level
predictions, the experimental data contains jets of hadrons,
and the edges of the jets are not distinct.
Figure~\ref{CDFLEGO} shows jet events in the
the CDF calorimeter. Jet identification in two jet
events is straight forward.  Jet identification
in multijet events, or in events in which the jets are
close to each other introduces ambiguities which 
are not modeled in the NLO parton level predictions.
For example, studies found that the 
experimental algorithm is more efficient at
separating nearby jets~\cite{RSEP} than the idealized
Snowmass algorithm. That is,
two jets would be identified even though their centroids were separated
by less than 2R. 
Specifically, two jets are separated 50\% of the time if they are 1.3R apart.
An additional parameter, $R_{sep}$, was introduced
in the QCD predictions to approximate the experimental
effects of cluster merging and
separation. Partons within $R_{sep}
\times R$ were merged into a jet, otherwise
they were identified as two individual jets. 
A value of $R_{sep}$ = 1.3 was found to
give the best agreement with cross section and jet
shape data\cite{RSEP}.

\begin{figure}
\centerline{\psfig{figure=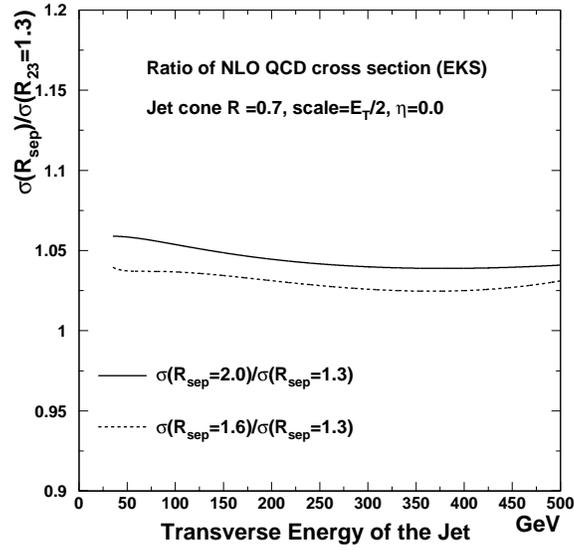,height=8cm,width=8cm}}
\caption{The variation of the inclusive jet cross section for different
$R_{sep}$
parameters. These calculations used the EKS program.}
\label{fig-eks-rsep}
\end{figure}

Figure~\ref{fig-eks-rsep} shows the change in the NLO QCD predictions 
for a range of 
$R_{sep}$ values.
The ratio of cross sections for $R_{sep}=1.3$ and $R_{sep}=2$ 
shows a 5-7\% normalization shift.
The cross section is smaller with smaller \RSep because it 
essentially uses smaller effective cone size. 
Naively, smaller cones would imply more jets and a larger cross section.
However, with the steeply falling spectrum, the higher energy 
obtained by merging jets is the dominant factor.
This result is consistent with the early results\cite{RSEP}
where the comparison used $\mu=E_t/4$ and different
parton distribution functions.
The NLO predictions in this paper from JETRAD and EKS 
follow the Snowmass algorithm with the additional parameter $R_{sep}$.
We use $R_{sep}=$1.3 unless otherwise indicated.

\subsection{Choice of the $\mu$ scale}

\begin{figure}
\centerline{
\psfig{figure=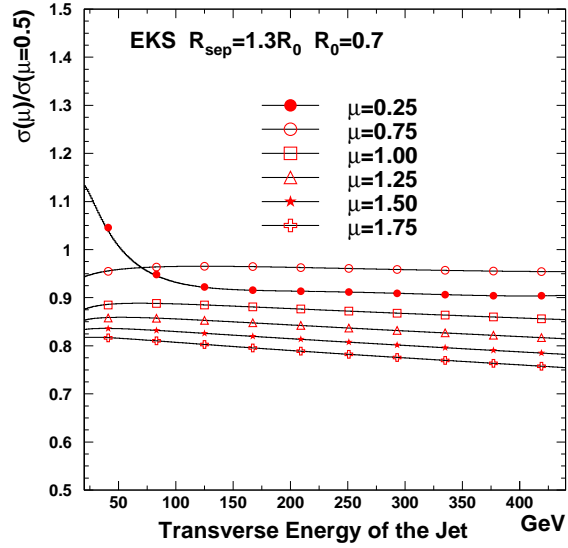,height=8cm,width=8cm}}
\caption{
Variation in theory predictions for different renormalization scales.}
\label{fig-qcd}
\end{figure}

The choice of $\mu$ is an intrinsic uncertainty in a fixed
order perturbation theory.
The effects of
higher order corrections are typically estimated by the sensitivity of the
predictions to variations in the choice of $\mu$.
Fig.\ref{fig-qcd} shows the
inclusive jet cross section where
the $\mu$ scale is varied from $2\,E_T$ to
$E_T/4$.  Above $E_T>70$ GeV these changes result only
in normalization changes of 5-20\%.

\begin{figure}
\centerline{\psfig{figure=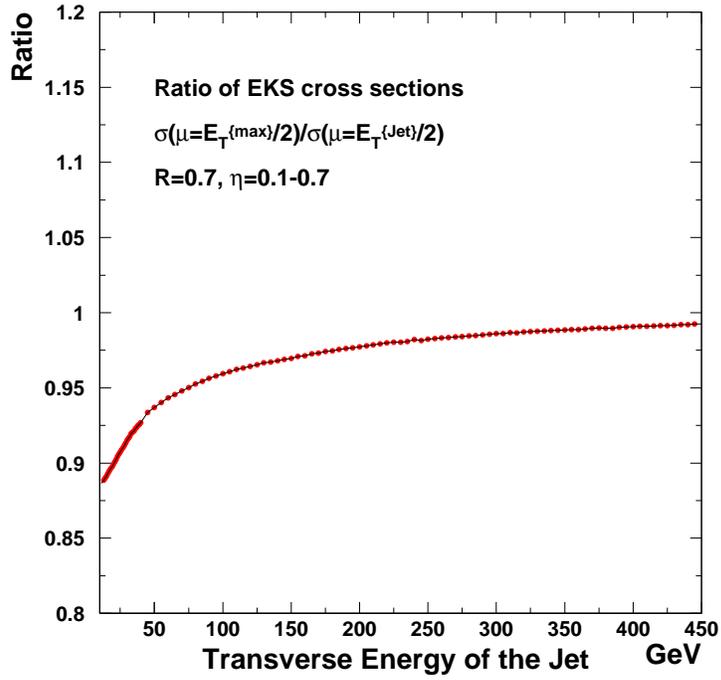,height=10cm,width=10cm}}
\caption{Comparison of NLO cross sections using $\mu=E_t^{max}/2$
and $\mu=E_t^{jet}/2$. The EKS program is used,
$R_{sep}$ = 2.0, and the PDF's are CTEQ3M.}
\label{fig-eks-max}
\end{figure}

As described earlier, the EKS and JETRAD programs
made different choices for the $\mu$ scale.  The EKS program 
calculates the cross section at a particular jet $E_T$, integrating over all 
configurations that contribute.  In contrast, for each event, 
the JETRAD program uses $E_T^{max}$, the $E_T$ of the maximum $E_T$ jet.
We have calculated
the inclusive jet cross section using both, $\mu=E_t^{max}/2$ and
$\mu=E_T^{jet}/2$ with the EKS program~\cite{DSoper-mu-scale}.
Figure \ref{fig-eks-max} shows the resulting ratio of the
cross sections.
The effect of using $\mu$=$E_T^{max}/2$ instead of $\mu$=$E_T/2$
ranges from $\approx$ 4\% at 100 GeV
to $<$ 1\% at 450 GeV.  The difference increases with decreasing $E_T$
because the second and third jets in the event constitute a 
larger (but still small) fraction of the jets in the bin.
As the $\mu$ scale used
in the $\mu=E_T^{jet}$ convention is less than or equal to the
maximum $E_T$ jet in an event, the cross section for the
$\mu=E_T^{jet}$ case
is slightly larger ($\alpha_s$ is larger).

\subsection{Parton distribution functions}

The momentum distributions of the partons in the protons and antiprotons
(the PDF's) are determined from global fits to data from different 
experiments and
different kinematic ranges.  
The information about the quark distributions
comes primarily from deep inelastic scattering (DIS) and Drell Yan processes.
DIS is observed at fixed target experiments such as NMC~\cite{NMC}
and Fermilab E665~\cite{E665},
and at colliding beam experiments such as H1~\cite{H1} and ZEUS~\cite{ZEUS}.
Drell-Yan is observed at Fermilab fixed target experiments (for example
E605~\cite{E605} and E866~\cite{E866})
and at colliding beam experiments (for example~\cite{CDF-Inclusive-Jet-96} 
and~\cite{Wasym}).
The center-of-mass
energy of most of these data is much lower than that of the Tevatron,
although the fraction of the proton momentum carried by the quarks 
is similar.  Information about the gluon distribution is derived
indirectly from scaling violations in the 
DIS experiments and directly from fixed target 
photon experiments and collider jet measurements.
The fixed target photon predictions suffer large 
uncertainties, which makes
them currently unreliable for 
inclusion in the global fits.
Data from fixed target and the e-p collider experiments have 
improved over the years and
the inclusion of new data into the PDF global fits has led to 
more precise PDF's.  
 
Uncertainties in the PDF's arise from uncertainties in the
data used in the global fits, uncertainty in the theoretical 
predictions for that data and from the extrapolation
of the fits (and uncertainties) to different kinematic ranges.
Recent studies have begun to quantify some of these uncertainties
by producing families of PDF's with different input parameters.
One of the early attempts to understand the flexibility
of the PDF's at high $x$ was motivated by the excess over
the theoretical predictions observed in Run 1A inclusive jet cross section.
Studies\cite{CTEQ4M} revealed that there was enough flexibility in the
gluon distribution  at high $x$ to give 
a significant increase in the jet cross section at
high $E_{T}$, while maintaining reasonable agreement with the 
other data used in the global fit.

\begin{figure}
\centerline{
\psfig{figure=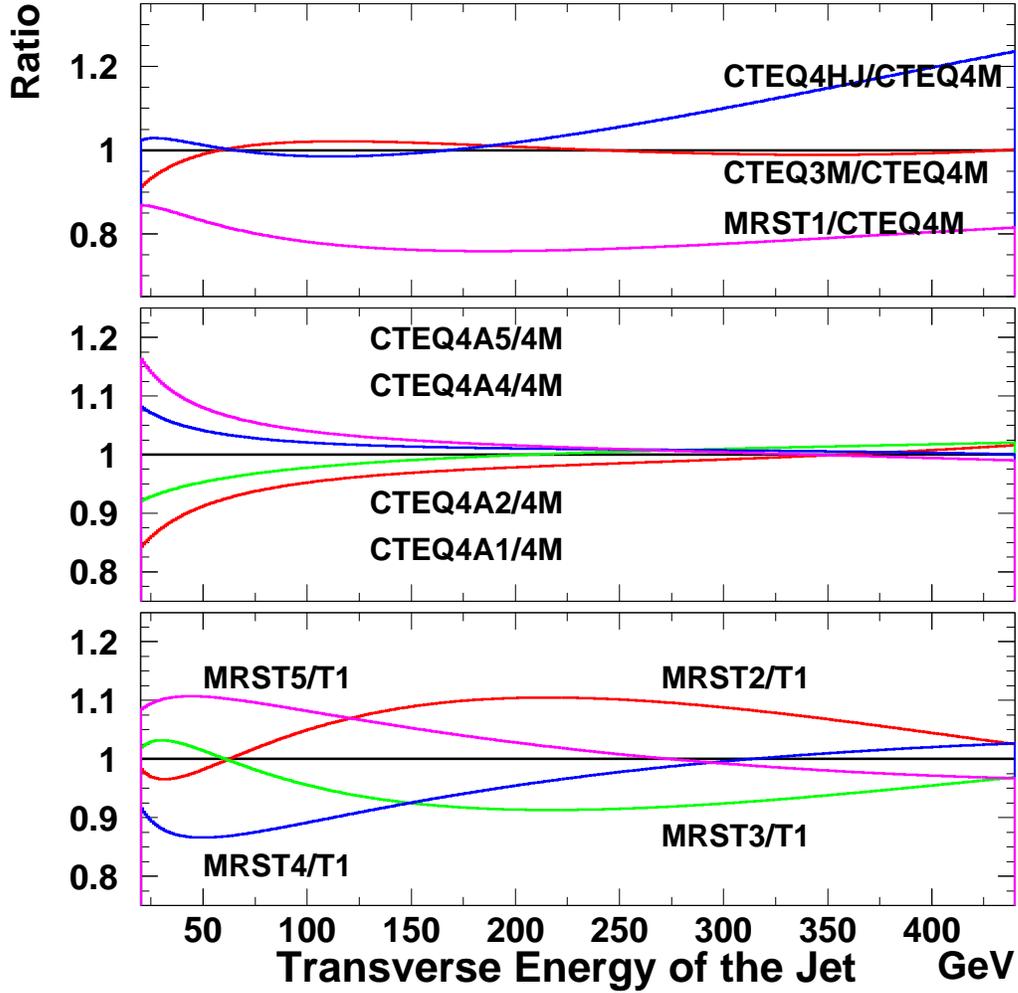,height=15cm,width=15cm}}
\caption{
Variation in theory predictions for different parton distribution functions.
In the top two plots the predictions have been normalized by CTEQ4M.  In the 
bottom plot the different predictions have
been divided by MRST.}
\label{fig-pdfs}
\end{figure}

Figure~\ref{fig-pdfs} shows the variation in the predictions of
the inclusive jet cross section for a variety of PDF's.  
The top plot shows the differences between calculations using CTEQ4M, 
CTEQ4HJ (which was derived with special emphasis on the
high $E_T$ CDF jet data) and MRST.  The middle plot shows the
variation in the family of CTEQ4M curves for a range of allowed values for
$\alpha_s$. The PDF with nominal $\alpha_s$ is called CTEQ4M, and in the
following Fig.s is referred to as CTEQ4Ma3.  
The lower plot shows the variation in the cross section 
for the MRST series.  Note that in the following figures 
MRST1 = MRST, MRST2 = MRST-g$\uparrow$, MRST3 = MRST-g$\downarrow$,
MRST4 = MRST-$\alpha_s\downarrow\downarrow$ and 
MRST5 = MRST-$\alpha_s\uparrow\uparrow$.
Details of these studies can be found in 
References~\cite{cteq-4,MRST}.  Briefly, MRST-g$\uparrow$ and 
MRST-g$\downarrow$ represent extreme variations in the contribution of
gluons and MRST-$\alpha_s\downarrow\downarrow$ and
MRST-$\alpha_s\uparrow\uparrow$ represent PDF's derived with extreme 
values of $\alpha_s(M^2_Z)$.  These are 0.1125 and 0.1225 respectively.

It should be noted that the variation in QCD predictions shown in
Fig.\,\ref{fig-pdfs} does not cover the full range of
uncertainties associated with the data used in the global analysis to
determine PDF's. In particular, the gluon distributions at high
$x$ are mainly determined by direct photon production experiments for the
MRST set and from jet data for the CTEQ set.
The QCD calculations for the photon production at
fixed target energies have 
a large 
scale dependence and require a resummation of the emission of soft gluons
for a direct comparison to experimental data.
The same is true for low $E_T$ photon production at the Tevatron, and
this data is not currently included in any PDF fit.
Proper inclusion of these uncertainties into a global analysis
is the subject of recent discussions~\cite{kt706}.

Recently, a reanalysis of DIS data has found that the uncertainty
in the quark distributions at high $x$ may be larger than previously thought
\cite{CTEQ98,YANG98}, due to nuclear binding effects 
which have not been included in any PDF
to date.

\subsection{Other theoretical uncertainties}

The inclusive jet cross section calculation does not include
other Standard Model processes $e.\,g.$ top production, $W^+W^-$ production,
however estimates of their contributions can be derived from 
measured quantities.
The top cross section~\cite{cdftopxsec} and the $E_T$ spectrum 
of the jets in these events indicate that top contamination of the jet
sample is less than 0.01\%.  The $W^+W^-$ contribution will be even smaller.

Higher order QCD corrections ($O(\alpha_s^4))$ are not available yet.
Soft gluon summation may lead to a
small increase in the cross section
at high $E_T$\cite{{Walter-pbarp},{Sterman-pbarp}}. A recent
calculation shows that the effect for di-jet mass distribution is
about 7\%\cite{Nason}.  
The actual size of effect might be
different for inclusive one-jet spectrum\cite{Sterman-pbarp}.

\subsection{Summary of theoretical uncertainties}

Table~\ref{tab-thy-err} shows a summary of the uncertainties associated
with the theoretical predictions.  For this table the
shifts observed in Figures~\ref{fig-eks-rsep}
to ~\ref{fig-pdfs} for
the various changes in parameters are taken as the theoretical uncertainty
and tabulated for three $E_T$ points.  
In the top half of the table the 
percent changes were 
calculated with respect to a reference prediction which
used the EKS program, CTEQ4M, $R_{sep}$ = 1.3 and $\mu = E_T^{jet}/2$.
The column labeled ``shape" indicates whether the shift in
the prediction increased (or decreased) smoothly as a function of $E_T$.
Both the CTEQ4 and MRST families show significant changes in the overall
shape of the spectrum.
The lower half of the table summarizes the changes within a particular 
PDF family.
From this table and the figures one concludes that 
the theoretical predictions are uncertain in both shape and normalization.
Normalization changes of up to 20\% are allowed from the 
typical choices of scale.  The difference between CTEQ4M and
MRST-g$\downarrow$ could be viewed as a 30\% shift in normalization combined
with a change in shape of roughly half that size, and quite 
comparable to the shape changes in the CTEQ4M series.
These issues will be discussed in more detail when the data is 
compared to the predictions.

\section{Comparison with predictions}

Below we present 
the comparison of the CDF data to the theoretical predictions.
The precision of the Run 1B data, the sensitivity of this 
measurement to PDF's and the potential for new physics have
motivated a detailed study of the best way to compare data and theory.
In this endeavor we deviate significantly from techniques used
for previous results and from other Run 1B high $E_T$ jet measurements
at CDF~\cite{dijetmass}.
The main difference is that we now compare the raw data
to theoretical predictions which have been smeared
with detector resolution effects rather than compare unsmeared
theoretical predictions to the corrected data.
Below we first show the comparisons with only uncorrelated uncertainties
on the data.  We then describe the  $\chi^2$ fitting technique 
which includes the
experimental uncertainties.  With these tools we 
quantify the degree to which
a particular theory prediction reproduces the
observed data.
To further exploit the power
of the data we introduce a $\Delta \chi^2$ technique to indicate
relative probabilities of the theoretical predictions. 

A number of different
methods have been used to compare the previous CDF measurements
of the inclusive jet cross section to theoretical predictions. 
Details of these techniques and the prescriptions for 
construction of the covariance
matrix (used in previous analyses) are included in Appendix A.
In contrast to the covariance matrix approach,  the fitting method 
used in the analysis of the Run 1B data
allows detailed study of the individual contributions of each systematic
uncertainty. In particular, we learn how the combination of 
the eight independent sources of uncertainty
interact in a fit.  Although the source of each uncertainty is
independent of the others, the $E_T$ dependence of the uncertainty curves
are quite similar. Consequently, in any fit the 
systematic uncertainties are correlated. More details on
this method are presented in Appendix B.

Figure~\ref{fig-cor-pdfs} shows 
the corrected 1B cross section compared
to QCD predictions using three current PDF's.  
Considering only the statistical uncertainties we see that 
the CTEQ4HJ curve provides the
best qualitative agreement with the data in overall shape and normalization;
CTEQ4M agrees well with the data at low $E_T$ but is lower than the data 
above $E_T$ $\approx$ 250 GeV; MRST disagrees in shape and normalization over
the full $E_T$ range.

\begin{figure}
\centerline{\psfig{figure=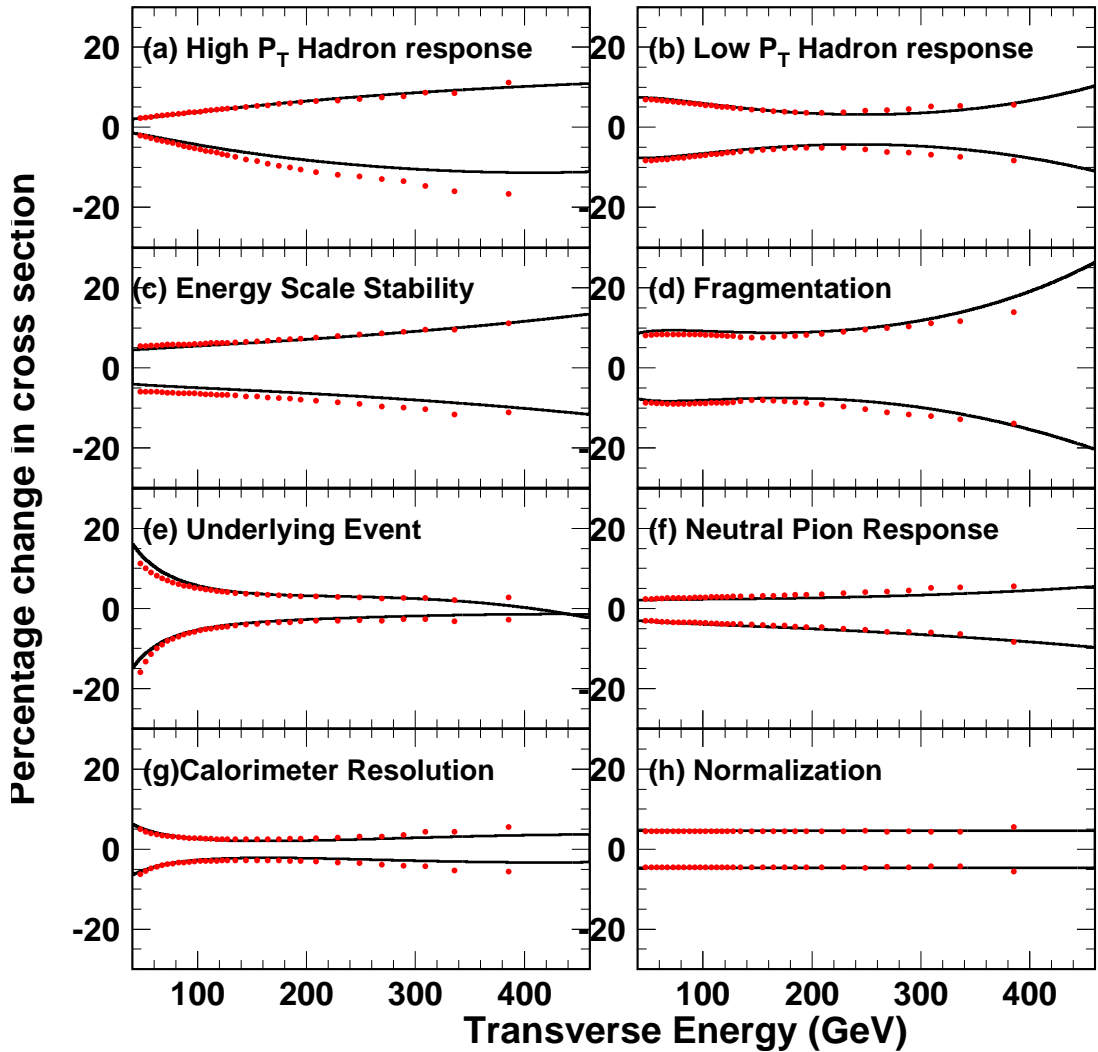,height=16cm,width=16cm}}
\caption{The fractional uncertainty on the raw CDF cross section (points)
compared to the fractional uncertainty
on the corrected CDF cross section (curves). 
The uncertainty on the corrected cross section
is affected by the statistical precision on CDF data and hence the curves
are not stable at very high $E_T$.}
\label{Fig-syserrors-cdfstd}
\end{figure}

Comparison of the 
smeared theoretical predictions with the observed data
rather than comparing corrected data to
unsmeared predictions, 
is a more rigorous,
although more cumbersome technique, but it has several
advantages over the more traditional methods.
First, 
the process of deriving the systematic uncertainty curves for the 
corrected cross section couples the systematic shift in the
cross section due to its uncertainty with the statistical 
uncertainty in the data.  Figure~\ref{Fig-syserrors-cdfstd} shows the
percent uncertainty from the corrected cross section (the curves)
compared to the uncertainty on the raw cross section (points).
The differences are quite small ($<$3\%) but with
statistical uncertainties of $\approx$1\% these differences can be important.
Second, the amount of smearing
depends on the shape of the initial spectrum.  Where the spectrum
is steep, more smearing will occur.  
Thus, for each theoretical prediction
it is necessary to derive the corresponding systematic uncertainty curves.

For comparisons of CDF jet data to theoretical predictions
we define the $\chi^2$ in terms of the raw number of
events and the smeared predictions as follows:

\begin{equation}
\chi_{t}^2= \sum_{i=1}^{nbin}\frac{(n_{d}(i)-n_{t}(i))^2}
{\sigma_{t}^2} +
                \sum_{k=1}s_{k,t}^2
\end{equation}
where $n_{d}$ is the observed number of jets in bin $i$ and
$n_{t}$ and $\sigma_{t}$ are the corresponding 
predicted number of jets and the
uncertainty on the prediction as described below for theoretical prediction
$t$. The $s_{k,t}$ is the
shift in the $k$th systematic for the $t$  theoretical  prediction.
The first term represents the uncorrelated scatter of the points around
a smooth curve, while the second represents the $\chi^2$ penalty from the
systematic uncertainties.  Later we refer to these two terms as
$\chi^2_{stat}$ and $\chi^2_{sys}$ respectively.

To calculate the predicted number of jets in a bin, we smear the
theoretical cross section using CDF detector response functions. The nominal
response function results in nominal prediction $n^0_{t}$. For
each systematic uncertainty $k$, a prediction is obtained using
corresponding response functions and denoted by  $n^k_{t}$. The
systematic uncertainty in bin $i$ is defined as
\begin{equation}
f^k_{t}(i)=n^k_{t}(i)-n^0_{t}(i).
\end{equation}
Using this nomenclature, the
predicted number of jets in a bin is given by
\begin{equation}
n_{t}(i) =  n^0_{t}+\sum_{k=1}^8 s_{k,t}\times f^k_{t}(i).
\end{equation}
Figure~\ref{Fig-syserrors-cdfstd} shows the fractional change in cross section
($f^k_{t}(i)/n^0_{t}(i)$) when the
CDF standard curve is used as the theory.

From the predicted number of entries in a bin, we calculate
the statistical (or uncorrelated) uncertainty as in the actual data 
by including the uncertainties from the trigger efficiency and
prescale factors (see section IV.D).
The parameters $s_k$ are chosen to minimize 
the total $\chi^2$
as above using the program MINUIT.
The results of the fit are given in
Table~\ref{table-best-theory-errors}. 

The systematic uncertainties are 
(1) high $P_T$ charged pion response,
(2) low $P_T$ charged pion response, (3) calorimeter energy scale
stability,
(4) fragmentation function,
(5) underlying event, 
(6) neutral pion response,
(7) energy resolution,
and (8) overall normalization.
From this table we conclude that the prediction with CTEQ4HJ PDF's 
provides the best description of the CDF inclusive jet cross section.
Appendix B discusses the correlated nature of these parameters and shows
graphically the effect of each shift on the comparison between data and theory.

\subsection{Using limited number of uncertainties}

In the fitting procedure described above, the combination of uncertainties 
which produces the smallest $\chi^2$ can be the result of precise 
cancelations between the eight effects. 
Although the sources of uncertainty are independent of each other,
they produce similar changes in shape in the cross section.
To interpret the values for the $s_k$ listed in Table~\ref{table-best-theory-errors}
we perform the fits using from 0 to 
eight systematic uncertainties at a time. All combinations are used.
The best $\chi^2$ using from 0 to eight 
systematic uncertainties are given in Table~\ref{Table-limited-sys-4hj}
for CTEQ4HJ predictions. We see that the total $\chi^2$ is reduced 
from 94.2 to 47.6 when
four systematic uncertainties are included. 
Also note that the sign of the shifts is such that they tend to cancel
any overall shift in normalization. 
The contribution from
systematic uncertainties is 6.9. Adding additional freedom
(the remaining four systematic uncertainties) reduces 
the $\chi^2$ by only 0.8.
The results for MRST predictions are given in
Table~\ref{Table-limited-sys-mrst1}. In this case, the $\chi^2$ is reduced from
11040 to 50.0 when 5
systematic uncertainties are allowed to contribute. 
Here the shifts tend to all go in the same direction, i.e. to reduce the
cross section so that it is in better agreement with the prediction.
The systematic
contribution is 9.6. Including the remaining sources, further reduces it
by 0.5. The results for other PDF's are given in Appendix C.

\subsection {Confidence levels and Probabilities}

To determine confidence levels from the $\chi^2$ results presented in 
Table~\ref{table-best-theory-errors} 
we must first 
determine the probability distributions associated with the
$\chi^2$ variable we have defined, as a priori it is
not necessarily distributed as a traditional
$\chi^2$ variable~\cite{NR}.
To do this we use a
large number of pseudo-experiments for each theoretical prediction 
which include the
effects of the systematic uncertainties.
The procedure is described below.
We use CTEQ4HJ as an example.

(1) We generate fake raw data (a pseudo experiment)
using CTEQ4HJ as the initial spectrum
and the systematic and statistical uncertainties described above.
A nominal prediction using the nominal smearing is
used to predict the nominal raw number of events per bin.
Then variations around this nominal prediction are generated
using 33+8 random numbers, one for the statistical fluctuations
of each data point and one for each systematic uncertainty.  
We assumed that the systematic uncertainties had
Gaussian distributions.  The widths of the distributions
are $E_T$ dependent as shown in Figure~\ref{Fig-syserrors-cdfstd}.

(2) Each pseudo-experiment is fit to the nominal prediction 
(the smeared CTEQ4HJ distribution)
using the $\chi^2$ definition above.

\begin{figure}
\centerline{\psfig{figure=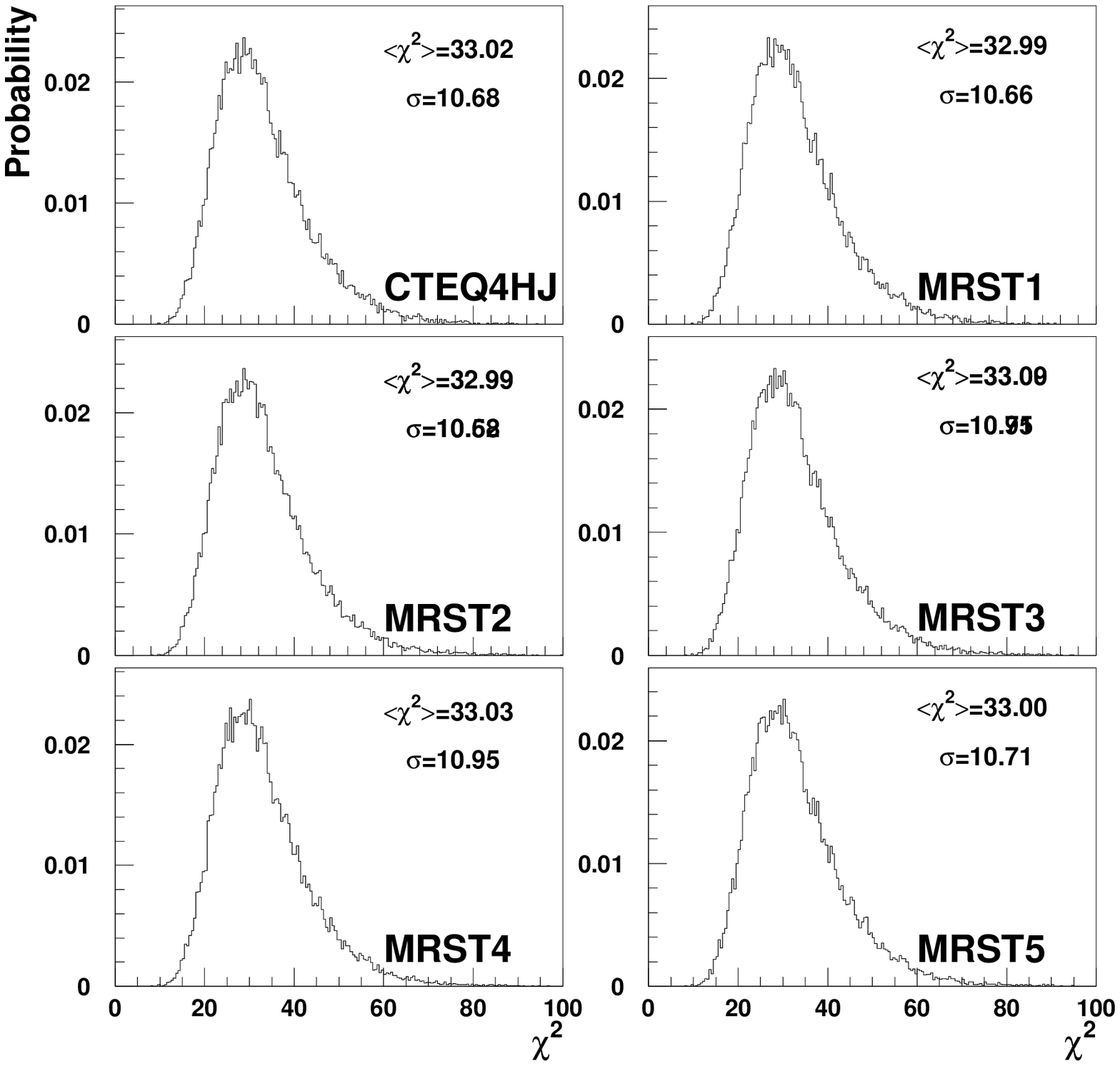,height=16cm,width=16cm}}
\caption{The $\chi^2$ distributions for 
pseudo-experiments using a variety of QCD predictions.
For each plot, the pseudo-experiments are generated and fit to 
QCD predictions using the same PDF's, e.g. for the upper left plot
CTEQ4HJ is used to generate the data samples and the samples are
fit to the nominal smeared CTEQ4HJ prediction.}
\label{Fig-chisqr-same-hj-mrst}
\end{figure}

\begin{figure}
\centerline{\psfig{figure=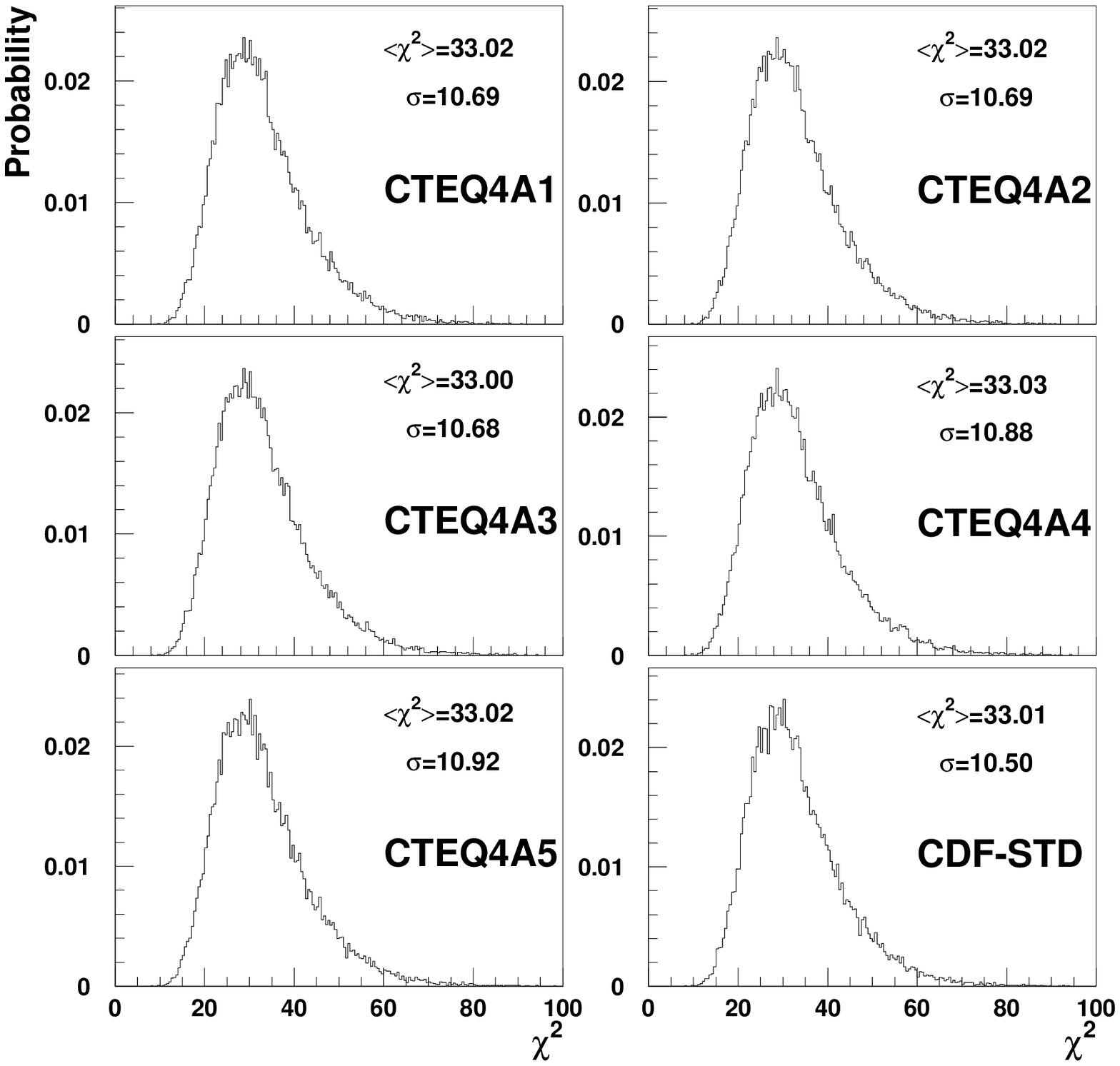,height=16cm,width=16cm}}
\caption{
The $\chi^2$ distributions for 
pseudo-experiments using a variety of QCD predictions.
For each plot, the pseudo-experiments are generated and fit to 
QCD predictions using the same PDF's
 e.g. for the upper left plot
CTEQ4A1 is used to generate the data samples and the samples are
fit to the nominal smeared CTEQ4A1 prediction}
\label{Fig-chisqr-same-cteq4a}
\end{figure}

(3) The $\chi^2$ distribution for each  pseudo-experiment
for CTEQ4HJ are shown in the upper left plot of 
Fig.~\ref{Fig-chisqr-same-hj-mrst}.  The other plots in 
Fig.~\ref{Fig-chisqr-same-hj-mrst} and the plots in 
Fig.~\ref{Fig-chisqr-same-cteq4a} show the distributions
when other PDF's are used to generate the pseudo experiments.
The spread in the distributions represents the fluctuations 
introduced in generating fake data.
The mean $\chi^2$ is approximately equal to the
number of data points, implying that it has some of
the features of a more conventional $\chi^2$ variable.

(4) We calculate the  $\chi^2$ between the CDF data and the nominal 
smeared CTEQ4HJ prediction. The integral of the $\chi^2$ distribution above this
value represents the CL that the initial distribution for
the data was CTEQ4HJ.  

The results for other the PDF's are given in Table~\ref{Table-dchisqr}.
The standard CDF curve has a CL of 16\%, CTEQ4HJ is 10\%,
and MRST is 7\%. 
All the other PDF's have CLs less than 5\%, but the
the differences between them are small.  However, as seen in 
Fig.~\ref{fig-cor-pdfs} 
the various levels of disagreement between the data
and predictions using different PDF's suggests 
a more sensitive test should be possible.

The $\chi^2$ statistic does not distinguish between scatter
and trend.  
We noted earlier (Section VII)  that the data have
a sufficient scatter that a smooth curve adjusted to
follow the trends in the data -- what we denote
as the CDF standard curve -- has a confidence level of
16\%.  Thus, no theoretical prediction will have a
better confidence level, and we expect that all will
appear less likely based on this statistic.  To
enhance our sensitivity to differences in the various
theoretical predictions, we use a 
$\Delta\chi^2$ technique.
We first establish the sensitivity of our measurement 
by comparing 
pseudo-experiments generated with a particular
theoretical prediction to the nominal predictions from 
different theories.  In other words, we try to answer the
question: 
do the systematic uncertainties wash out the sensitivity to the
differences in the theoretical predictions?
Then we find where the data falls on the distributions and
extract relative probabilities for a pair of theoretical predictions.
For these comparisons we pick CTEQ4HJ as the reference prediction.
Thus, all the probabilities will be
relative to this distribution.  

To be specific we compare the theoretical prediction with 
MRST to the prediction with CTEQ4HJ.
First, the pseudo experiments are generated as described above
for CTEQ4HJ. For each pseudo experiment the following are calculated:
(1)the $\chi^2$ with the nominal MRST distribution, $\chi_{MRST}^2$;
(2) the $\chi^2$ with the nominal CTEQ4HJ distribution 
$\chi^2_{4HJ}$ (this
will be smaller on average than $\chi_{MRST}^2$ since it is what the 
pseudo experiments were generated with).
The distribution
$\Delta\chi^2 =\chi^2_{MRST}-\chi^2_{4HJ}$ is plotted and 
finally, the procedure is repeated using pseudo experiments
generated from MRST as the initial theory.

\begin{figure}
\centerline{\psfig{figure=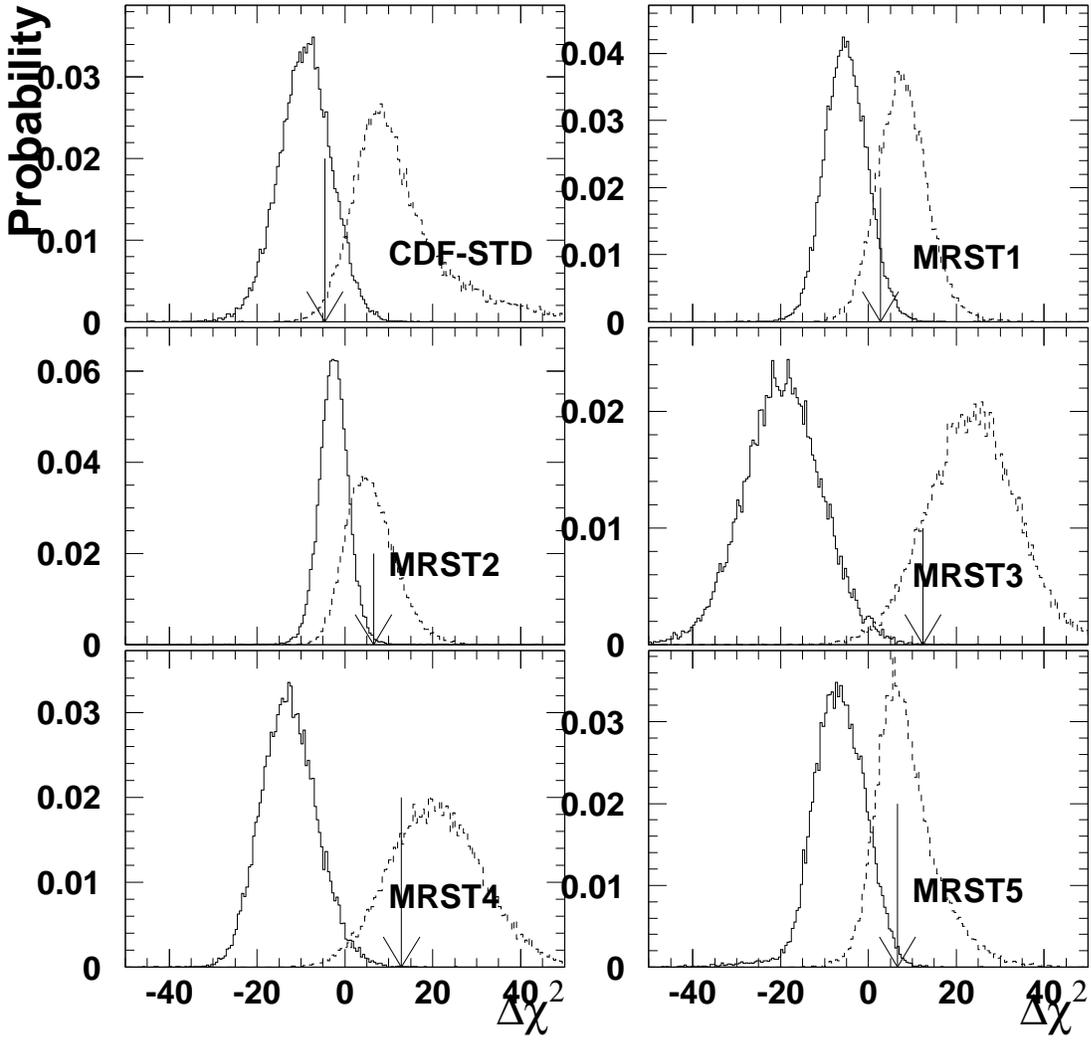,height=16cm,width=16cm}}
\caption{The $\Delta \chi^2$ distributions for 
CTEQ4HJ compared to the CDF standard curve, and theoretical predictions with
the MRST series as described in the text. The arrows indicate the 
$\Delta \chi^2$ of the CDF data.}
\label{Fig-cteq4hj1}
\end{figure}

These $\Delta\chi^2$ distributions are shown in the upper
right plot of Fig.~\ref{Fig-cteq4hj1}.
The distribution to the right of zero is when 
CTEQ4HJ is used as the initial distribution
for the pseudo experiments and the distribution to the
left is from using MRST as the source for the pseudo experiments.
The two distributions are separated indicating that a larger $\chi^2$
will result if the initial distribution and the distribution used
to generate the pseudo experiments are different.  If the 
two distributions completely overlapped it would indicate that systematic
and statistical uncertainties had washed out the ability to 
discriminate between the two predictions.

The $\Delta\chi^2$ for the actual data, e.g. the difference
between the $\chi^2$ to CTEQ4HJ
and the $\chi^2$ to MRST is indicated on the plot by the arrow.
Note that it falls well within the peak which was derived from 
CTEQ4HJ and on the tail of the distribution which was derived
from MRST indicating that the data is more likely to have 
an initial distribution similar to CTEQ4HJ than MRST.
To quantify the relative probability for the two initial distributions
we take the ratio of the heights of the distributions
where the measured data falls~\cite{lyons}.
Note that where the two distributions intersect, it is
not possible, based on this statistic, to
indicate which initial distribution is more
likely to be the correct one.

For CTEQ4HJ compared to MRST, the
$\Delta\chi^2$ is 2.7.  The height of the CTEQ4HJ curve is
0.026 while for the
MRST  curve it is 0.012, a ratio of 0.5. Thus, 
the data favors CTEQ4HJ over MRST by a factor of 2.

\begin{figure}
\centerline{\psfig{figure=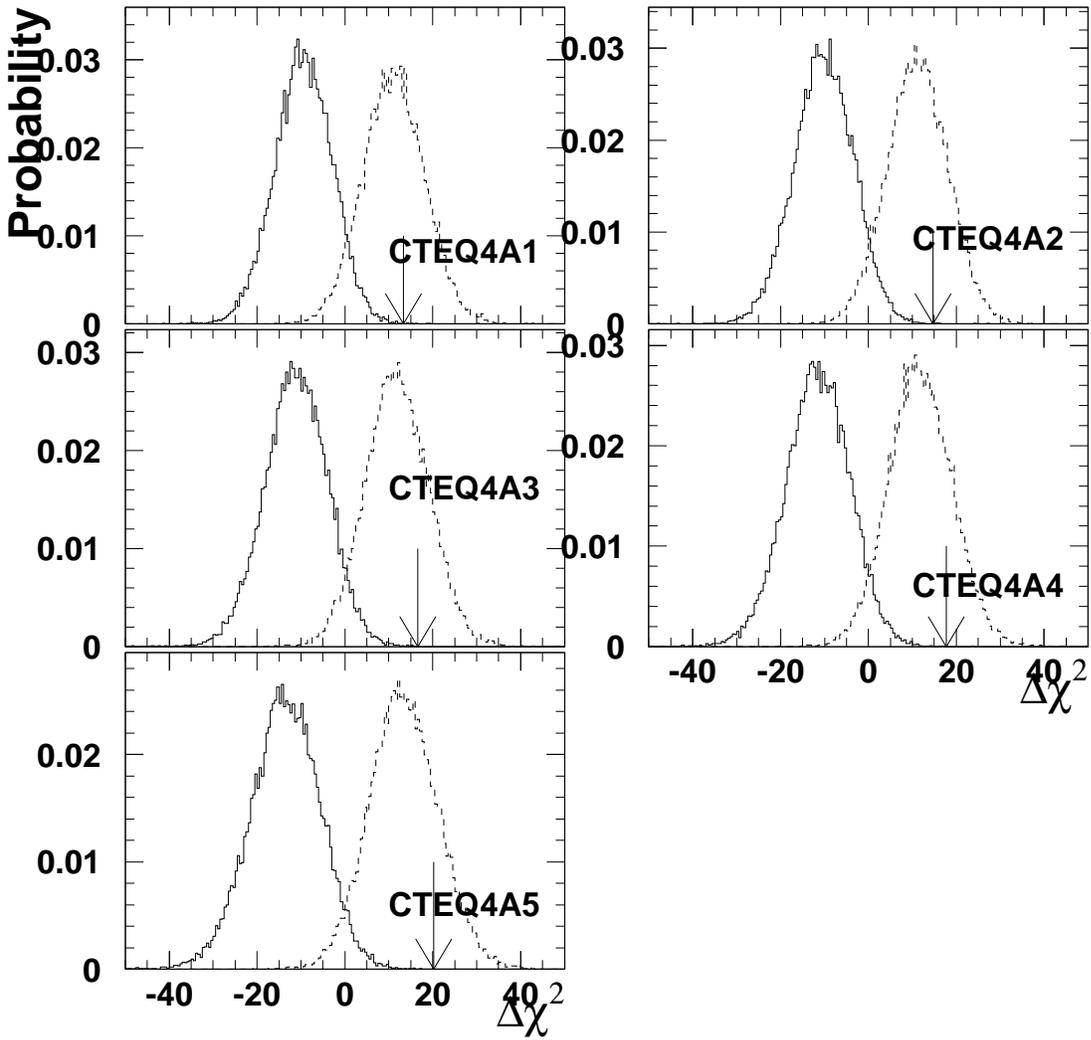,height=16cm,width=16cm}}
\caption{The $\Delta \chi^2$ distributions for CTEQ4HJ and 
theoretical predictions with the CTEQ4M series
as described in the text. The arrows indicate the 
$\Delta \chi^2$ of the CDF data.}
\label{Fig-cteq4hj2}
\end{figure}

Results for predictions using other PDF's are shown in the 
other panels of Fig.~\ref{Fig-cteq4hj1} and in  Fig.~\ref{Fig-cteq4hj2}.
The $\Delta\chi^2$ for the data, e.g. the differences
between the $\chi^2$ to CTEQ4HJ
and the $\chi^2$ to distributions with other PDF's,
are listed in Table~\ref{Table-dchisqr} and indicated in
by arrows in Fig. ~\ref{Fig-cteq4hj1} and Fig.~\ref{Fig-cteq4hj2}.
The probability relative to CTEQ4HJ for 
each PDF to be the initial distribution for the data
(ratio of the heights of the curves at the CDF data $\Delta\chi^2$) is given 
in the last column of Table~\ref{Table-dchisqr}.
Note that a set of PDF's which gave a prediction like the CDF standard curve
would be favored by a factor of about 10 compared with the CTEQ4HJ prediction,
which in turn is favored over most of the other PDF's by a factor of more than
100.

\section{Summary and Conclusions}

\begin{figure}
\centerline{
\psfig{figure=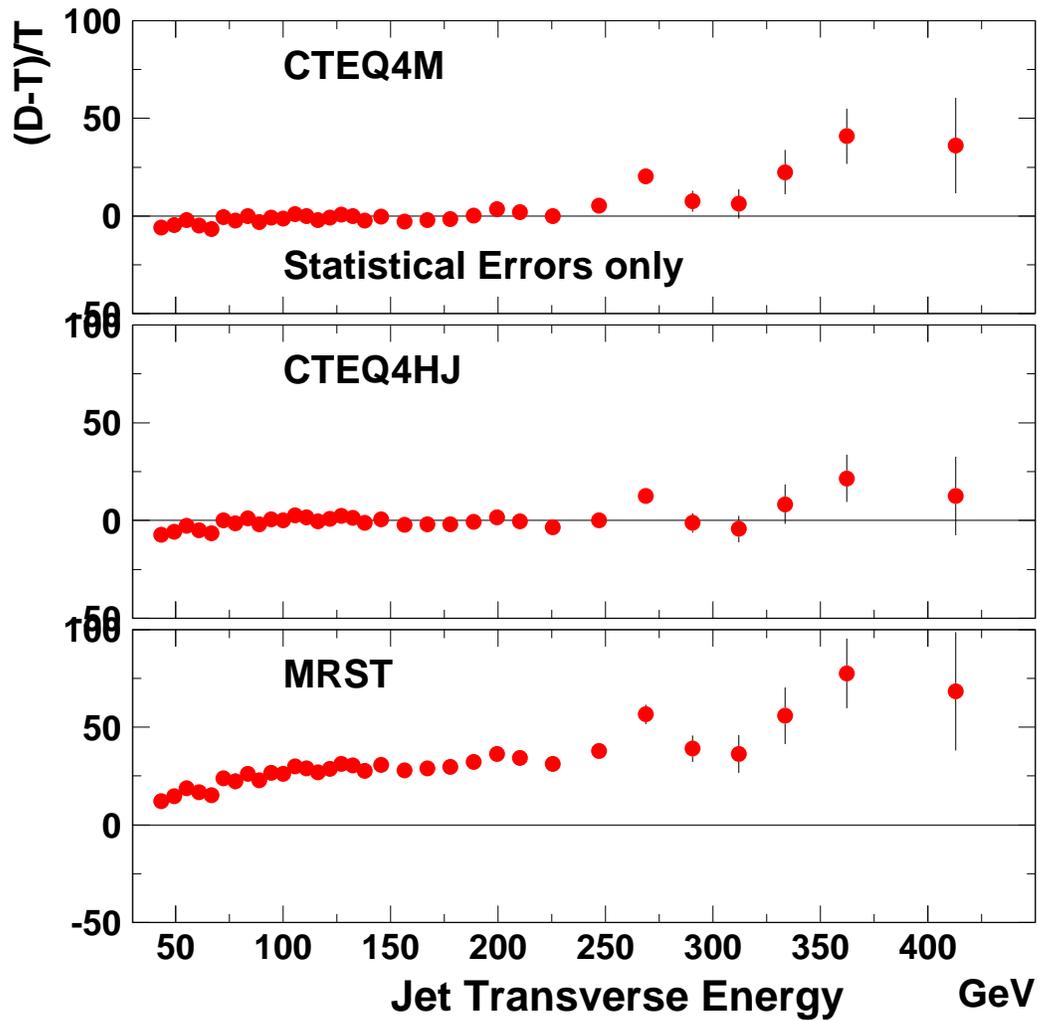,height=15cm,width=15cm}}
\caption{Run 1B data compared to QCD predictions 
(EKS, $\mu$=$E_T$/2, $R_{sep}$=1.3) using 
the CTEQ4M, CTEQ4HJ and MRST PDF's.  Only statistical 
uncertainties are shown on the data points.}
\label{fig-cor-pdfs}
\end{figure}

Comparison of the CDF data to
theoretical predictions with CTEQ4M, CTEQ4HJ and MRST parton distribution
functions are presented in Fig.~\ref{fig-cor-pdfs}. 
The predictions using CTEQ4HJ have the best agreement with the data 
in both shape and normalization without consideration
of systematic uncertainties.  When these are included our 
analysis finds that combinations of systematic uncertainties 
tend to balance against each other 
and produce only small overall changes in the shape of the 
inclusive jet $E_T$ spectrum.
The total $\chi^2$ and confidence level for CTEQ4HJ 
are 46.8 and 10.1\% for 33 degrees of freedom.
When only statistical uncertainties are considered,
the CTEQ4M predictions agree well with the CDF data in shape and 
normalization at
low $E_T$, but diverge from the data at high $E_T$.  The statistical
precision of the data and the smooth, generally 
monotonic $E_T$ dependence of the 
systematic uncertainties result in a poor fit to 
the CTEQ4M prediction.  The abrupt change 
in agreement with the data 
between 200 and 250 GeV can not be accounted for through the systematic
uncertainties resulting in a $\chi^2$ of 63.1 and confidence level of 1\%.
As shown in Fig.~\ref{fig-cor-pdfs}, the 
predictions using MRST do not agree with the CDF data 
in shape or normalization when only statistical uncertainties are considered.
The fitting technique developed
in this paper makes it possible to see how the systematic uncertainties
combine to accommodate this disagreement. In contrast to the fits to
CTEQ4M and CTEQ4HJ, with MRST the systematic uncertainties tend to all shift
in the same direction, decreasing the cross section. 
The monotonically increasing disagreement between the prediction and the data
is similar in shape to the $E_T$ dependence of some
of the systematic uncertainties.  With MRST, the total $\chi^2$ of 49.5 and 
confidence level
of 7\% falls between the results for CTEQ4M and CTEQ4HJ.

Fig.~\ref{fig-cor-pdfs} illustrates that
a quantitative representation of the level of agreement between
the data and the different predictions should indicate
significant differences between the different PDF's.  However,
the resulting $\chi^2$s and confidence levels do not.  
To enhance the discriminating power
of the data we employ a new $\Delta\chi^2$ technique.  This method
results in relative probabilities between two predictions.  Using 
this technique we find that the CTEQ4HJ prediction
is favored over the MRST prediction by a factor of two
and over most of the other predictions by a factor of more than
100.

In conclusion, we have measured the inclusive jet cross section in
the $E_T$ range 40-465 GeV.  The statistical precision of the
data are significantly better than the systematic
uncertainty in the measurement and in the theoretical
predictions.  The CDF Run 1B data is consistent with the Run 1A
result and with the D0 measurement.
Our result is also consistent with NLO QCD
predictions over seven orders of magnitude in jet production rates
if the flexibility allowed by current knowledge of the proton
parton distributions is included in the calculation.

\section{Acknowledgments}

     We thank the Fermilab staff and the technical staffs of the
participating institutions for their vital contributions.  This work was
supported by the U.S. Department of Energy and National Science Foundation;
the Italian Istituto Nazionale di Fisica Nucleare; the Ministry of Education,
Science, Sports and Culture of Japan; the Natural Sciences and Engineering 
Research Council of Canada; the National Science Council of the Republic of 
China; the Swiss National Science Foundation; the A. P. Sloan Foundation; the
Bundesministerium fuer Bildung und Forschung, Germany; the Korea Science 
and Engineering Foundation, and the Comision Interministerial de Ciencia y
Tecnologia, Spain.


\appendix
\section{}
For the results from the 1987 run~\cite{CDF-Inclusive-Jet-87}  and the
associated compositeness limits a covariance matrix was constructed from
the quadrature sum of the systematic uncertainties.  In subsequent analyses
~\cite{CDF-Inclusive-Jet-96,CDF-Inclusive-Jet-92,CDF-XT-analysis}
to better take into account the independence of the eight components of
systematic uncertainty, a covariance matrix was constructed as follows:

\begin{equation}
cov(i,j) = \sum_{k=1}^{8} \rho_{ij}\sigma_k(i) \sigma_k(j) + 
\delta(i,j)stat(i)^2,
\end{equation}
where $\rho_{ij}$ are correlation coefficients 
($=1.0$ for the 100\% correlation of our uncertainties),
$\sigma_k(i)$ and $\sigma_k(j)$ represent the uncertainty 
from source $k$ in bins $i$ and $j$,
the sum is over the eight systematic uncertainties 
in Fig.~\ref{fig-sys-uncertainties}, and $\delta$ is $1$ when
$i=j$ and $0$ otherwise.
For the Run 1B analysis we have decided to average the positive and
negative side uncertainties to determine
$\sigma_k(i)$ and $\sigma_k(j)$.
For the Run 1A analysis and previous analyses, the positive or negative
side uncertainty was chosen depending on whether the data was above or below 
the theoretical prediction.  
Since the uncertainties are almost symmetric, the results are insensitive to 
this choice.

\begin{figure}
\centerline{\psfig{figure=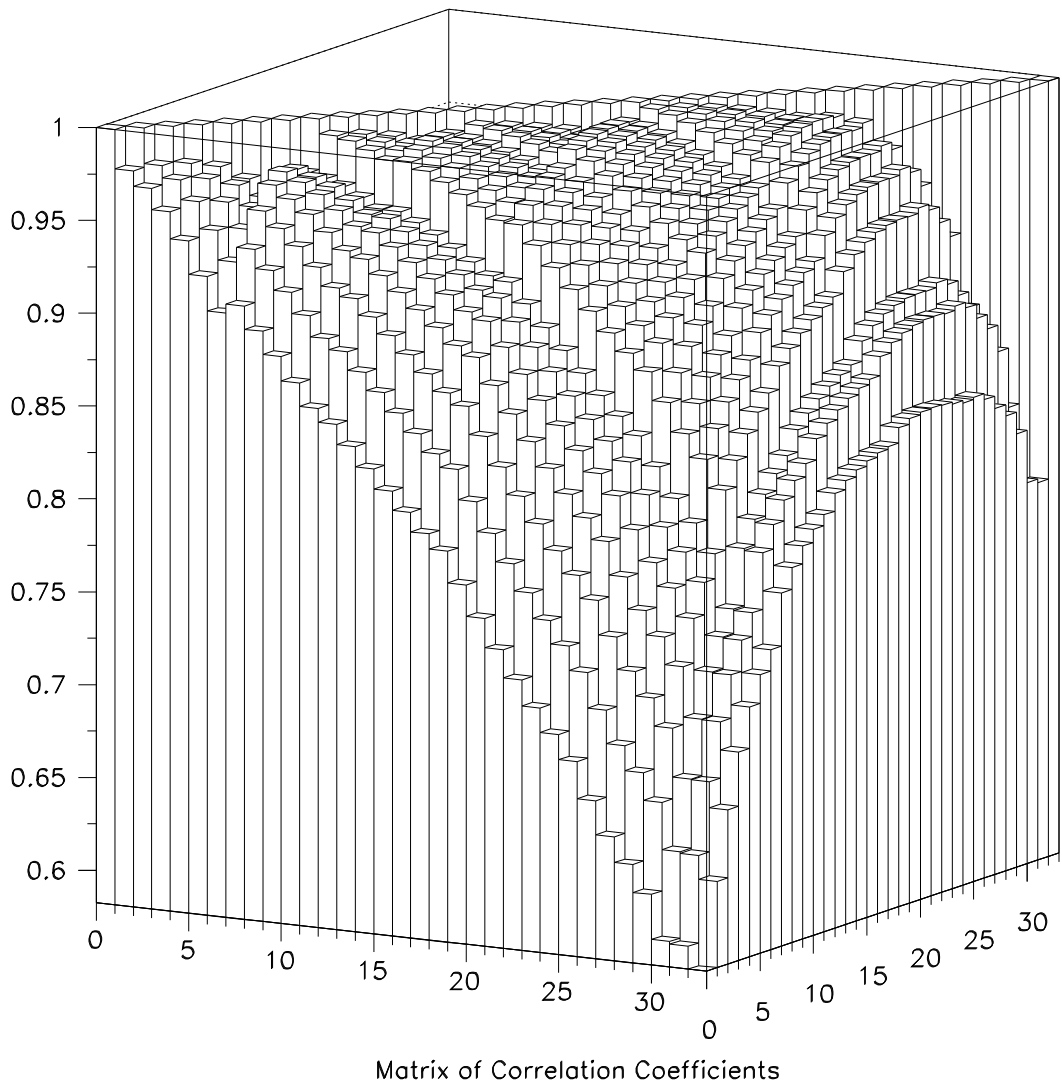,height=16cm,width=16cm}}
\caption{Matrix of correlation coefficients as defined in the text. 
Note the suppressed zero. }
\label{fig-cormat}
\end{figure}

\begin{figure}
\centerline{\psfig{figure=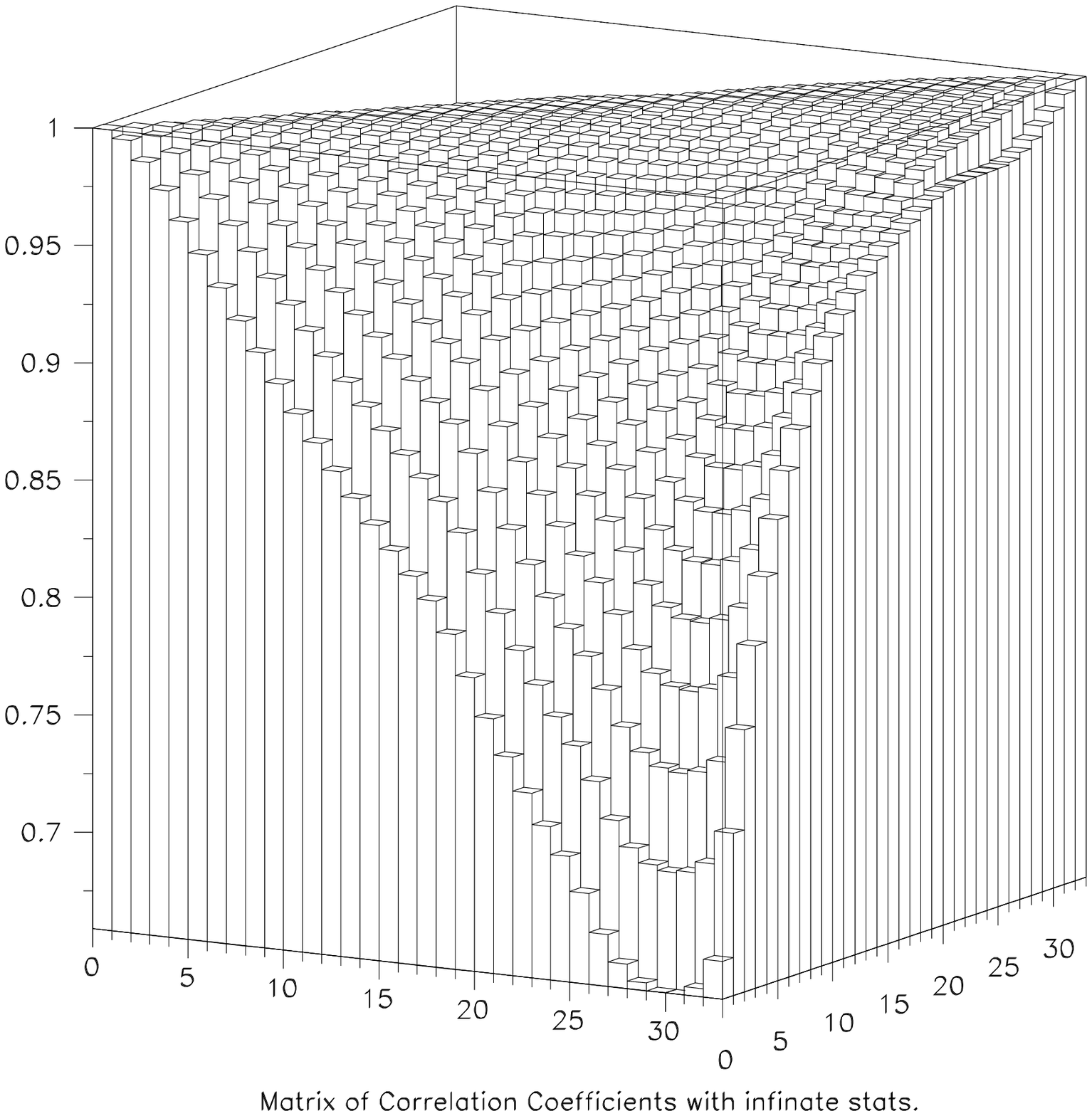,height=16cm,width=16cm}}
\caption{Matrix of correlation coefficients for infinite statistics 
as defined in the text.}
\label{fig-cormat-inf}
\end{figure}

The associated matrix of correlation coefficients can be formed from the
covariance matrix:

\begin{equation}
cor(i,j) = \frac{cov(i,j)}{\sqrt{cov(i,i)cov(j,j)}}.
\end{equation}
Figure~\ref{fig-cormat} shows the correlation matrix for the Run 1B
data and systematic uncertainties.  The steps in the distribution are
from the different trigger samples and relative normalization uncertainties.
Although the eight independent uncertainties are each 100\% correlated from
bin to bin, the combination results in 
the lowest and highest $E_T$ points being
only 60\% correlated.  This is due primarily to the 
statistical uncertainty on the high $E_T$ points.  In addition, 
the underlying event
uncertainty allows shifts in the low $E_T$ region without 
affecting the high $E_T$ region and the high $P_T$ pion response uncertainty
which allows shifts at high $E_T$ with only small changes at low $E_T$.
In the limit of infinite statistics in each
bin, these correlations become larger, particularly for the high $E_T$ 
points.  
Figure~\ref{fig-cormat-inf} shows the matrix of correlation coefficients
for infinite statistics.

The agreement between data and a prediction can be expressed as

\begin{equation}
\chi^2 = \sum_{i=1}^{N}\chi_i,
\end{equation}
where
\begin{equation}
\chi_i = \sum_{j=1}^{N} (D-T)_i cov^{-1}(i,j) (D-T)_j,
\end{equation}
N is the number of bins, $\Delta_i$ and $\Delta_j$ are the 
difference between data and theory for bins $i$ and $j$, and $cov^{-1}(i,j)$ is
the inverse of the covariance matrix.

As an initial study, we calculate the $\chi^2$ of 
the corrected data to the nominal curve. 
In this case inclusion of systematic uncertainties is irrelevant because the
curve already is a good fit to the shape of the data.

\begin{figure}
\centerline{\psfig{figure=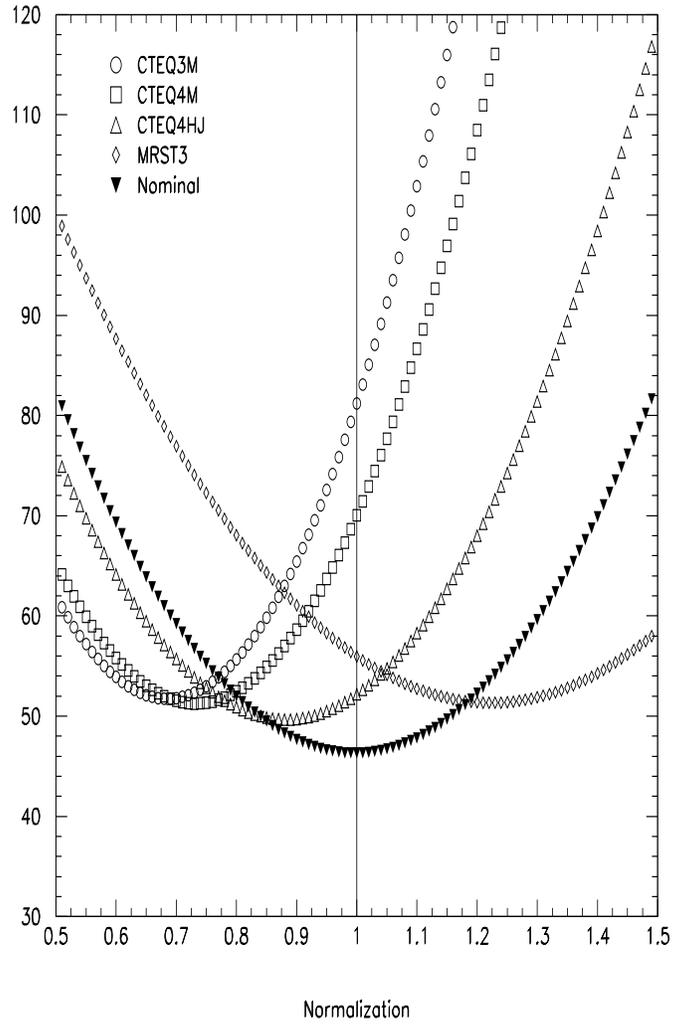,height=15cm,width=10cm}}
\caption{Covariance matrix $\chi^2$ as a function of theory normalization factor
for predictions with different PDF's.}
\label{fig-chi2-normed}
\end{figure}

Many of the theoretical uncertainties can be characterized
primarily by a change in normalization.
To investigate the effects of different normalizations
we perform the fits with a range of normalization factors.
Figure~\ref{fig-chi2-normed} shows the $\chi^2$ as a function of the
theory normalization factor.
Note that if the normalization were completely unconstrained, 
all the PDF's would give similar agreement with the data.

To illustrate the effect of individual systematic uncertainties
we calculate the covariance matrix and $\chi^2$ 
with only one systematic uncertainty.
Table~\ref{Table-chi2-1sys} shows
the $\chi^2$ for  MRST-g$\downarrow$ and CTEQ4HJ.  We chose these two theory predictions
for comparisons since they are have the most discrepant shapes.
For MRST-g$\downarrow$ , the single 
most effective systematic uncertainty
is the jet energy scale since a 1$\sigma$ shift produces a slope similar
to the disagreement between the prediction and the data.
For CTEQ4HJ the most effective uncertainty is the underlying event since
it allows a change of shape at low $E_T$ without affecting the agreement
at high $E_T$.

\subsection{Details and problems with the covariance matrix}

It can be shown that the covariance matrix is equivalent to
the fitting method described in the main text if the following
definition of the $\chi$ is used:
\begin{equation}
 \chi^2 = \sum_i  \frac { (T_i - F_i - Y_i)^2}{(\Delta Y_i)^2} +
\sum_k S_k^2, 
\end{equation}
where
\begin{equation}
F_i = \sigma_{CDFSTD}*(\sum_i f_i^k S_k). 
\end{equation}

Here the systematic shifts are implemented as an additive rather than
a multiplicative factor (the corrections to our data are derived as 
multiplicative factors).  
In this definition, the shifts can be seen as modifying either
the data or the theoretical predictions.
If one views this definition as shifting the data, this definition has
the unfortunate feature that the sum of the percentage
shifts (the $f_i$) enter the cross section calculation by multiplying by the
standard curve rather than the actual corrected cross section.
This effectively reduces the statistical scatter
of the data around the smooth curve.  

On the other hand, if one views the $F_i$ term in
this $\chi^2$ as modifying the theory to give
better agreement with the data, then a more correct estimate of
the uncertainty on the theory would be to scale the sum of the shifts by
the theoretical prediction.  This requires a different covariance matrix
for each theoretical curve.
A more formal discussions of these problems with 
the covariance matrix is presented in reference~\cite{covprob}.

\section{}

Here we expand the procedures developed for comparisons of
data sets to include comparison to theoretical predictions.
In contrast to the analysis presented in the main text, this section
compares the corrected cross section to  
the theoretical predictions rather than comparing the uncorrected
data (number of events/bin) to theoretical predictions which have 
been smeared by detector resolution effects.

As discussed in the main text,  
the $\chi^2$ between data and theory is defined as:

\begin{equation}
\chi^2 = \sum_i  \frac { (T_i/F_i - Y_i)^2}{(\Delta Y_i)^2} +
\sum_k S_k^2,
\end{equation}
where
\begin{equation}
F_i = \Pi(1+f_i^k S_k),
\end{equation}
and
\begin{equation}
f_i^k= |C_i^k-C_i^{STD}|/C_i^{STD}, 
\end{equation}

and $Y_i$ are the data, $\Delta Y_i$ are the statistical
uncertainty in the cross section, $T_i$ are the theory predictions,
$C_i^k$ are the curves for each of the $k$ systematic uncertainties
(in cross section), evaluated
for the $ith$ bin.  The $S_k$ are the eight parameters that
are adjusted in the fit to give good agreement between the data $Y_i$ and
the theory curve, $T_i$.
Figure~\ref{fig-sys-uncertainties}
shows the systematic uncertainty curves, e.g. the $f_i^k$.

\subsection{Results: comparison of corrected data to predictions}

Table~\ref{Table-chi1-derr} shows the results of
the best fit for a variety of PDF's.  All calculations used $\mu=E_T/2$ 
and the EKS program with $R_{sep}$=1.3.
The parameters resulting from the fit (i.e. the factors multiplying the
systematic uncertainty curves) are shown in Table~\ref{Table-parm-derr}.

\begin{figure}
\centerline{
\psfig{figure=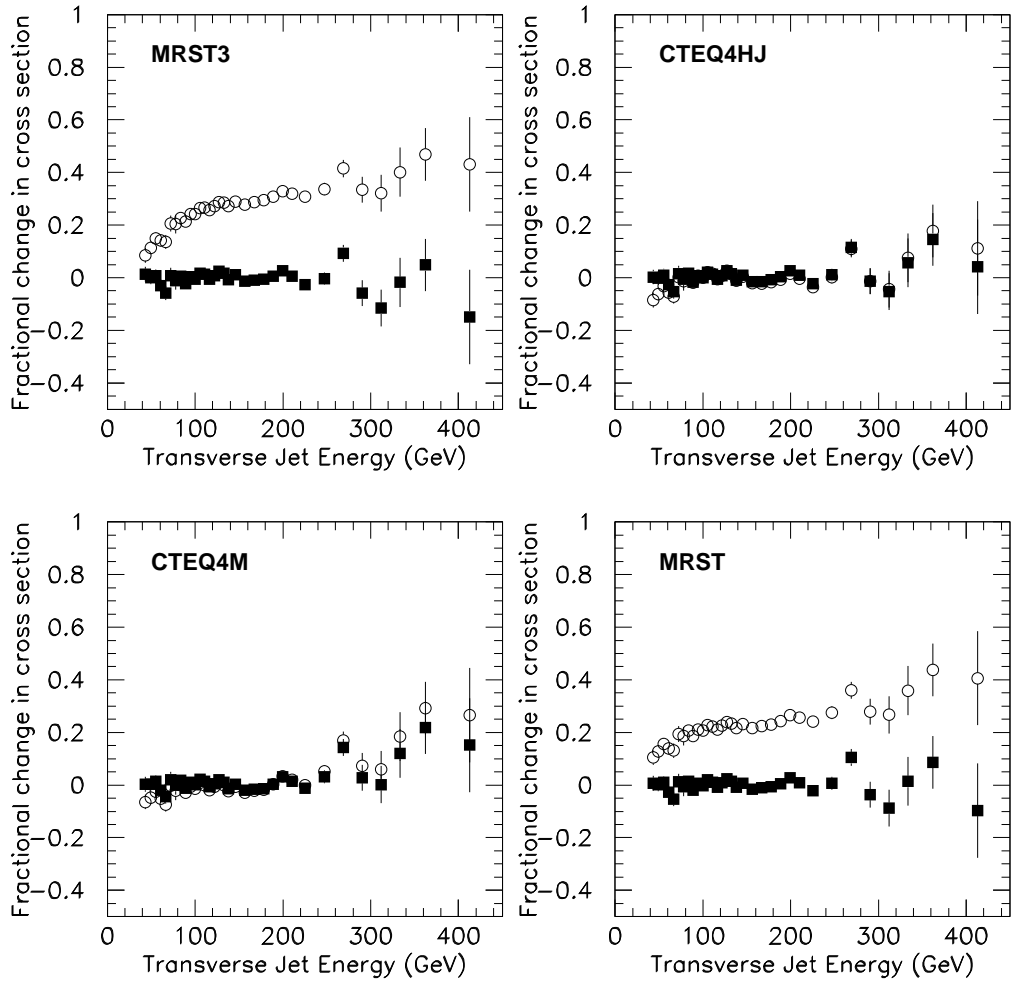,height=15cm,width=15cm}}
\caption{Data compared to theory before (open) and after (solid) shifts
for four theoretical predictions.}
\label{fig-sys-all-a}
\end{figure}

\begin{figure}
\centerline{
\psfig{figure=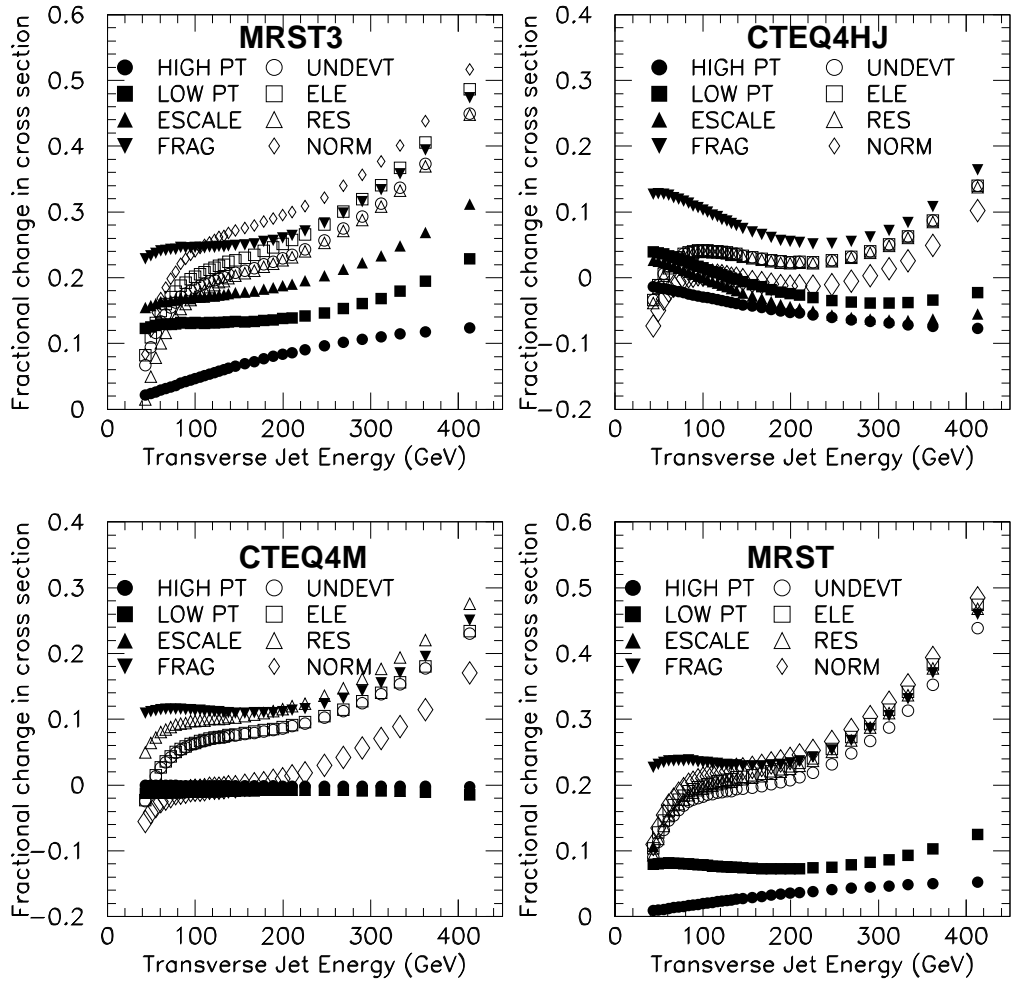,height=15cm,width=15cm}}
\caption{Sequential sum of the fitted shifts.}
\label{fig-sys-all-c}
\end{figure}

Figure~\ref{fig-sys-all-a} shows plots of (data-theory)/data with the
solid points and (data-scaled theory)/data as the open circles,
where scaled theory is the $T_i/F_i$ from above. 
Comparisons are shown for predictions using CTE4HJ, CTEQ4M, MRST and 
MRST-g$\downarrow$ .
To illustrate the size of each shift another series of plots have been made.
In these, the individual curves are multiplied by the associated fit parameter
shown in Table~\ref{Table-parm-derr}.  
In Fig.~\ref{fig-sys-all-c}  the sum
of the shifts is shown sequentially starting from
the upper left of the list of parameters and working down.
First the fit parameter multiplied by the high-pt pion curve
is plotted, then hipt + lowpt, then hipt+lowpt+escale, etc.
The total scale factor is thus labeled NORM, since this is the final uncertainty
in the list.

Since the shapes of the systematic uncertainty curves are very similar,
there are different solutions which can each give similar $\chi^2$.
In effect the systematic uncertainties can compensate for each other,
and the resulting fit parameters are highly correlated with each other.
For example, a pseudo-theory curve can be
created which is simply the standard curve
plus a 1$\sigma$ shift in the high Pt pion response.  When this curve is fit,
the results are not 1$\sigma$ for high pt pion and negligible shifts for
the other systematics.  Rather, the $\chi^2$ penalty is spread over all
the systematics, with a total contribution of 0.5 instead of 1.0.
This suggests that the
individual fit parameters are not extremely meaningful.

This whole procedure ignores the theoretical uncertainties, which
we previously established as primarily normalization but with some shape 
as well.  The procedure above was repeated but the normalization was allowed to
be a free parameter.  The results are shown in Tables~\ref{Table-chi1-derr-n}
and~\ref{Table-parm-derr-n}.

\section{}

As discussed in Section IX.A, 
Tables~\ref{Table-limited-sys-mrst3}
and ~\ref{Table-limited-sys-cteq4a3} show the
results of the fits between the raw jet cross section
and the smeared theoretical predictions when a limited number of 
systematic uncertainties are used.
The combination of uncertainties 
which produces the smallest $\chi^2$ can be the result of precise 
cancelations between the eight effects. 
Although the sources of uncertainty are independent of each other,
they produce similar changes in shape in the cross section.
The fits are performed using 0 to eight systematic uncertainties.
The best $\chi^2$s from all combinations of systematic uncertainties 
are given in the Tables.


%

\vspace{2cm}

\vspace{2cm}






%
%

\begin{table}
\begin{center}
\caption{Coverage the CDF calorimeter components.}
\label{table-calo}
\small
\vspace{0.2cm}  
\begin{tabular}{c|c|c}
\multicolumn{3}{c}{Central}\\
\hline
Name & Rapidity & $\phi$-$\eta$ Segmentation \\
\hline
 CEM & 0.0- 1.1 & \\
 CHA & 0.0- 0.9 & $15^0 \times 0.1$ \\
 WHA & 0.7 - 1.3        & \\
\hline
\multicolumn{3}{c}{Forward}\\
\hline
Name & Rapidity & $\phi$-$\eta$ Segmentation \\
\hline
 PEM & 1.1-2.4 & \\
 PHA & 1.3-2.4 & $5^0\times 0.1$ \\
 FEM & 2.2-4.2 & \\
 FHA & 2.3-4.2 & \\
\end{tabular}
\end{center}
\end{table}

\begin{table}
\begin{center}
\caption{Trigger requirements for Run 1B jet data}
\label{table-treq}
\small
\vspace{0.2cm}  
\begin{tabular}{c|l|c|l|c}
Offline $E_T$ (GeV) &
L2 $E_T$ (GeV) &
L3 $E_T$ (GeV) &
PS &
Efficiency \\
\hline
 40-45 & & & & 96.3 $\pm$ 2\%  \\
 45-50 & & & & 98.5 $\pm$ 1\%  \\
 50-55 & & & & 99.3 $\pm$ 1\%  \\
 55-60 & Single Jet $>$ 20 &Single Jet $>$ 10 & 967 & 99.7 $\pm$ 0.5\%  \\
 60-65 & & & & 99.9 $\pm$ 0.1\%  \\
 65-70 & & & & 100.0 \\
 70-75 & & & & 100.0 \\
\hline
 75-80 & & & & 94.7 $\pm$ 0.8\% \\
 80-85 & & & & 98.0 $\pm$  0.6\% \\
 85-90 & Single Jet $>$ 50 & Single Jet $>$35 & 39.5 & 94.7 $\pm$ 0.6\% \\
 90-95 & & & & 94.7 $\pm$ 0.6\% \\
 95-100 & & & & 94.7 $\pm$ 0.7 \% \\
\hline
 100-105 & & & & 96.7 $\pm$ 0.3\%   \\
 105-110 & & & & 98.3 $\pm$ 0.3\%   \\
 110-115 & Single Jet $>$ 70 & Single Jet $>$55 & 8.11 & 98.9 $\pm$ 0.3\%   \\
 115-120 & & & & 99.0 $\pm$ 0.3\%   \\
 120-125 & & & & 99.3 $\pm$ 0.3\%   \\
 125-130 & & & & 99.5 $\pm$ 0.3\%   \\
\hline
 130-440 & Sum Jet $>$ 175   & Single Jet $>$80 & 1 & $100^{+0.0}_{- 0.5}\%$    \\
\end{tabular}
\end{center}
\end{table}

\begin{table}
\begin{center}
\caption{Bins in leading jet $E_T$ 
for comparison of event parameters to HERWIG + detector 
simulation}
\label{tab-raw-plots}
\small
\vspace{0.2cm}  
\begin{tabular}{l|l}
$E_T$ (GeV) & Trigger name \\
\hline
 100-130  & jet-70 \\
 130-150  & \\
 150-200  & \\
 200-250  & jet-100 \\
 250-300  & \\
 300-500  & \\
\end{tabular}
\end{center}
\end{table}

\begin{table}
\begin{center}
\caption{Underlying Event Energy: Raw $E_T$ in a cone of R = 0.7 in 
minimum bias data as a function of the number of found vertices}
\label{table-ue}
\small
\vspace{0.2cm}  
\begin{tabular}{c|c }
Vertices & $E_T$ in Cone(GeV) \\ 
\hline
0	&	0.48 \\
1	&	1.27 \\
2	&	2.18 \\
3	&	3.01 \\
4	&	3.78 \\
$>$4	&	4.98 \\
\end{tabular}
\end{center}
\end{table}

\begin{table}
\begin{center}
\caption{CDF inclusive jet cross section and uncorrelated uncertainty 
from Run 1B data}
\label{tab-cor-xsec}
\scriptsize
\begin{tabular}{c|c|c}
Bin & $E_T$  (GeV) & cross section  (nb/GeV)\\
\hline
 1&$  43.3$&$ (0.576\pm 0.016)\times 10^{+2}$\\
 2&$  49.3$&$ (0.290\pm 0.007)\times 10^{+2}$\\
 3&$  55.2$&$ (0.160\pm 0.004)\times 10^{+2}$\\
 4&$  61.0$&$ (0.893\pm 0.021)\times 10^{+1}$\\
 5&$  66.7$&$ (0.528\pm 0.014)\times 10^{+1}$\\
 6&$  72.3$&$ (0.355\pm 0.011)\times 10^{+1}$\\
 7&$  77.9$&$ (0.226\pm 0.008)\times 10^{+1}$\\
 8&$  83.5$&$ (0.154\pm 0.002)\times 10^{+1}$\\
 9&$  89.0$&$ (0.102\pm 0.001)\times 10^{+1}$\\
10&$  94.5$&$ (0.729\pm 0.010)\times 10^{+0}$\\
11&$ 100.0$&$ (0.513\pm 0.008)\times 10^{+0}$\\
12&$ 105.5$&$ (0.378\pm 0.007)\times 10^{+0}$\\
13&$ 110.9$&$ (0.274\pm 0.003)\times 10^{+0}$\\
14&$ 116.3$&$ (0.199\pm 0.002)\times 10^{+0}$\\
15&$ 121.7$&$ (0.151\pm 0.002)\times 10^{+0}$\\
16&$ 127.1$&$ (0.116\pm 0.002)\times 10^{+0}$\\
17&$ 132.5$&$ (0.877\pm 0.014)\times 10^{-1}$\\
18&$ 137.9$&$ (0.659\pm 0.012)\times 10^{-1}$\\
19&$ 145.7$&$ (0.466\pm 0.003)\times 10^{-1}$\\
20&$ 156.4$&$ (0.281\pm 0.002)\times 10^{-1}$\\
21&$ 167.2$&$ (0.178\pm 0.001)\times 10^{-1}$\\
22&$ 177.9$&$ (0.115\pm 0.001)\times 10^{-1}$\\
23&$ 188.7$&$ (0.763\pm 0.009)\times 10^{-2}$\\
24&$ 199.5$&$ (0.520\pm 0.008)\times 10^{-2}$\\
25&$ 210.2$&$ (0.344\pm 0.006)\times 10^{-2}$\\
26&$ 225.4$&$ (0.195\pm 0.003)\times 10^{-2}$\\
27&$ 247.1$&$ (0.968\pm 0.023)\times 10^{-3}$\\
28&$ 268.8$&$ (0.535\pm 0.017)\times 10^{-3}$\\
29&$ 290.5$&$ (0.236\pm 0.012)\times 10^{-3}$\\
30&$ 312.1$&$ (0.117\pm 0.008)\times 10^{-3}$\\
31&$ 333.6$&$ (0.685\pm 0.064)\times 10^{-4}$\\
32&$ 362.2$&$ (0.322\pm 0.032)\times 10^{-4}$\\
33&$ 412.9$&$ (0.630\pm 0.113)\times 10^{-5}$\\
\end{tabular}
\end{center}
\end{table}

\begin{table}
\begin{center}
\caption{Parameters for systematic error curves described in Equation 12 and
shown 
in Figure~\ref{fig-sys-uncertainties}.}
\label{Table-syserr}
\small
\vspace{0.2cm}  
\begin{tabular}{l|c|c|c|c|c|c|c}
   & ($P_0$)$\times10^{+07}$ & $P_1$ & $P_2$ & $P_3$ & $P_4$ & $P_5$ & $P_6$ \\
\hline
Standard Curve& 0.14946 & -2.9228& 4.4881& -4.9447& 1.7891& -0.2297& 5.6147\\
\hline
\multicolumn{8}{c}{Positive Systematic Uncertainties}\\
\hline
 High Pt Hadron & 0.11521&-2.7511& 4.4129&-4.9487& 1.7989&-0.2325& 5.3079\\
 Low Pt hadron  & 0.16445&-2.9824& 4.4867&-4.9415& 1.7911&-0.2287& 6.3165\\
 Stability      & 0.15275&-2.9176& 4.4883&-4.9449& 1.7889&-0.2297& 5.4732\\
 Fragmentation  & 0.17922&-3.0070& 4.4857&-4.9406& 1.7917&-0.2285& 6.5970\\
 Und. Event     & 0.02392&-2.2945& 4.4609&-4.9923& 1.7764&-0.2228& 5.8629\\
 Neutral Pion   & 0.14852&-2.9146& 4.4884&-4.9451& 1.7888&-0.2298& 5.4920\\
 Resolution     & 0.10392&-2.8451& 4.4958&-4.9455& 1.7878&-0.2304& 5.4340\\
\hline
\multicolumn{8}{c}{Negative Systematic Uncertainties}\\
\hline
 High Pt Pion& 0.12506&-2.7639& 4.3972&-4.9442& 1.8030&-0.2324& 5.6243\\
 Low Pt  Pion & 0.13604&-2.8651& 4.4891&-4.9479& 1.7870&-0.2306& 4.9412\\
 Stability    & 0.14757&-2.9299& 4.4878&-4.9444& 1.7892&-0.2296& 5.7798\\
 Fragmentation & 0.12561&-2.8404& 4.4904&-4.9487& 1.7865&-0.2308& 4.6655\\
 Und. Event & 0.34976&-3.1079& 4.4710&-4.9422& 1.7923&-0.2279& 6.3048\\
 Neutral Pion  & 0.15065&-2.9332& 4.4877&-4.9443& 1.7893&-0.2296& 5.7700\\
 Resolution& 0.20458&-2.9888& 4.4814&-4.9441& 1.7901&-0.2291& 5.7412\\
\end{tabular}
\end{center}
\end{table}

\begin{table}
\begin{center}
\caption{Estimates of theoretical uncertainty for
three values of jet $E_T$. The \% difference
between various predictions is shown in 
Figures~\ref{fig-eks-rsep} to~\ref{fig-pdfs}.}
\label{tab-thy-err}
\small
\vspace{0.2cm}  
\begin{tabular}{l|c|c|c|c }
Source & \multicolumn{3}{c}{\% difference} & Shape \\
\hline
& 50 GeV  & 150 GeV & 400 GeV & \\
\hline
Clustering ($R_{sep}=2.0)$& 5.2 & 4.8 & 4.0 & Monotonic\\
Scale: $E_T^{jet}$ vs. $E_T^{max}$ & 6.0 & 3.0 & 1.0 & Monotonic \\
Scale:$\mu=C*E_T^{jet}$, C=0.5 - 2.0 & 20 & 20 & 20 & Flat \\
\hline
PDFs & & & & \\
\hline
 CTEQ4 series (CTEQ4M ref.) & 10 & 3 & 2 & Monotonic \\
 CTEQ4HJ (CTEQ4M ref.)     & 1 & 1 & 20 & Not monotonic \\ 
 MRST series (MRST ref.) & 15 & 20 & 6 & Not monotonic \\
 MRST vs CTEQ4M & 15 & 30 & 20 & Not monotonic \\ 
\end{tabular}
\end{center}
\end{table}

\begin{table}
\begin{center}
\small
\caption{Results of the fit described by Equation 19.
$\chi^2_{stat}$ represents the scatter of the points around
a smooth curve, while the $\chi^2_{sys}$ represents 
the $\chi^2$ penalty from the
systematic uncertainties.  $\chi^2_{tot}$ is the sum of the two terms.
The systematic shift columns show the individual $s_k$ for each systematic
as defined in the text.}
\label{table-best-theory-errors}.
\begin{tabular}{c|c|c|c|c|c|c|c|c|c|c|c}
PDF & $\chi^2_{tot}$ & $\chi^2_{stat}$ & $\chi^2_{sys}$ &
Hi-Pi & Lo-Pi & Sta. & Frg. & UE & $\pi^0$  & Res. & Norm. \\
\hline                                                   
 CDFSTD                              &  42.3& 41.3&  1.0& -0.380& -0.223& -0.285&   0.791& -0.141& -0.140&  0.056& -0.278\\
 CTEQ4M                              &  63.4& 48.2& 15.2& -0.395& -0.411& -0.500&   2.350& -1.443&  0.168&  0.937& -2.467\\
 CTEQ4HJ                             &  46.8& 40.7&  6.1&  0.329& -0.741& -0.549&   1.686& -1.235& -0.166& -0.053& -0.872\\
 CTEQ4A1                             &  60.1& 47.1& 13.0& -0.001& -0.670& -0.560&   2.401& -0.877&  0.075&  0.875& -2.219\\
 CTEQ4A2                             &  61.5& 47.4& 14.1& -0.083& -0.667& -0.604&   2.404& -1.126&  0.073&  0.833& -2.358\\
 CTEQ4A3                             &  63.4& 48.2& 15.2& -0.395& -0.411& -0.500&   2.350& -1.443&  0.168&  0.937& -2.467\\
 CTEQ4A4                             &  64.5& 48.8& 15.7& -0.365&  0.061& -0.732&   2.270& -1.555&  0.026&  0.866& -2.597\\
 CTEQ4A5                             &  67.0& 49.8& 17.2& -0.490&  0.214& -0.751&   2.264& -1.723& -0.068&  0.911& -2.719\\
 MRST                                &  49.5& 40.8&  8.7&  0.743&  0.756&  0.684&   2.123& -1.508&  0.485& -0.293&  0.210\\
 MRST-g$\uparrow$                    &  53.3& 43.3& 10.0&  0.773& -0.314&  0.166&   2.677& -1.014&  0.283&  0.030& -1.005\\
 MRST-g$\downarrow$                  &  59.2& 45.7& 13.5&  0.687&  1.726&  1.166&   1.741& -1.879&  0.699& -0.692&  1.068\\
 MRST-$\alpha_s\downarrow\downarrow$ &  59.7& 41.4& 18.3&  2.436& -0.050&  0.581&   2.604& -1.302&  0.362& -1.234&  1.391\\
 MRST-$\alpha_s\uparrow\uparrow$     &  53.4& 43.9&  9.5& -0.221&  1.413&  0.508&   1.922& -1.640&  0.440&  0.309& -0.731\\
\end{tabular}
\end{center}
\end{table}

\begin{table}
\begin{center}
\caption{The effect of including limited systematic uncertainties in the fit
to QCD predictions using CTEQ4HJ PDF's.
The first column indicates the number of systematic uncertainties included
(e.g. the first row is with no systematic uncertainties).
The next three columns indicate the total $\chi^2$, the contribution
from the uncorrelated scatter of the points around
a smooth curve, $\chi^2_{stat}$, and the penalty from the
correlated shifts from the systematics uncertainties $\chi^2_{sys}$ .
The remaining eight columns represent the $s_k$ which result from
the fit for the eight systematic uncertainties
as described in the text. }
\label{Table-limited-sys-4hj}
\small
\begin{tabular}{c|c|c|c|c|c|c|c|c|c|c|c}
& $\chi^2_{total}$ & $\chi^2_{stat}$  & $\chi^2_{sys}$   &
Hi-Pi & Lo-Pi & Stab. & Frg & UE & $\pi^0$ & Res. & Norm. \\
\hline                 
0& 94.2&  94.2&   0.00& 0.000&  0.000&  0.000&  0.000&  0.000&  0.000&   0.000&   0.000\\
1& 79.0&  79.0&   0.0& 0.000&  0.000&  0.000&  0.000& -0.200&  0.000&   0.000&   0.000\\
2& 62.9&  59.5&   3.4& 0.000&  0.000&  0.000&  0.500&  0.000&  0.000&  -1.787&   0.000\\
3& 49.1&  43.3&   5.8& 0.000& -1.459&  0.000&  1.412& -1.304&  0.000&   0.000&   0.000\\
4& 47.6&  40.7&   6.9& 0.000& -1.301&  0.000&  1.729& -1.255&  0.000&   0.000&  -0.821\\
5& 47.1&  40.4&   6.7& 0.000& -0.950& -0.583&  1.883& -1.213&  0.000&   0.000&  -0.686\\
6& 46.9&  40.7&   6.2& 0.339& -0.782& -0.585&  1.664& -1.259&  0.000&   0.000&  -0.868\\
7& 46.9&  40.7&   6.2& 0.338& -0.749& -0.557&  1.682& -1.261& -0.169&   0.000&  -0.860\\
8& 46.9&  40.7&   6.2& 0.329& -0.741& -0.549&  1.686& -1.234& -0.166&  -0.053&  -0.871\\
\end{tabular}
\end{center}
\end{table}

\begin{table}
\begin{center}
\caption{As in previous table except the 
QCD predictions use MRST PDF's.}
\label{Table-limited-sys-mrst1}
\small           
\begin{tabular}{c|c|c|c|c|c|c|c|c|c|c|c}
& $\chi^2_{total}$ & $\chi^2_{stat}$  & $\chi^2_{sys}$   &
Hi-Pi & Lo-Pi & Stab. & Frg & UE & $\pi^0$ & Res. & Norm. \\
\hline              
0& 11039.8& 11039.8&      0.00&  0.000&  0.000&  0.000& 0.000&  0.000& 0.000&    0.000&   0.000\\
1&  141.1&    124.4&     16.7&  0.000&  0.000&  4.083& 0.000&  0.000& 0.000&    0.000&   0.000\\
2&   73.2&     48.0&     25.2&  0.000&  0.000&  0.000& 4.486& -2.259& 0.000&    0.000&   0.000\\
3&   53.4&     39.8&     13.6&  0.000&  0.931&  0.000& 3.270& -1.433& 0.000&    0.000&   0.000\\
4&   50.8&     40.2&     10.6&  1.065&  1.151&  0.000& 2.382& -1.584& 0.000&    0.000&   0.000\\
5&   50.0&     40.4&      9.6&  0.887&  0.827&  0.780& 2.194& -1.657& 0.000&    0.000&   0.000\\
6&   49.8&     40.5&      9.3&  0.840&  0.735&  0.711& 2.134& -1.656& 0.499&    0.000&   0.000\\
7&   49.6&     40.7&      8.9&  0.800&  0.771&  0.723& 2.140& -1.496& 0.502&   -0.322&   0.000\\
8&   49.5&     40.8&      8.7&  0.743&  0.756&  0.684& 2.123& -1.508& 0.485&   -0.293&   0.210\\
\end{tabular}
\end{center}
\end{table}

\begin{table}
\begin{center}
\caption{Comparison of CDF Run 1B data to various theoretical predictions
using the $\chi^2$ and the $\Delta\chi^2$ statistics.}
\label{Table-dchisqr}
\vspace{2mm}
\begin{tabular}{c|c|c|c|c}
PDF & $\chi^2$ & $CL (\%)$ & $\chi^2-\chi^2_{cteq4hj}$ & Prob. Rel. to CTEQ4HJ\\
\hline
 CDFSTD    & 42.3 & 16 & -4.5 &10\\
 CTEQ4HJ   & 46.8 & 10 &  0.0 &1 \\
 MRST      & 49.6 & 7.4&  2.7& 0.5 \\
 MRST-g$\uparrow$     & 53.3 & 4.6 & 6.5& 0.06 \\
 MRST-g$\downarrow$          & 59.2 & 2.4 & 12.4 & 0.01 \\
 MRST-$\alpha_s\downarrow\downarrow$ & 59.8 &2.0&12.9&$<10^{-4}$\\
 MRST-$\alpha_s\uparrow\uparrow$      & 53.4 & 4.8&6.6& 0.07 \\
 CTEQ4A1   & 60.1 &2.1 & 13.3&$<10^{-4}$\\
 CTEQ4A2   & 61.6 &1.8 & 14.7&$<10^{-4}$\\
 CTEQ4M    & 63.4 &1.4 & 16.6&$10^{-3}$\\
 CTEQ4A4   & 64.5 &1.3 & 17.7&$10^{-3}$\\
 CTEQ4A5   & 67.0 &1.0 & 20.2&$<10^{-4}$ \\
\end{tabular}
\end{center}
\end{table}

\begin{table}
\begin{center}
\caption{Covariance matrix $\chi^2$ comparison for various theoretical 
predictions for Run 1B jet data.
The $\chi^2$ for the nominal curve is 46.3 for 33 bins with only statistical 
uncertainty and when the systematics uncertainty is included.}
\label{Table-chi2-nom}
\small
\vspace{0.2cm}  
\begin{tabular}{c|r|r }
PDF  & Stat. only & Stat. and Sys. \\
\hline
 CTEQ3M & 227.0   & 81.2 \\
 CTEQ4M & 119.9 & 70.0   \\
 CTEQ4HJ& 85.4  & 52.2 \\
 MRST-g$\downarrow$ & 12204.0  & 56.0 \\
 MRST-g$\uparrow$ & 4363.0   & 54.6 \\
\end{tabular}
\end{center}
\end{table}

\begin{table}
\begin{center}
\caption{Minimum value of the covariance matrix $\chi^2$ and 
corresponding theory 
normalization factor.} 
\label{Table-chi2-norm}
\small
\begin{tabular}{c|r|r|r|r }
\multicolumn{1}{c}{}	&
\multicolumn{2}{c}{Sys. unc. OFF}  &
\multicolumn{2}{c}{Sys. unc. ON} \\
\hline
PDF & $\chi^2$ & Norm. & $\chi^2$ & Norm. \\
\hline
 CTEQ3M & 118.9 & 0.97 & 51.7 & 0.68 \\
 CTEQ4M & 101.6 & 0.99 & 51.3 & 0.74 \\
 CTEQ4HJ& 75.3 & 0.99 & 49.6 & 0.88 \\
 MRST-g$\downarrow$ & 569.0 & 1.38 & 51.3 & 1.22 \\
 MRST-g$\uparrow$ & 90.8 & 1.19 & 52.2 & 0.88 \\
\end{tabular}
\end{center}
\end{table}

\begin{table}
\begin{center}
\caption{Covariance matrix
$\chi^2$ comparison for various theoretical predictions for 1B jet 
data where only the indicated systematic uncertainties is included.}
\label{Table-chi2-1sys}
\small
\begin{tabular}{c|r|r }
Sys. Uncertainty & MRST-g$\downarrow$  $\chi^2$ & CTEQ4HJ $\chi^2$ \\
\hline
Hi-Pi & 248.6 &77.2 \\
Low-Pi& 1330.0 & 75.2 \\
Stability & 127.9 & 76.1 \\
Fragmentation & 382.1 & 75.9 \\
UE & 3630.0 & 69.6 \\
Neutral Pi & 179.5 & 76.2 \\
Resolution & 1952.0 & 71.0 \\
Normalization & 359.6 & 75.2 \\
\end{tabular}
\end{center}
\end{table}

\begin{table}
\begin{center}
\caption{Results of fits to various PDF's. The first line shows the $\chi^2$
when only the uncorrelated errors on the data points are included.
The next two rows show the contribution to the total $\chi^2$ from
the data - theory term and the $\sum{S_K}$ term.}
\label{Table-chi1-derr}
\small
\begin{tabular}{c|r|r|r|r|r|r|r }
PDF     & CDFSTD & CTEQ3M & CTEQ4M & CTEQ4HJ & MRST & MRST-g$\uparrow$ & MRST-g$\downarrow$\\
\hline
stat.only&44.16    & 220.0 &116.5  & 83.5   & 8119.3 & 4394.9 & 12271.5\\
1st term & 43.66    & 67.75 & 60.52 & 46.33  & 42.70 & 48.05 & 43.13 \\
$\sum{S_K}$ &  1.63e-2 & 14.07 & 9.74  & 4.33   &  4.90 &  6.87 & 9.52  \\
total   &  43.68   & 81.82 & 70.27  & 50.57 & 47.61 & 54.92& 52.64  \\
\end{tabular}
\end{center}
\end{table}

\begin{table}
\begin{center}
\caption{Individual fit parameters for fit results in
Table~\ref{Table-chi1-derr}}
\label{Table-parm-derr}
\small
\begin{tabular}{c|r|r|r|r|r|r|r }
PDF     & STD & CTEQ3M & CTEQ4M & CTEQ4HJ & MRST & MRST-g$\uparrow$ & MRST-g$\downarrow$\\
\hline
Hi-Pi   &  0.0057& 4.23e-7& 0.0020  & 0.710   & -0.478 & 0.0078 & -1.13  \\
Lo-Pi    & -0.0048& 0.861  & 0.159   & -0.702  & -0.937 & -0.722 & -1.35  \\
Stab.   & 0.0023 & -0.0086& -6.2e-6 & 0.288   & -0.629 & -0.227 & -0.741 \\
Frag.    & -0.0053& -1.365 & -1.44   & -1.192  & -1.433 & -2.046 & -0.879 \\
UE   & 0.0086 & 0.5998 & 0.926   & 1.119   & 0.950  & 0.695  & 1.121  \\
Neutral Pi& 3.34e-3& -0.245 & -0.0049 & 1.3e-4  & -0.534 & -0.279 & -0.559 \\
Res.    & 3.83e-3& -1.878 & -1.235  & 0.0071  & 0.180  & -0.752 & 1.131  \\
Norm.     & 1.63e-4  &  2.74& 2.29    & 0.761   & -0.354 & 0.987  & -1.494 \\
\end{tabular}
\end{center}
\end{table}

\begin{table}
\begin{center}
\caption{Results of fits to various PDF's with 
normalization as a free parameter.
The first line shows the $\chi^2$
when only the uncorrelated errors on the data points are included.
The next two rows show the contribution to the total $\chi^2$ from
the data - theory term and the $\sum{S_K}$ term.}
\label{Table-chi1-derr-n}
\small
\begin{tabular}{c|r|r|r|r|r|r|r }
PDF     & STD & CTEQ3M & CTEQ4M & CTEQ4HJ & MRST & MRST-g$\uparrow$ & MRST-g$\downarrow$\\
\hline
stat.only& 43.8    & 113.0 & 99.9   & 74.1  &  216.0 & 86.5 & 575.0 \\
1st term   & 43.66    & 46.63  & 45.76 & 43.90  & 43.01 & 45.16 & 42.88 \\
$\sum{S_K}$    &  1.63e-2 &  9.59 & 8.11  & 4.81   &  4.26 &  7.19 & 4.08  \\
total   &  43.68   & 56.22 & 53.88  & 48.71 & 47.27 & 52.35 & 46.96  \\
\end{tabular}
\end{center}
\end{table}
\begin{table}
\begin{center}
\caption{Individual fit parameters for fit results in
Table~\ref{Table-chi1-derr-n}. Normalization is a free parameter.}
\label{Table-parm-derr-n}
\small
\begin{tabular}{c|r|r|r|r|r|r|r }
PDF     & STD & CTEQ3M & CTEQ4M & CTEQ4HJ & MRST & MRST-g$\uparrow$ & MRST-g$\downarrow$\\
\hline
Hi-Pi   &  0.057& -0.813 & -0.398 & 0.444  & -0.346 & -0.846e-4& -0.586  \\
Lo-Pi    & -0.048& 0.863  & -1.068 & -1.23 & -0.689 & -1.21  & -0.329  \\
Stab.   & 0.023 & -1.114 & -0.882 & -0.054& -0.555 & -0.662 & -0.449 \\
Frag.    & -0.053& -2.247 & -2.205 & -1.405 & -1.434 & -2.10  & -0.943 \\
UE   & 0.086 & 0.070  & 0.537  & 0.997  & 1.026  & 0.576  & 1.441  \\
Neutral Pi& .327e-2& -0.947 & -0.764 & -0.081& -0.493 & -0.579 & -0.411 \\
Res.    & .392e-2& -0.997 & -0.554 & 0.341  & 0.041 & -0.477 & 0.540 \\
Norm     & .842e-3&  9.22 & 7.363 & 2.57 & -.964 & 2.53  & -3.86 \\
\end{tabular}
\end{center}
\end{table}

\begin{table}
\begin{center}
\caption{The effect of including limited systematic uncertainties in the fit
to QCD predictions using MRST-g$\downarrow$ PDF's.
The first column indicates the number of systematic uncertainties included
(e.g. the first row is with no systematic uncertainties).
The next three columns indicate the total $\chi^2$, the contribution
from the uncorrelated scatter of the points around
a smooth curve, $\chi^2_{stat}$, and the penalty from the
correlated shifts from the systematics uncertainties, $\chi^2_{sys}$.
The remaining eight columns represent the results of the
fit (the $s_k$) eight systematic uncertainties
as described in the text. }
\label{Table-limited-sys-mrst3}
\small
\begin{tabular}{c|c|c|c|c|c|c|c|c|c|c|c}
& $\chi^2_{total}$ & $\chi^2_{stat}$  & $\chi^2_{sys}$   &
Hi-Pi & Lo-Pi & Stab. & Frg & UE & $\pi^0$ & Res. & Norm. \\
\hline              
0& 18044.1&  18044.1&      0.0&  0.000&  0.000&  0.000&  0.000&  0.000& 0.000&   0.000&  0.000\\
1&   268.0&    242.9&     25.1&  0.000&  5.010&  0.000&  0.000&  0.000& 0.000&   0.000&  0.000\\
2&   103.7&     52.5&     51.2&  0.000&  0.000&  6.784&  0.000& -2.282& 0.000&   0.000&  0.000\\
3&    69.2&     49.0&     20.2&  0.000&  2.178&  0.000&  3.449& -1.884& 0.000&   0.000&  0.000\\
4&    64.2&     45.4&     18.7&  0.000&  2.000&  0.000&  2.729& -1.988& 0.000&   0.000&  1.809\\
5&    61.5&     45.4&     16.1&  0.000&  1.515&  1.393&  2.265& -2.143& 0.000&   0.000&  1.473\\
6&    60.4&     45.6&     14.8&  0.789&  1.770&  1.208&  1.806& -2.216& 0.000&   0.000&  1.186\\
7&    59.7&     45.6&     14.1&  0.740&  1.850&  1.252&  1.822& -1.884& 0.000&  -0.681&  1.116\\
8&    59.2&     45.7&     13.5&  0.686&  1.726&  1.166&  1.741& -1.879& 0.699&  -0.692&  1.069\\
\end{tabular}
\end{center}
\end{table}

\begin{table}
\begin{center}
\caption{As in previous table except with CTEQ4M PDF's.}
\label{Table-limited-sys-cteq4a3}
\small
\begin{tabular}{c|c|c|c|c|c|c|c|c|c|c|c}
& $\chi^2_{total}$ & $\chi^2_{stat}$  & $\chi^2_{sys}$   &
Hi-Pi & Lo-Pi & Stab. & Frg & UE & $\pi^0$ & Res. & Norm. \\
\hline                       
0& 138.5&    138.5&      0.00&  0.000&  0.000&  0.000&  0.000&  0.000& 0.000&    0.000&   0.000\\
1& 110.9&    110.8&      0.1&  0.000&  0.000&  0.000&  0.000& -0.269& 0.000&    0.000&   0.000\\
2&  90.4&     89.6&      0.8&  0.000&  0.000&  0.000&  0.336& -0.812& 0.000&    0.000&   0.000\\
3&  66.9&     52.6&     14.3&  0.000&  0.000&  0.000&  1.969& -0.841& 0.000&    0.000&  -3.116\\
4&  65.1&     50.6&     14.5&  0.000&  0.000&  0.000&  1.876& -1.159& 0.000&    0.603&  -3.043\\
5&  63.97&     48.7&     15.2&  0.000&  0.000& -0.655&  2.182& -1.483& 0.000&    0.974&  -2.622\\
6&  63.7&     48.7&     15.0& -0.255&  0.000& -0.730&  2.315& -1.350& 0.000&    0.870&  -2.551\\
7&  63.5&     48.3&     15.2& -0.399& -0.372& -0.465&  2.376& -1.451& 0.000&    0.945&  -2.456\\
8&  63.4&     48.2&     15.2& -0.396& -0.412& -0.501&  2.350& -1.443& 0.168&    0.937&  -2.467\\
\end{tabular}
\end{center}
\end{table}


\begin{thebibliography}{999}
\bibitem{CDF-Detector}
F. Abe {\it et al.} (CDF Collaboration), Nucl. Instrum. Methods
{\bf  A271}, 387 (1988); Nucl. Instrum. Methods
{\bf  A268},75 (1988) and references therein. 
\bibitem{CDF-Inclusive-Jet-96}
F. Abe {\it et al.} (CDF Collaboration), Phys. Rev. Lett. {\bf  77} 438(1996).
\bibitem{CDF-Inclusive-Jet-92}
F. Abe {\it et al.} (CDF Collaboration), Phys. Rev. Lett. {\bf  68} 1104(1992).
\bibitem{CDF-Inclusive-Jet-87}
F. Abe {\it et al.} (CDF Collaboration), Phys. Rev. Lett. {\bf  62} 613(1989).
\bibitem{Eichten}E. Eichten, K. Lane, and M. Peskin, Phys. Rev. Lett.
{\bf  50}, 811 (1983).
\bibitem{cteq-jhuston}
J. Huston {\em et al.}, Phys. Rev. Lett. {\bf  77}, 444(1996).
\bibitem{walter-98}
W. Giele and S. Keller Preprint hep-ph/9803393, FNAL-Pub-98/082-T.
\bibitem{cteq-4}
J. Huston {\em et al.}, Phys. Rev. Lett. {\bf  77}, 444(1996).
\bibitem{CDFLUM} F.~Abe {\it et al.} (CDF Collaboration), Phys. Rev. {\bf  D50}
5550, (1994), D. Cronin-Hennessy et al. (CDF Collaboration), Nucl. Instrum. 
Methods {\bf A433} (2000) 37 and 
M.~Albrow, {\it et al.} CDF NOTE 4844, FERMILAB-TM-2071
(1999).
\bibitem{D0inc}
B. Abbot {\it et al.} (D0 Collaboration), 
Phys.Rev.Lett. {\bf 82}, 2451 (1999). 
\bibitem{bj1971}
S.M. Berman, M. Jacob, Phys. Rev. Lett. {\bf 25} 1683 (1970);
S.M. Berman, J.D. Bjorken and J.B. Kogut, Phys. Rev. 4D 3388 (1971);
R.P. Feynman: Photon hadron interactions. New York: Benjamin 1972.
\bibitem{ISR-1}
M.G. Albrow {\em et al.}, Nucl Phys. {\bf B160} 1 (1979);
A.L.S. Angelis {\em et al.}, Phys Scr. 19 116 (1979);
A.G. Clark {\em et al.}, Nucl. Phys. {\bf B160} 397 (1979);
D. Drijard {\em et al.}, Nucl. Phys. {\bf B166} 233 (1980).
\bibitem{UA2-1}
M. Banner {\em et al.} (UA2 Collaboration), Phys. Lett. {\bf 118B} 203 (1982).
\bibitem{UA1-1} G. Arnison {\em et al.} (UA1 Collaboration), Phys. Lett. 
{\bf 123B} 115 (1983).
\bibitem{ISR-2} T. Akesson {\em et al.} (AFS Collaboration), Phys. Lett. 
{\bf 118B}
185 (1982) ; A.L.S. Angelis {\em et al.}, Phys. Lett. {\bf 126B} 132 (1983).
\bibitem{UA2-6}
J.A. Appel {\em et al.} (UA2 Collaboration), Phys. Lett. {\bf 165B} 441 (1985).
\bibitem{bfthesis} 
F. Abe {\em et al.} (CDF Collaboration), Phys. Rev. {\bf 44D} 601 (1991).
\bibitem{kellisqcd}
R.K. Ellis, W.J.Stirling and B.R. Webber "QCD and Collider Physics"
Cambridge Monographs on Particle Physics, Nuclear Physics and Cosmology,
Cambridge Univ. Press, United Kingdom (1996).
\bibitem{LO}
B.L. Combridge, J. Kripfganz and J. Ranft, Phys. Lett. {\bf 70B} 234 (1977).
\bibitem{Greco}F. Aversa, P. Chiappetta, M. Greco, P. Guillet,
Phys. Lett. B {\bf  210}, 225 (1988); {\bf  211}, 465 (1988); Nucl. Phys.
{\bf  B327}, 105 (1989).
\bibitem{EKS}S. Ellis. Z. Kunszt, and D. Soper, Phys. Rev. Lett.
{\bf  62} 2188 (1989); Phys. Rev. Lett. {\bf  64} 2121 (1990);
Phys. Rev. D {\bf  40} 2188 (1989).
\bibitem{RSEP}
 S.K. Ellis, Z. Kunszt, D. Soper, Phys. Rev. Lett. {
\bf 69} 3615 (1992) and S. Ellis CERN--TH--6861--93
In proceedings of 28th Rencontres
de Moriond: QCD and High Energy Hadronic Interactions,
Les Arcs, France, 20-27 Mar 1993.
hep-ph/9306280.
\bibitem{SNOWMASS-me} B. Flaugher and K. Meier ``Proceedings 1990 Summer 
Study on High
Energy Physics, ed. E Berger. Singapore: World Scientific, 128
(1992).
\bibitem{UA2-2}
P. Bagnaia {\em et al.} (UA2 Collaboration), Z. Phys. {\bf C20} 117 (1983).
\bibitem{UA1-2} G. Arnison {\em et al.} (UA1 Collaboration), Phys. Lett. {\bf 132B}
 214 (1983).
\bibitem{UA2-3}
P. Bagnaia {\em et al.} (UA2 Collaboration), Phys. Lett. {\bf 138B} 430 (1984).
\bibitem{ISAJET}
F.E. Paige and S.D. Protopopescu, ISAJET, BNL 31987.
\bibitem{comp}
E. Eichten {\em et al.}, Phys. Rev. Lett. 50 811 (1983).
M. Abolins {\em et al.}, Proc. DPF summer study on
elementary particle physics, Snowmass, Co. 274 (1982).
\bibitem{UA1-3} G. Arnison {\em et al.} (UA1 Collaboration), Phys. Lett. 
{\bf 136B} 294 (1984).
\bibitem{UA1-6} G. Arnison {\em et al.} (UA1 Collaboration), Phys. Lett. 
{\bf 177B} 244 (1986) .
\bibitem{UA2-5}
J.A. Appel {\em et al.} (UA2 Collaboration), Phys. Lett. {\bf 160B} 349 (1985) .
\bibitem{UA2-4}
P. Bagnaia {\em et al.} (UA2 Collaboration), Phys. Lett. {\bf 144B} 283 (1984) .
\bibitem{UA2-7}
P. Bagnaia {\em et al.} (UA2 Collaboration), Phys. Lett. {\bf 186B} 452 (1987) .
\bibitem{UA1-4} G. Arnison {\em et al.} (UA1 Collaboration), Phys. Lett. 
{\bf 158B} 494 (1985) .
\bibitem{UA1-7} G. Arnison {\em et al.} (UA1 Collaboration), Z. Phys. {\bf C36}
 33 (1987).
\bibitem{UA2-8}
J.A. Appel {\em et al.} (UA2 Collaboration), Z. Phys. {\bf C30} 341 (1986).
\bibitem{UA1-5} G. Arnison {\em et al.} (UA1 Collaboration), Phys. Lett. 
{\bf 172B} 461 (1983).
\bibitem{altarelli}
G. Altarelli "QCD at the Collider"
Presented at Inst. School of Subnuclear Physics, Erice, Italy, Aug 3-14, 1983.
CERN TH-3733 (1983).
\bibitem{PYTHIA}
H.U. Bengtsson and T. Sjostrand, PYTHIA, Comput. Phys. Commun. {\bf 46} 
43 (1987).
\bibitem{HERWIG}
G. Marchesini and B. R. Webber, Nucl. Phys. {\bf  B310}, 461 (1988).
\bibitem{UA2-10}
J. Alitti {\em et al.} (UA2 Collaboration), Phys. Lett. {\bf 257B} 232 (1991).
\bibitem{JETSET}
T. Sjostrand and H.U. Bengtsson, JETSET, Comput. Phys. Commun. {\bf 43} 
(1987).
\bibitem{SNOWMASS} J.~Huth {\it et al.} ``Proceedings 1990 Summer 
Study on High
Energy Physics, ed. E Berger. Singapore: World Scientific, 134
(1992).
\bibitem{walter-alphas}
Walter T. Giele (Fermilab). FERMILAB-CONF-97-240-T, Jul 1997,
Proceedings of 5th International Workshop on Deep Inelastic 
Scattering and QCD (DIS 97), Chicago, IL, 14-18 Apr 1997,
hep-ph/9707300;
and W. T. Giele, E.W.N. Glover and J. Yu,
FERMILAB-PUB-127-T, DTP/95/52.  hep-ph/9506442 (1995).
\bibitem{JETRAD}
W. T. Giele, E.W.N. Glover and D.A. Kosower, Nucl. Phys. {\bf  B403}
633 (1993).
\bibitem{cdf-dijet1}
F. Abe {\it et al.} (CDF Collaboration), Phys. Rev. Lett. {\bf 62} 3020 (1989).
\bibitem{cdf-dijet2}
F. Abe {\it et al.} (CDF Collaboration), Phys. Rev. Lett. {\bf 64} 157 (1990).
\bibitem{cdf-dijet3}
F. Abe {\it et al.} (CDF Collaboration), Phys. Rev. {\bf D41} 1722 (1990).
\bibitem{cdf-dijet4}
F. Abe {\it et al.} (CDF Collaboration), Phys. Rev. Lett. {\bf 69} 2896 (1992).
\bibitem{cdf-dijet5}
F. Abe {\it et al.} (CDF Collaboration), Phys. Rev. {\bf D48} 998 (1993).
\bibitem{cdf-dijet6}
F. Abe {\it et al.} (CDF Collaboration), Phys. Rev. Lett. {\bf 71} 2542 (1993).
\bibitem{cdf-dijet7}
F. Abe {\it et al.} (CDF Collaboration), Phys. Rev. Lett. {\bf 74} 3538 (1995).
\bibitem{cdf-dijet8}
F. Abe {\it et al.} (CDF Collaboration), Phys. Rev. Lett. {\bf 77} 5336 (1996).
\bibitem{cdf-dijet9}
F. Abe {\it et al.} (CDF Collaboration), Phys. Rev. {\bf D55} 5263 (1997).
\bibitem{CDF-XT-analysis} F. Abe {\it et al.} (CDF Collaboration),
Phys. Rev. Lett. {\bf  70} 1376 (1993) .
\bibitem{CDF-Clustering}F. Abe {\it et el.} (CDF Collaboration),
Phys. Rev. {\bf  D45} 1448 (1992).
\bibitem{SUMET} F.~Abe {\it et al.} (CDF Collaboration),
Phys.Rev.  {\bf  D56} 2532 (1997); Phys.Rev. {\bf  D54} 4221 (1996);
Phys.Rev.Lett.{\bf  75} 608 (1995); 
Phys.Rev. {\bf  45} 2249 (1992).
\bibitem{SVX-det} C. Haber {\it et al.} Nucl. Instrum. Methods A289 388 (1990).
%
\bibitem{topmass} F. Abe {\it et el.} (CDF Collaboration),
Phys.Rev.{\bf  D50} 2966 (1994).
\bibitem{frag} F. Abe {\it et el.} (CDF Collaboration),
Phys.Rev. Lett. {\bf  65} 968 (1990).
\bibitem{WZXSEC} F.~Abe {\it et al.} (CDF Collaboration),
Phys.Rev. {\bf  59} 052002 (1999).
\bibitem{zpt} F.~Abe {\it et al.} (CDF Collaboration),
Phys.Rev.Lett.{\bf 84} 485 (2000). 
\bibitem{d0ue}
This uncertainty is not included in the 
inclusive jet cross section reported by the D0 collaboration.
\bibitem{topmass2} F. Abe {\it et el.}
(CDF Collaboration), Phys.Rev.{\bf  D80} 5720 (1998).
\bibitem{MINUIT} 
'MINUIT' A System for Function Minimization and Analysis of the 
Parameter Errors and Correlations.
F. James, M. Roos Comput.Phys.Commun.{\bf 10} 343 (1975).
\bibitem{D0inc1}
The D0 data is presented in~\cite{D0inc}.  The curve was provided
by Private Communication with Bob Hirosky (D0 Collaboration).
\bibitem{annrev} G. Blazey and B. Flaugher "Inclusive Jet 
and Dijet production at the Tevatron" to be published in Annu.
Rev. Nucl. Part. Sci. 1999; FERMILAB-PUB-99/038-E, hep-ex/9903058.
\bibitem{DSoper-mu-scale} Z.Kunszt and D.Soper, Phys. 
Rev. {\bf D46} 192 (1992),
 and http://zebu.uoregon.edu/~soper/soper.html, program version 3.4.
\bibitem{NMC} M.\ Arneodo {\em et al.} (NMC Collaboration), Nucl.\ Phys.\ 
\textbf{B483} 3 (1997);
M. Arneodo {\em et al.}, Phys. Lett. {\bf B364}, 107  (1995).
%
\bibitem{E665}  M.R.\ Adams {\em et al.} (E665 collaboration), Phys.\ Rev.\ 
\textbf{D54} 3006 (1996).
%
\bibitem{H1}   S.\ Aid {\em et al.} (H1 collaboration), Nucl. Phys. \textbf{%
B470} 3 (1996); C.\ Adloff {\em et al.}, Nucl. Phys. \textbf{B497} 3 (1997);
C.\ Adloff {\em et al.}, Z. Phys. \textbf{C72} 593 (1996);
S. Aid {\em et al.}, Nucl. Phys. {\bf B439}, 471 (1995).
%
\bibitem{ZEUS} M.\ Derrick {\em et al.} (ZEUS collaboration), Z. Phys. 
\textbf{C69} (1996) 607; M.\ Derrick {\em et al.}, Z. Phys. \textbf{C72}
399 (1996); 
J.~Breitweg {\it et al.}, Eur. Phys. J. {\bf C7}, 609 (1999),
hep-ex/9809005;
J. Breitweg {\em et al.}, Phys. Lett. 
\textbf{B407} 402 (1997); Paper N-645 presented at International Europhysics
Conference on High Energy Physics, HEP97, Jerusalem 1997.
M. Derrick {\em et al.}, Z.~Phys. {\bf C65}, 379 (1995).
%
\bibitem{E605}  G.\ Moreno {\em et al.} (E605 collaboration), Phys.\ Rev.\ 
\textbf{D43} 2815 (1991) 
%
\bibitem{E866}   E.A.\ Hawker {\em et al.} (E866 collaboration), Phys.\ Rev.\
Lett. \textbf{80} 3715 (1998), hep-ex/9803011.
%
\bibitem{Wasym}
F. Abe {\em et al.} (CDF Collaboration), Phys. Rev. Lett. {\bf 74}, 850 (1995);
F.~Abe \emph{et al.} (CDF Collaboration), Phys. Rev. Lett., {\bf 81}, 
5754 (1998), hep-ex/9809001.
\bibitem{CTEQ4M}
H.~L.~Lai {\it et al.}, Phys.\ Rev.\ D {\bf  55}, 1280 (1997).
\bibitem{MRST} A.D. Martin, R.G. Roberts, W.J. Stirling and T
 hep--ph/9803445, Eur.\ Phys.\ J. C {\bf  C 4}, 463 (1998)
\bibitem{kt706}
C. Bromberg {\it et al.} (E706 Collaboration), 
Influence of Parton k(t) on High-p(t) Particle Production and 
Determination of the Gluon
Distribution Function, 
In Proceedings of the 29th International Conference on High-Energy 
Physics (ICHEP 98), Vancouver, British Columbia, Canada, 23-29 Jul 1998.
Vol. 1 867.
\bibitem{CTEQ98} J. Huston {\it et al.}, Phys.Rev. {\bf  D58} 114034 (1998),
hep--ph/9801444.
%
\bibitem{YANG98} U.K Yang, A. Bodek, 
Phys.Rev.Lett. {\bf 82} 2467 (1999).
\bibitem{cdftopxsec}  F. Abe {\it et el.}
(CDF Collaboration), Phys.Rev.{\bf  D80} 2773 (1998).
%
\bibitem{Walter-pbarp}Walter Giele,
Next to leading order PQCD calculations,
10th Topical Workshop in proton-antiproton Collider Physics,
Fermilab, (1995). FERMILAB-CONF-95-169-T.
%
\bibitem{Sterman-pbarp} 
E. Laenen, G. Oderda, G. Sterman 
Phys. Lett. {\bf B438} 173 (1998).
\bibitem{dijetmass}
F. Abe {\em et al.} (CDF Collaboration), Phys. Rev. {\bf D61:091101} (2000);
F. Abe {\em et al.} (CDF Collaboration), in preparation.
%
\bibitem{NR}
W.H. Press, B.P. Flannery, S.A. Teukolsky, and W.T. Vetterling
"Numerical Recipies, The Art of Scientific Computing"
Cambridge University Press (1986).
%
\bibitem{Nason} S. Cantani, M.L. Mangano, P. Nason, L. Trentadue
Nucl. Phys. {\bf B478} 273 (1996).
\bibitem{lyons} Louis Lyons "Seleting between two hypotheses" OUNP-99-12.
\bibitem{covprob} G.D'Agostini "Probaility and measurement Uncertainty 
in Physics - a Bayesain primer" hep/ph/9512295v2.
%
%
\end{thebibliography}
\end{document}